\documentclass[12pt,a4paper,oneside, titlepage=false]{scrreprt}
\usepackage{mathrsfs}
\usepackage[hmargin=2.5cm,vmargin=3cm]{geometry}
 \usepackage{graphicx}
\usepackage{amsmath,amssymb,amsthm,mathtools}
\usepackage{bm}
  \usepackage{enumerate}
\usepackage{cite}
\usepackage{mathabx}
\usepackage{url}
\usepackage{hhline}
\newcommand{\bey}{\begin{eqnarray}}
\newcommand{\eey}{\end{eqnarray}}
\usepackage{etoolbox}
\usepackage{tocloft}
\theoremstyle{plain}

\theoremstyle{definition}
\newtheorem{definition}{Definition}[chapter]

\newtheorem{exercise}{Exercise}[chapter]

\theoremstyle{remark}

\newcommand{\po}{\prec}
\usepackage{lipsum}

\let\OLDthebibliography\thebibliography
\renewcommand\thebibliography[1]{
  \OLDthebibliography{#1}
  \setlength{\parskip}{0pt}
  \setlength{\itemsep}{2pt plus 0.3ex}
}
 \usepackage{enumitem}
\setlist[itemize]{leftmargin=1.2em}
\setlist[enumerate]{leftmargin=1.2em}
\newcommand{\op}[1]{\widehat{#1}}
\newcommand{\R}{\mathbb{R}}
\newcommand{\C}{\mathbb{C}}

\newcommand{\DM}{\mathrm{Diff}(\M)}
\newcommand{\DS}{\mathrm{Diff}(\Sigma)}
\newcommand{\M}{{\cal M}}
\newcommand{\E}{\mathcal E}
\newcommand{\HH}{\mathcal H}
\newcommand{\X}{\mathcal X}
\newcommand{\PP}{\mathcal P}
\newcommand{\oph}[1]{(\op{#1}_{t_1},\op{#1}_{t_2},\ldots,\op{#1}_{t_n})}
\newcommand{\T}{\mathbb{T}}
\newcommand{\EE}{\mathscr{E}}

\begin{document}
\title{Time and causality in quantum gravity}
\author {Charis Anastopoulos \\
{\small Department of Physics, University of Patras, 26500 Greece}\\
{\small Email: anastop@upatras.gr}
}
 
\publishers{%
\raisebox{-3cm}{%
\parbox{0.85\textwidth}{%
\centering\small
  These notes are based on lectures delivered at the
Training School \emph{Time, Causality and Memory in Quantum Physics}, held at
Stockholm University in May 2026, and organized by the COST Action ``Relativistic Quantum Information".

\medskip 

 Sara Butler edited the original manuscript and brought it into alignment with the lectures as delivered.
}}}

\date{\scriptsize \today}
 
\maketitle
\thispagestyle{empty}

\vfill

\noindent\makebox[\textwidth][c]{%
    \includegraphics[width=0.25\textwidth]{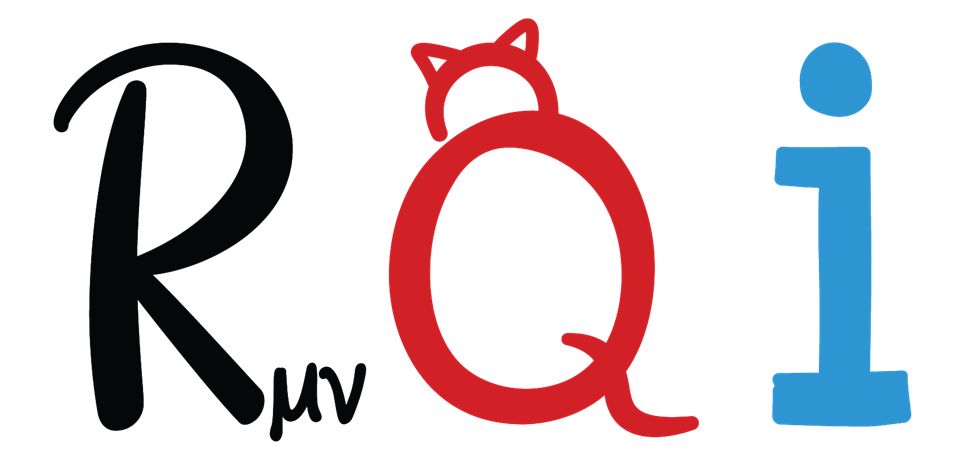}%
}

\newpage
\vspace*{3cm}
 \begin{abstract}

 \noindent\textbf{Abstract.} 
The problem of time in quantum gravity is often presented as a consequence of applying quantum theory to general relativity. These notes adopt a broader perspective. We first examine how time enters classical mechanics, relativity, quantum theory, and quantum field theory, distinguishing three of its principal aspects: causal order, temporal duration, and the present. This analysis shows that the conceptual difficulties associated with time do not arise only upon quantizing gravity; tensions between causal structure, clocks, observables, and measurement are already present in the theories from which quantum gravity is constructed.
We then survey the   responses to the problem of time from the full range of quantum gravity approaches. This includes canonical and Wheeler–DeWitt quantisation, path-integral and histories formulations, perturbative and background-dependent approaches, causal sets, and twistor theory. These programmes are compared according to whether they retain an external time, render time problematic through the quantisation procedure, or incorporate causal or temporal structure into their fundamental description.
Particular emphasis is placed on questions arising from quantum foundations and relativistic quantum information. We conclude by examining linearized gravity and weak-gravity quantum systems. This regime shows that the problem of time is not confined to Planck-scale physics but arises already in the attempt to provide a predictive and operationally meaningful quantum description of spacetime near flat geometry.
\end{abstract}

\setcounter{tocdepth}{1}
\clearpage
\begingroup
\small
\linespread{0.9}\selectfont
\setlength{\cftbeforesecskip}{1pt}
\setlength{\cftbeforesubsecskip}{0pt}
\setlength{\cftbeforetoctitleskip}{-10pt}
\setlength{\cftaftertoctitleskip}{7pt}

\tableofcontents
\endgroup

 \chapter{Introduction}
\section{Scope and perspective of these notes}

The problem of time in quantum gravity has been discussed extensively over the past several decades, and there are already several excellent reviews and monographs on the subject, most notably those by Isham \cite{Isham92}, Kucha\v r \cite{Kuchar}, and Anderson \cite{Anderson17}. The present notes do not attempt to replace these works. Instead, they pursue a different line of exposition, intended for a mixed audience of graduate students and researchers working in quantum gravity, quantum foundations, relativity, and quantum information theory.

The discussion avoids excessive technical detail. Quantum gravity remains a collection of competing programmes and ideas rather than a definitive theory, and technicalities can obscure the conceptual choices that distinguish them. The aim is to provide a broad overview of the principal approaches, ask how each treats time, and assess which confront the problem most directly and which largely set it aside, perhaps to their detriment. Although the literature is dominated by canonical quantum gravity, Wheeler--DeWitt quantisation, and related approaches, the discussion also encompasses histories formulations, causal sets, twistor theory, path-integral methods, and perturbative and background-dependent theories.

The problem of time is often presented as a technical consequence of canonical quantisation. These notes take a broader view: it also reflects the different roles assigned to time in classical mechanics, quantum theory, and general relativity. Problems involving time, causality, events, measurement, and spacetime observables already arise in the theories from which quantum gravity is constructed, including in weak-gravity regimes. Particular attention will therefore be given to quantum foundations, relativistic causality, and the operational description of measurements and events. From this perspective, the problem of time is not confined to Planck-scale physics.

A central theme of the notes is the distinction between temporal ordering and the measurement of temporal duration by clocks. Temporal order is conceptually distinct from duration and, in many respects, more primitive. Aristotle's classic definition of time as ``the number of motion according to before and after'' already presupposes an ordering relation. Newtonian physics incorporates temporal order and duration into a single background structure---the real line of absolute time---even though the order structure of the real line is mathematically distinct from its metric structure.

General relativity likewise represents causal ordering and proper time through the Lorentzian metric, but the two notions remain conceptually distinguishable. Indeed, the analysis of Ehlers, Pirani, and Schild \cite{EPS72} suggests that the causal structure of spacetime is operationally more primitive than its metric structure: the conformal light-cone structure determines causal relations before lengths and durations are specified. A systematic analysis of time at the frontiers of physics should therefore begin by distinguishing the structures that established theories gather together under the single name of time.

\subsubsection*{Why quantum gravity begins in Chapter 5}

A reader consulting the contents may wonder why quantum gravity first appears explicitly in Chapter~5. The preceding chapters establish a common framework for comparing programmes that begin from different assumptions about geometry, quantisation, observables, and causality. Chapter~4, in particular, develops the features of classical general relativity that make its quantisation distinctive: diffeomorphism invariance, constrained dynamics, hypersurface deformations, and the geometrical character of time.

This preparation also allows two central themes---the distinction between clocks and causal order, and the unavoidable role of events---to be formulated in classical physics, quantum theory, and relativity before asking how they change when geometry becomes quantum mechanical. Quantum theory itself raises problems concerning the measurement of time, the construction of quantum clocks, the temporal ordering of measurements, and the causal description of quantum operations in quantum field theory. These are not secondary complications unrelated to gravity; they form part of the conceptual background against which the problem of time must be understood.

With this framework in place, Chapter~5 divides approaches to quantum gravity into four broad categories: spin-2 and quantum-field-theoretic approaches, quantisation of geometry, theories retaining a fundamentally classical gravitational field, and structural reformulations of geometry or quantum theory. This classification permits a comparison not only of their mathematical methods, but also of their assumptions about time and of the questions each programme resolves or leaves open.

 \section{The trinity of time and its offspring}

The notion of time in physics comprises several logically distinct elements that are often conflated because they are often represented by a single mathematical object. A central theme of these notes is that some of the conceptual difficulties of quantum gravity arise precisely from this conflation. Broadly speaking, time has three distinct components.

\begin{enumerate}

\item {\em Order.}

The first and most primitive aspect of time is order: which events occur before others and which can influence which. Temporal order makes succession and causal influence possible, but it does not require a notion of duration or metric scale. One may know that one event precedes another without assigning a numerical interval between them. Mathematically, temporal order may be represented by the ordering induced by Newtonian time, by the causal partial order defined by relativistic light cones, or by more abstract structures in approaches to quantum gravity.

\item {\em Measure.}

The second aspect is measure: the assignment of durations to temporal intervals. In Newtonian mechanics, durations are represented by differences in a universal time parameter. In relativity, they are given by proper times along worldlines. This metric aspect of time makes it possible to compare intervals, define frequencies and energies, and formulate dynamical laws quantitatively. It presupposes more structure than temporal order alone.

\item {\em The present.}

The third aspect is the present, or, more precisely, temporal passage: the apparent transition of events from the future, through the present, into the past. This is perhaps the most elusive aspect of time. Classical dynamical equations typically describe complete histories and are, at the fundamental level, often time-reversal invariant, whereas experience presents the world through an evolving ``now''. Whether temporal passage corresponds to an objective physical structure, emerges from more fundamental processes, or belongs only to the standpoint of observers remains one of the deepest questions in the foundations of physics.

\end{enumerate}

Order, measure, and the present constitute the threefold structure of time that will guide these notes. Different physical theories relate these elements in different ways. Newtonian mechanics incorporates order and measure into a single universal time parameter, while general relativity encodes both in the causal and metric structure of spacetime, without introducing an observer-independent present. Quantum theory introduces further difficulties through superposition, measurement, and the tension between externally prescribed time and dynamically treated physical systems. When the word ``time'' is used in these notes, it should therefore be understood as referring not to a single concept, but to this threefold structure.

This trinity generates several closely related concepts that may be viewed as the ``offspring'' of time.

\begin{itemize}

\item {\em Causation.}
If A causes B, then A must be prior to B.  

\item {\em Irreversibility.} Macroscopic material processes exhibit a thermodynamic arrow of time.

\item {\em Locality.} Information propagates at finite speed.

\item {\em Memory and records.}
Physical systems contain records of the past but not of the future.  

\item {\em Prediction and retrodiction.}
Physics apparently distinguishes between predicting future events and reconstructing past ones, even when the microscopic laws are time-reversal invariant.

\item {\em Persistence and identity.}
The notion that an object persists through time presupposes an ordering structure connecting different events into a single history.

\end{itemize}

I will not be dealing with time's offspring  in these notes. This would require a set of lectures with a strong emphasis on the foundations of thermodynamics.

\section{Events, causal order and clocks}

The most primitive notion underlying any physical description of time is arguably not that of a clock, but that of an {\em event}. Physics does not describe time in isolation; rather, it describes relations between occurrences. Clocks, durations, and temporal coordinates are ultimately abstractions constructed from correlations between events. For this reason, it is useful to begin with a general description of events and the structures that may be defined on them.

By an ``event'' we mean a uniquely identifiable occurrence with definite physical characteristics. In classical mechanics, events are often represented geometrically as intersections of worldlines. For example, a particle-detection event may correspond to the intersection between the worldline of a particle and the worldline of  a detector. More generally, one may define an event as the first intersection of a worldline with a specified timelike hypersurface. A familiar example is the crossing of the finish-line world tube by a runner in a marathon.

In quantum theory, however, trajectories are not themselves observables. Definite properties are attributed only to measurement outcomes. For this reason, events are more naturally identified with macroscopic records. A particle-detection event, for example, is the ``click'' of a detector, or more generally, a localized irreversible change in a macroscopic apparatus. In this operational sense, events correspond to physical records whose occurrence can be unambiguously identified.

Suppose that a physical system admits a set $E$ of possible events. We denote events by Greek letters $\alpha,\beta,\gamma,\ldots$. The trinity of time can then be represented by mathematical structures defined on the space $E$.

\subsection*{Causal order}

The most primitive temporal structure on $E$ is causal order. We write
\[
\alpha \prec \beta
\]
if the event $\alpha$ occurs prior to the event $\beta$. A causal order on $E$ is therefore an assignment of the relation $\prec$ to pairs of events.

The relation $\prec$ satisfies the properties of a partial order:
\begin{enumerate}
\item {\em Irreflexivity:} $\alpha \nprec \alpha.$

\item {\em Asymmetry:}
if $\alpha \prec \beta$, then $\beta \prec \alpha$ is false.

\item {\em Transitivity:}
if $\alpha \prec \beta$ and $\beta \prec \gamma$, then $\alpha \prec \gamma$.
 
\end{enumerate}

A partial order does not require all pairs of events to be comparable. Physically, we distinguish two possibilities for unrelated events:
\begin{itemize}
\item events may be {\em simultaneous}, written $\alpha \sim \beta$, 
 
\item or they may be {\em causally disconnected}, written $\alpha | \beta$.
 
\end{itemize}

By a {\em causal order} we mean the partial order $\prec$ for precedence  together with an additional possibilities of  simultaneous and causally disconnected pairs of events.

We denote the set of all causal orders on $E$ by $CO(E)$. For example, if $E=\{\alpha,\beta\}$
there are four possible causal structures:
\begin{itemize}
\item $M_1=\{\alpha \prec \beta\}$,
\item $M_2=\{\beta \prec \alpha\}$,
\item $M_3=\{\alpha | \beta\}$,
\item $M_4=\{\alpha \sim \beta\}$.
\end{itemize}

A causal order defines a {\em time order} if no pair of events is causally disconnected. We denote the set of all time orders on $E$ by $TO(E)$, so that
$
TO(E)\subset CO(E)$.

\subsection*{Clocks and duration}

Causal ordering alone does not define temporal duration. A second structure is required in order to measure intervals between events. This role is played by clocks.

There is one important sense in which time behaves as a continuum: clocks assign numerical values to durations. Strictly speaking, all physical clocks produce discrete outputs, since measurements occur in multiples of some elementary period. However, by assuming that this period may become arbitrarily small, durations are represented mathematically by real numbers.

\begin{definition}
A clock $T$ is a pair of functions $(T^{(1)},T^{(2)})$.
\\
The function
$T^{(1)}:E\rightarrow \mathbb R^+$
assigns a duration $T^{(1)}(\alpha)$ to each event $\alpha$.
\\
The function
$
T^{(2)}:E\times E\rightarrow \mathbb R^+
$
assigns a time interval $T^{(2)}(\alpha,\beta)$ to each pair of events satisfying $\alpha \prec \beta$.
\end{definition}

In many situations, the space $E$ is generated by elementary instantaneous events, for which $T^{(1)}(\alpha)=0$.
In this case, only the interval function is needed, and we write simply
$
T(\alpha,\beta)
$.

It is then natural to require the additivity property
\bey
\textbf{Clock additivity.}
\;\;
\mbox{If }
\alpha\prec\beta
\mbox{ and }
\beta\prec\gamma,
\mbox{ then}
\;\;
T(\alpha,\gamma)
=
T(\alpha,\beta)
+
T(\beta,\gamma).
\label{additive}
\eey
A clock in this sense corresponds to one worldline. As in special
relativity, the reading along a curve is the proper time of an observer traversing it. Additivity   expresses   that
splitting a given path into segments gives additive readings.

This definition of a clock presupposes the existence of a causal order. In this sense, causal ordering appears conceptually prior to duration. However, one may reverse the logic and define causal order directly from clock readings.

To this end, we extend the clock function to
$
\bar T:E\times E\rightarrow \mathbb R\cup\{N\},
$
where $N$ denotes ``no reading'',
\bey
\bar T(\alpha,\beta)
=
\left\{
\begin{array}{cl}
T(\alpha,\beta),
&
\alpha\prec\beta,
\\
-T(\alpha,\beta),
&
\beta\prec\alpha,
\\
0,
&
\alpha\sim\beta,
\\
N,
&
\alpha|\beta.
\end{array}
\right.
\eey

Thus, causally connected events are assigned opposite signed durations depending on orientation, simultaneous events correspond to zero duration, and causally disconnected events admit no clock reading.

This motivates a more abstract definition.

\begin{definition}
A clock on a space $E$ of elementary events is a map
\[
T:E\times E\rightarrow \mathbb R\cup\{N\},
\]
such that:
\begin{enumerate}
\item if $T(\alpha,\beta)\neq N$, then
$
T(\beta,\alpha)=-T(\alpha,\beta),
$

\item if $T(\alpha,\beta)=N$, then
$
T(\beta,\alpha)=N,
$

\item if
$
T(\alpha,\beta),T(\beta,\gamma)\geq 0,
$
then
$
T(\alpha,\gamma)
=
T(\alpha,\beta)+T(\beta,\gamma).
$
\end{enumerate}
\end{definition}

With this definition, every clock determines a unique causal order. We denote the space of clocks on $E$ by $C(E)$.

Two clocks $T_1$ and $T_2$ are said to be equivalent if they define the same causal order. Hence,
\[
CO(E)\simeq C(E)/\sim.
\]
In this sense, causal order may be viewed as the structure common to all equivalent clocks.
It is important to keep in mind that the clock may give no reading. If two events are
not comparable, they have no time duration between them; this is why we introduce the
symbol $N$. More generally, whenever time is associated to an event as an observable, we
must always account for the possibility that the event will not happen at all.

\subsection*{The present}

Besides order and duration, our intuitive notion of time also involves the notion of a moving present. At any given moment, events appear to be divided into those that have already occurred, those occurring now, and those that have yet to occur. As time ``passes'', the set of future events decreases while the set of past events increases.

The boundary between past and future defines the \emph{present}. Given a causal structure that includes a simultaneity relation, the set of possible presents may be identified with the quotient space
$
E/{\sim},
$
whose elements are equivalence classes of simultaneous events.

A central question is whether the present corresponds to physical
structure beyond causal ordering. Unlike temporal order and duration,
there is no agreed mathematical or operational representation of a
``moving now'' in fundamental physics. There have nevertheless been
attempts to assign the present a physical role. Heisenberg associated
quantum measurement with the transition from the possible to the
actual \cite{Heisenberg}. Later authors have identified this transition more explicitly
with the present \cite{Stapp77, BlanchardJadczyk1995, Saunders95, Dyson04, ElRo10}. From this perspective, the creation of
definite records through measurement provides a possible physical
interpretation for the present.

In these notes, we will focus primarily on order and duration, which enter more directly into the problem of time in quantum gravity. We will refer occasionally to the notion of the present, but we note that---if the Heisenbergian interpretation is correct---its proper domain of investigation is the quantum measurement problem.

 \chapter{Time in classical physics}

 \section{Spacetime and events}
 
Newton's idea of an {\em absolute time} is one of the foundational pillars of his mechanics. As he wrote in the {\em Principia Mathematica}:

\begin{quote} {\small
Absolute, true, and mathematical time, in and of itself and of its own nature,
flows uniformly without reference to anything external, and by another name
is called duration.

Relative, apparent, and common time is any sensible and external measure of
duration by means of motion (whether accurate or not), such as an hour, a day,
a month, a year, and is commonly used instead of true time.}
\end{quote}

Newton treats time is an objective structure of the world, 
  neither a relation between physical systems nor a feature of human perception.
  The same holds for space.

\begin{quote} {\small
Absolute space, in its own nature, without relation to anything external,
remains always similar and immovable.

Relative space is some movable dimension or measure of the absolute spaces,
which our senses determine from the positions of bodies and which is commonly
taken for immovable space; for example, a space defined by its position
relative to the earth.}
\end{quote}

We will now formalize Newton's intuitions in a way that is compatible with our modern understanding, informed by general relativity. 

We define the spacetime $\M$ as a manifold, with elements $x = (x^0, x^1, x^2, x^3)$, where $x^0$ is a temporal and $x^i$ are spatial  coordinates.  In Newtonian physics, all physical events are represented in spacetime. This means that there exist embedding maps ${\cal X}: E \rightarrow S(\M)$, where $S(\M)$ is the set of   subsets of $M$\footnote{More precisely, $S(\M)$ is the set of Borel subsets of $\M$, that is, $S(\M)$ is generated from the open subsets of $\M$ by through countable unions, countable intersections, and complements.}. Hence, ${\cal X}(\alpha)$ is a  subset of $\M$, the {\em spacetime support} of the event $\alpha$. 

We often  assume that some events in $E$ are {\em elementary}. The intrinsic designation of such events may vary from theory to theory. The key point is that the spacetime support of an elementary event $\alpha$ consists of single spacetime points, that is, ${\cal X}(\alpha) = \{x\}$. If $E$ consists only of elementary events, then we can view ${\cal X}$ as a map from $E$ to $M$, so that ${\cal X}(\alpha) = x$.

The key idea of Newton is that the spacetime $\M$ carries absolute structures which generate the temporal properties in any space of events through the embedding maps ${\cal X}$. These absolute structures are the following.
\begin{itemize}
\item The {\em causal structure}: a partial order $\po$ on $\M$, denoting before and after. 
Then, we can define the {\em causal curves} of $M$ as all continuous maps $\lambda: [0,1]\rightarrow \M$, such that $\lambda(s_1) \po \lambda(s_2)$ for all $s_1 < s_2$. We denote the set of causal curves on $\M$ by ${\cal P}_{\po}(\M)$.

\item The {\em local clocks}: a map $\tau: {\cal P}_{\po}(\M) \rightarrow \R^+$ that assigns to each causal curve its duration; $\tau(\lambda)$ is interpreted as the reading of an ideal clock of an observer traversing the curve. 

\item The {\em global time}: a map $t: \M \rightarrow \R$, which assigns a time parameter to each spacetime point, so that $t(x_1) < t(x_2)$ if $x_1 \po x_2$.

\end{itemize}

The causal order and duration of events are defined as pullbacks of those structures through the map ${\cal X}$. For $E$ consisting solely of elementary events:

\begin{itemize}
\item $\alpha \po \beta$ if ${\cal X}(\alpha) \po {\cal X}(\beta)$.

\item Any global time function $t$, defines a clock function $T(\alpha, \beta) = t({\cal X}(\beta)) - t({\cal X}(\alpha))$. 

\end{itemize}
 
\begin{exercise}
How can one use the spacetime causal order and global time, to define the causal order of non-elementary events and associated clock functions?
\end{exercise}

\subsubsection*{Newtonian spacetime}
In Newton's theory, the spacetime $\M= \R^4$, and there is a preferred global time, the {\em absolute time}, defined by $t(x) = x^0$.
\begin{itemize}

\item  Causal ordering is defined in terms of the absolute time: $x_1 \po x_2$ if $t(x_1) < t(x_2)$. 

\item Two  spacetime points are instantaneous if $t(x_1) = t(x_2)$. Hence, 
 all spacetime points with the same value of $t$ define an instant of time. 

\item Local clocks measure the change in absolute time. For any curve $\lambda: [0,1]\rightarrow \M$, $\tau(\lambda) = t(\lambda(1)) - t(\lambda(0))$.

\end{itemize}

To summarize, the temporal relations in any set of events are determined by (i) the spacetime temporal structures and (ii) the embedding maps ${\cal X}$. In Newtonian theory---and as we will see, also in special relativity---the former are absolute and unchanged. However, the latter are not: in most cases they are dynamical variables. To understand this, we must first analyze the structure of the space of events. 

\subsubsection*{Historical note.}
In relativity, it has become standard to designate spacetime points as “events.” At first sight, this terminology appears to conflict with the usage adopted here. However, the apparent tension reflects a historical shift in usage and semantics, and not a fundamental conceptual conflict.

Minkowski had a fully geometric conception of spacetime, but he did not use the word ``events" for spacetime points \cite{Minkowski}. Einstein does occasionally use the compound term “Punktereignis” (point-event) in his foundational paper \cite{Einstein}. However, this expression simply connects physical occurrences to the points of the spacetime continuum used to represent them.  
%\begin{quote}
%{\small I will call a point in space at a given time, i.e. a system of values $x, y, z, t$
%a worldpoint. The manifold of all possible systems of values $x, y, z, t$ will be
%called the world.} 
%\end{quote}  

%In his popular exposition of relativity \cite{Einstein2}, Einstein uses the term “event”   in a more flexible manner. At times, events are treated as points of spacetime, %specified by four coordinates and forming the elements of a continuum. Elsewhere, Einstein emphasizes their physical meaning, describing events as occurrences—coincidences of %material systems, such as the meeting of a train and a signal or the emission of a light flash. At the same time, he draws a clear distinction when he writes:
%\begin{quote}
%{\small Every physical description resolves itself into a number of statements, each of which refers to the space-time coincidence of two events A and B.}
%\end{quote}
%This formulation makes explicit that, at a fundamental level, events are understood as physical coincidences, even when represented geometrically.
 
  The term “event” was introduced systematically by Eddington \cite{Eddington}, who wrote:
\begin{quote}
{\small A point in this space-time, that is to say a given instant at a given place, is called an “event.” An event in its customary meaning would be the physical happening which occurs at and identifies a particular place and time. However, we shall use the word in both senses, because it is scarcely possible to think of a point in space-time without imagining some identifying occurrence.}
\end{quote}
A similar viewpoint is found in the early (1941) textbook of Landau and Lifshitz \cite{LL1}:
\begin{quote}
{\small We shall frequently use the concept of an event. An event is described by the place where it occurred and the time when it occurred. [...] It is frequently useful [...] to use a fictitious four-dimensional space [...] In this space events are represented by points, called world points.}
\end{quote}
The operational interpretation persists up to Misner, Thorne, and Wheeler (1973) \cite{MTW}, who assert that the manifold structure of spacetime arises as an idealization of the set of identifiable physical events.

A decisive shift occurs at the same year, when Hawking and Ellis write \cite{HE73}:
\begin{quote}
{\small The mathematical model we shall use for space-time, i.e. the collection of all events, is a pair $(\mathcal{M}, g)$ where $\mathcal{M}$ is a connected four-dimensional manifold.}
\end{quote}
Here, no distinction is made between physical occurrences and abstract points: the spacetime manifold is {\em defined} as the collection of all events. This usage becomes standard in later textbooks, for example, Schutz  \cite{Schutz} and  Wald  \cite{Wald}.

Thus, the modern identification of events with arbitrary points of the spacetime manifold is not part of the original conceptual framework of relativity, but the result of a gradual reinterpretation. The physics does not depend on the usage we adopt, provided we are clear about what we mean. Here, I employ the older relativity sense of the term: 
 an event is a happening and spacetime is built from the geometric structure
carrying these events.

 \section{Hamiltonian mechanics}
 We cannot describe events purely abstractly, we need to know what they are events of, that is, we need to know their ontology. Physicists up to the nineteenth century used to think in terms of particles, with all their properties being motional/geometrical. Then, the idea of fundamental fields gained ground, and it was eventually realized that fields have properties with no spatial representation, the so called {\em internal degrees of freedom}. These correspond to gauge symmetries and the conserved charges of high energy physics. However, all these classical ontologies can be described within the general framework of Hamiltonian mechanics and symplectic geometry.
 
The main elements of this description are the following. 
 
 \begin{enumerate}
 \item The  {\em state space} of a classical system is a manifold $\Gamma$, which may be either finite- or infinite-dimensional. Points of $\Gamma$---denoted by $\xi$ correspond to {\em microstates}, i.e.,  complete specification of the system's properties at a moment of time. 
 \item {\em Single-time observables} correspond to functions $f: \Gamma \rightarrow \R$. An idealized instantaneous measurement determines the specific value $f$ of $f(\xi)$.
 \item {\em Single-time propositions} correspond to subsets $C$ of $\Gamma$. We call them propositions because the logical connectives for propositions of the form ``the system is in a microstate $\xi \in C$" mirror set-theoretic operations between subsets of $\Gamma$.
\end{enumerate}
 In classical mechanics, the state space is  symplectic manifold, that is, it is equipped with a {\em symplectic form} $\omega$.
 \begin{definition}
A symplectic form on a manifold $\Gamma$ is a 2-form
\[
\omega=\frac12\,\omega_{ab}(\xi)\, d\xi^a \wedge d\xi^b
\]
that satisfies
\begin{enumerate}
\item non-degeneracy: $\det(\omega_{ab})\neq 0$, and 
\item closedness: $\partial_i \omega_{jk}
+\partial_j \omega_{ki}
+\partial_k \omega_{ij} = 0$ ($d \omega = 0$).
\end{enumerate}
\end{definition}
 
 For a finite dimensional manifold $\Gamma$, the local structure of the symplectic form is provided by {\em Darboux's theorem}. There exist local coordinates 
$\xi^a = (q^1,\ldots,q^n,p_1,\ldots,p_n)$ such that $\omega = \sum_{i=1}^n dp_i \wedge dq^i$. Hence, we can split the coordinates of a symplectic manifold into configuration variables and their conjugate momenta. Evidently, only even-dimensional manifolds are symplectic.

\subsubsection*{Examples}
\begin{enumerate}
\item On $\mathbb{R}^{2n}$ with coordinates $(q^i,p_i)$, $i = 1, \ldots, n$,  the canonical symplectic form is
$
\omega =   dp_i \wedge dq^i$. 
A system of $N$ particles in three dimensions is described by the state space $\Gamma = \R^{6N}$.

\item  Let $(\theta,\phi)$ be spherical coordinates on the unit sphere $S^2$.  
A symplectic form is given by the area form
\[
\omega = k \sin\theta \, d\theta \wedge d\phi ,
\]
rescaled by a constant $k$. The sphere is the state-space for the classical analogue of the spin.

 \item Let $\Gamma_1$ and $\Gamma_2$ be manifolds with symplectic forms $\omega_1$ and $\omega_2$, respectively. Then, their Cartesian product 
 $\Gamma_1 \times \Gamma_2$ is symplectic with symplectic form $\omega = \omega_1 \otimes I + I \otimes \omega_2$. In local Darboux coordinates,  $(q_i, p_i)$ for $\Gamma_1$
and $(Q_j, P_j)$ for $\Gamma_2$, 
\bey
\omega
=
  dp_i \wedge dq^i
+
  dP_j \wedge dQ^j . \nonumber 
\eey

 \item  
Let $Q$ be a smooth manifold with local coordinates $q^i$.  
The cotangent bundle $T^*Q$ carries a canonical symplectic structure.
On $T^*Q$, introduce coordinates $\xi^a  = (q^i,p_i)$, where $p_i$ are the 
components of a covector in the cotangent space.  

The canonical 1-form (Liouville form) is an one-form on $\Gamma$, defined by
\bey
\theta = \theta_a d \xi^a = p_i \, dq^i,
\eey
The symplectic form is then defined by $\omega = d \theta = dp_i\wedge dq^i$. This is the most common type of symplectic manifold in physics: if $Q$ is the configuration space, then $T^*Q$ is the Hamiltonian state space.

\item Darboux's theorem implies that the symplectic form can always be written locally as $\omega = d \theta$, where $\theta = p_i \, dq^i$. However, this definition does not work globally, unless the second cohomology group of $\Gamma$ vanishes. This is the case for $\Gamma = T^*Q$ and $\Gamma = \R^{2n}$ discussed above, but not for the sphere.
\end{enumerate}

\begin{exercise}
Show that a complex Hilbert space is a symplectic manifold, the symplectic form defined through the inner product.
\end{exercise}
 
 The most important property of the symplectic form is that it enables a duality relation between observables and transformations on the state space. The relevant transformations $\Gamma$ in a manifold are diffeomorphisms, which are generated by vector fields. We define the integral curves of a vector field $X = X^a \partial/\partial \xi^a$ on $\Gamma$ as the solutions $\xi^a(s)$ of the equation
 \bey
 \frac{d}{ds} \xi^a(s) = X^a(\xi(s)),
 \eey
 such that $\xi^a(0) = \xi^a$. The map $\sigma^X_s: \xi \rightarrow \xi(s)$ is an one-parameter of diffeomorphisms generated by the vector field $X$. 
 
 In a symplectic manifold $\Gamma$, we define the {\em Hamiltonian vector field} $X_f$ associated to the function $f$ on $\Gamma$ by 
 \bey
 \omega_{ab} X^a_f = \partial_b f.
 \eey
 Writing by $\omega^{ab}$ the inverse matrix of $\omega_{ab}$, we write
 $X^a_f = \omega^{ab}\partial_b f$, or in Darboux coordinates
 \bey
 X_f = \frac{\partial f}{\partial p_i} \frac{\partial}{\partial q^i} - \frac{\partial f}{\partial q^i} \frac{\partial}{\partial p_i}.
 \eey
 The diffeomorphisms generated by Hamiltonian vector fields are called {\em canonical transformations} or {\em symplectomorphisms}.
 
 This allows us to define the Poisson bracket of two functions $f$ and $g$ as
 \bey
 \{f, g\} = - \{g, f\} = X_g(f) = \omega^{ab} \partial_af \partial_bg,
 \eey
 and in Darboux coordinates
 \bey
  \{f, g\} = \frac{\partial f}{\partial q^i} \frac{\partial g}{\partial p_i} - \frac{\partial f}{\partial p_i} \frac{\partial g}{\partial q^i}.
 \eey
By definition, $\{\xi^a, \xi^b\} = \omega^{ab}$, or in Darboux coordinates
\bey
\{q^i, q^j\} = 0, \;\; \{p_i, p_j\} = 0, \;\; \{q^i, p_j\} = \delta^i_j.
\eey
 
 \begin{exercise}
 Show that $[X_f, X_g] = X_{\{f, g\}}$.
 \end{exercise}

 \begin{exercise}
Show that ${\cal L}_{X_f}\omega = 0$, where ${\cal L}_X$ is the Lie derivative with respect to the vector field $X$.
\end{exercise}

The Newtonian idea of an absolute time is implemented through the following axiom.
\\ \\
\noindent \textbf{ Time-evolution principle:} A closed system evolves 
  with respect to the absolute time $t$ through an one-parameter group of canonical transformations $\sigma^{X_h}_t$, where   $h$ the system's Hamiltonian, a function on $\Gamma$ that generates time translations.
\\ \\
 This means that the one-parameter group of diffeomorphisms generated by a function $h$ on $\Gamma$ satisfies Hamilton's equations,
 \bey
 \frac{d\xi^a}{dt} = X^a_h(\xi) = \{\xi^a, h\} = \omega^{ab}\partial_b h, 
 \eey
 or equivalently,
 \bey
 \frac{dq^i}{dt} = \frac{\partial h}{\partial p_i}, \; \;\; \frac{dp^i}{ds} = - \frac{\partial h}{\partial p_i}.
 \eey
The absolute time is the preferred parameter for all paths in the classical state space. If we use any alternative parameter $s$ such that  $t = t(s)$, the resulting evolution equations 
 \bey
 \frac{d\xi^a}{ds} = \dot{t}(s) \omega^{ab}\partial_b h
 \eey
 involve essentially a time-dependent Hamiltonian $\dot{t}(s) h$. 
 
 This point is crucial for our later discussion of general relativity. The parameter of the canonical transformation generating time evolution is, physically, the Newtonian time—an important physical
input. If we insist on a closed system, the Hamiltonian must not be time-dependent,
so that we have an one-parameter group of time translations.
But this is only
guaranteed if we use a specific time coordinate. If we use any other coordinate, we are
not guaranteed a time-independent Hamiltonian. 
 
Whenever we have ignorance about the initial conditions, we describe the system by a probability distribution $\rho$ on $\Gamma$. This satisfies {\em Liouville's equation} with respect to the absolute time $t$,
\bey
\frac{d\rho}{dt} = \{h, \rho\}.
\eey
 
 \section{Classical histories and the action principle}
Hamiltonian mechanics represents time-evolution in terms of microstates evolving in time. An alternative description describes evolution in terms of histories: we identify all possible evolution scenarios of the system, and then we select the ones the physically realized ones by some criterion.

In this set-up, histories are defined as paths $\gamma: \R \rightarrow \Gamma$. In general, it is not necessary to take the path parameter $s$ to coincide with Newtonian time $t$. We will denote the space of such paths by $\Pi$. History observables are functions $F: \Pi \rightarrow \R$, and history propositions correspond to subsets of $\Pi$. 

%Note that each function $f$ on the state space $\Gamma$ generates an one parameter family $F_t$ of history observables, defined by
%\bey
%F_t[\xi(\cdot)] = f[\xi(t)].
%\eey

The selection of physically realized paths is traditionally achieved through the least-action principle. Whenever the Liouville form $\theta$ can be defined globally, the action is a function on $\Pi$, 
\bey
S[\xi(\cdot)] =  \int \lambda(t) \left[\theta_a \dot{\xi}^a - h(\xi) \right]dt, \label{depar}
\eey
where the path are parameterized by the Newtonian time $t$, and $\lambda(t)$ is an arbitrary function of time that vanishes at $\pm \infty$.

Variation of the action yields
\bey
\delta S =  \int dt \lambda(t) \frac{d}{dt}\left(\theta_a \delta \xi^a\right)    + \int dt \lambda(t)\left(\omega_{ab}\dot{x}^a - \partial_b h \right)\delta \xi^b, \label{vars}
\eey
where $\omega_{ab} = \partial_a \theta_b - \partial_b \theta_a$. To obtain Hamilton's equations of motion, $\omega_{ab}\dot{x}^a = \partial_b h$, from the condition $\delta S = 0$, we need to choose $\lambda(t)$ and restrict the possible variations of the action so that the first term in Eq. (\ref{vars}) vanishes. 

The standard choice is to take $\lambda(t)$ the characteristic function of an interval $[t_1, t_2]$. Then,
\bey
\int dt \lambda(t) \frac{d}{dt}\left(\theta_a \delta \xi^a\right) = \theta_a(\xi(t_2))\delta \xi^a(t_2) - \theta_a(\xi(t_1))\delta \xi^a(t_1) = p_i(t_2)\delta q^i(t_2) - p_i(t_1) \delta q^i(t_1) \nonumber
\eey
and it suffices to take the variations $\delta q^i(t)$ to vanish at $t = t_1$ and $t = t_2$. 

Alternatively, we can choose as $\lambda(t)$ a smooth function that switches the action on and off. Then, the first term for $\delta S$ becomes $-\int dt \dot{\lambda}(t)\theta_a \delta \xi^a$, and Hamilton's equations of motion are obtained at the adiabatic limit where $\dot{\lambda} \rightarrow 0$. For example, taking $\lambda(t) = \frac{1}{\cosh(at)}$ for $a > 0$ yields $\dot{\lambda}(t) = -a \sinh(at)/\cosh^2(at)$, and the adiabatic limit corresponds to $a \rightarrow 0$. 

\subsubsection*{Remarks}
\begin{enumerate}
\item We can choose to parameterize the paths by an arbitrary parameter $s$ rather than the Newtonian time $t$. Then, the action becomes a function also of the function $t(s)$
\bey
S[\xi(\cdot), t(\cdot)] =  \int_{s_1}^{s_2} \left[\theta_a \dot{\xi}^a - \dot{t} h(\xi) \right]ds. \label{param1}
\eey
It is straightforward to show that variation of the action with respect to $\xi(\cdot)$ leads to Hamilton's equation, while variation with respect to $t(\cdot)$ yields energy conservation:  $\frac{d}{ds}h = 0$.

\item We can interpret the action (\ref{param1}) as a function on paths in an extended state space $\bar{\Gamma}$ with elements $(\xi^a, t, p_t)$, with symplectic form $\bar{\omega} = \omega + dp_t\wedge dt$. Then, the action 
\bey
S[\xi(\cdot), t(\cdot), p_t(\cdot)] =  \int_{s_1}^{s_2} \left(\theta_a \dot{\xi}^a + p_t \dot{t} \right)ds. \label{param2}
\eey
describes a system on $\bar{\Gamma}$ with zero Hamiltonian, but subject to the constraint $p_t + h(\xi) = 0$. We will discuss constraints in more detail in Chapter 4. The action (\ref{param2}) is known as the {\em parameterized form} of the action (\ref{depar}).

\item If a global symplectic potential $\theta$ does not exist, then the phase space action cannot be defined globally. In this case, one may define a form of the action using a Wess--Zumino method \cite{HenneauxTeitelboim}.  Let $\xi(t)$ be a trajectory in $\Gamma$ with fixed endpoints, and introduce an auxiliary parameter $\tau\in[0,1]$ defining a two--dimensional surface $\Sigma \subset \Gamma$ through an extension $\xi^a(t,\tau)$ such that $\xi^a(t,1)=\xi^a(t)$ and $\xi^a(t,0)=\xi^a_{ref}(t)$ for some reference curve. The action is then defined by
\bey
S[\xi(\cdot)] = \int_{\Sigma} \omega - \int_{t_1}^{t_2} h(\xi(t))\, dt
     = \int_{t_1}^{t_2} dt \int_0^1 d\tau \, \omega_{ab}(\xi)\,
       \partial_{\tau} x^a \partial_t x^b
       - \int_{t_1}^{t_2} h(\xi(t))\, dt .
       \eey
Different choices of the surface $\Sigma$ with the same boundary change the action by the integral of $\omega$ over a closed surface, which is irrelevant for the classical variational principle. When $\omega = d\theta$ globally, Stokes' theorem reduces this expression to the standard phase--space action.

\item If the state space is a contangent bundle $T^*Q$, we can write the action in an interval $[t_1, t_2]$ as $S = \int_{t_1}^{t_2} dt (p_a\dot{q}^a - h)$. Assuming 
that only $q^a$ is constrained at the endpoints, variation over $p_a$ yields $\dot{q}^a = \partial h/\partial p_a$. If we can solve this equation, to express $\dot{q}^a$ in terms of $p_a$, we can remove the dependence of the action on $p_a$, and express it as a functional over paths over the configuration space $Q$. Then,
\bey
S = \int_{t_1}^{t_2} dt L(q, \dot{q}),
\eey
where $L$ is the Lagrangian, defined as a function on the tangent bundle $TQ$. Variation of this action yields the Euler-Lagrange equations.
\bey
\frac{d}{dt} \frac{\partial L}{\partial \dot{q}^a} - \frac{\partial L}{\partial q^a} = 0.
\eey
The action and the Lagrangian are well defined globally only if a global canonical form exists. Otherwise, they are defined only in local patches. However, the equations of motion remail well defined globally. 
\end{enumerate}

 \begin{exercise}
In $\R^{2n}$, we can choose for $\theta = (1 - a) p_i dq^i - a q^idp_i$ for an arbitrary $a \in \R$. How does this choice affect the least-action principle?
 \end{exercise}
 
 \subsubsection{Symplectic histories formulation}
An alternative approach to classical histories, due to Savvidou \cite{Sav99, Sav10}, proceeds from the introduction of a symplectic form on the space $\Pi$ of histories,
\bey
\Omega[\xi(\cdot)] = \int dt \; \omega_{ab}[\xi(t)] \delta \xi^a_t \wedge \delta \xi^b_t. \label{hisav}
\eey
For any function $f : \Gamma \rightarrow \R$, we can define a family of functions $F_t: \Pi \rightarrow \R$, by
\bey
F_t[\xi(\cdot)]\ = f[\xi(t)].
\eey
The symplectic form generates a Poisson bracket among history observables 
\bey
\{\xi^a_t, \xi^b_{t'}\} = \omega^{ab} \delta(t, t'). \label{histpo}
\eey
 It is then straightforward to show that Hamilton's equations of motion are obtained from the condition
 \bey
 \{S, \xi^a_t\} = 0, \label{heqm}
 \eey
 where $S[\xi(\cdot)] = \int  dt \left(\theta_a \dot{\xi}^a - h  \right)$. In this picture, classical paths remain invariant under the symplectic transformations generated by thee action.
 
 A crucial point  is that each component of the action, $V = \int dt \theta_a \dot{\xi}^a $ and $H = \int dt h$,  generates an one parameter group of symplectomorphisms that correspond to time translations. We straightforwardly calculate
 \bey
 \{V, \xi^a_t\} = \dot{\xi}^a_t
 \eey
  which implies that $V$ generates the symplectic transformation $\xi^a_t \rightarrow \xi^a_{t+s}$.  
  
  Hence, the two components of the action each
generate their own time transformation, and both are interpreted as time translations on
the space of paths: the first ($V$ ) is purely kinematical, shifting the time label; the second
($H$) implements the dynamical canonical transformation.
  
   We also evaluate
  \bey
   \{H, \xi^a_t\} = \omega^{ab}\partial_b h(\xi_t).
 \eey
The corresponding symplectic transformation does not affect $t$: it is implemented as $\xi_t \rightarrow f_s(\xi_t)$, where $f_s$ is the canonical transformation generated by the Hamiltonian.  Eq. (\ref{heqm}) implies that the two notions of time translation coincide on physical paths.

 \section{Classical events, time and ordering as observables}
 
\begin{figure}[tb] 
\includegraphics[height=6.5cm]{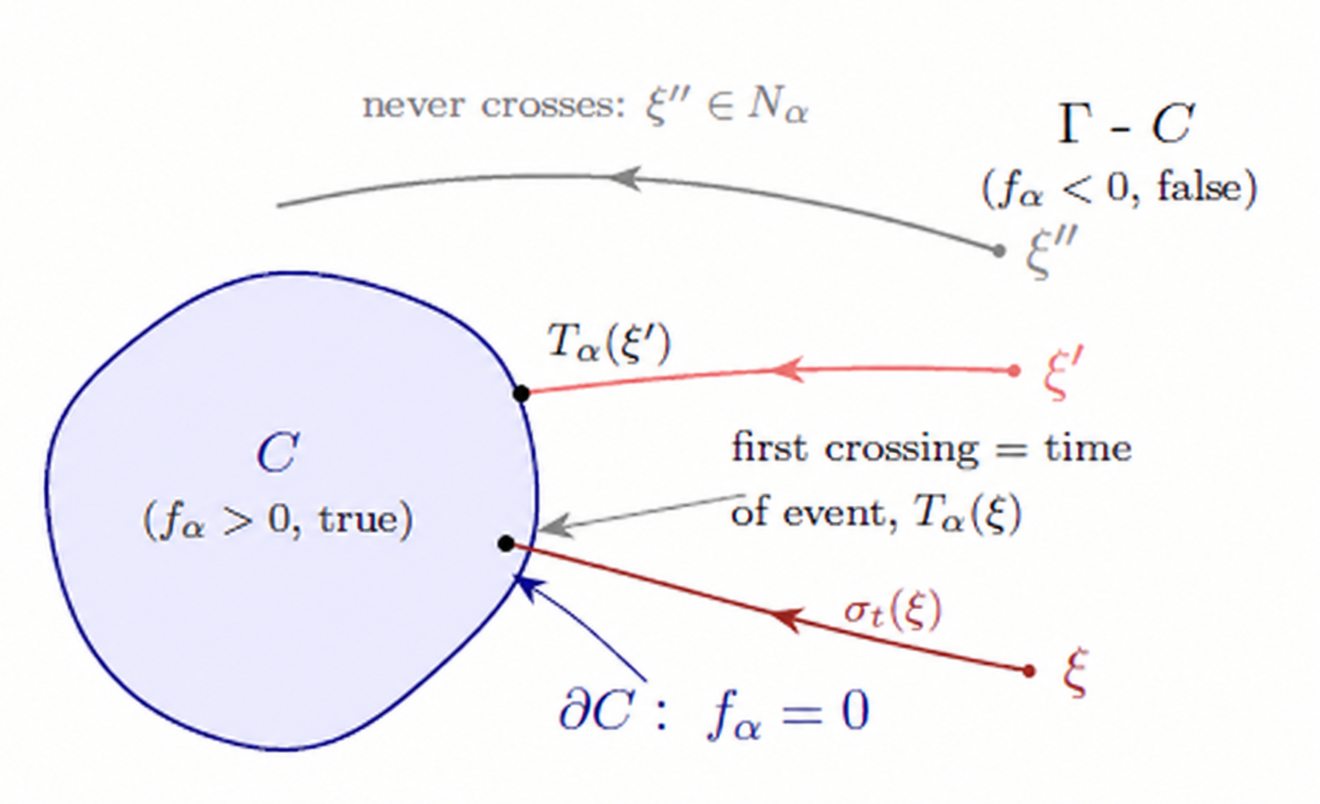} \caption{ A classical event $\alpha$ as the first crossing of a codimension-one surface $\partial C$ ($f_{\alpha} = 0$) in the
state space $\Gamma$. The proposition is true inside $C$ ($f_{\alpha} > 0$) and false outside ($f_{\alpha} < 0$). The
trajectory $\sigma_t(\xi)$ starting at $\xi$ realises the event the first time it meets $\partial C$; that instant
defines the time of the event $T_{\alpha}(\xi)$. Starting at a different initial condition $\xi'$, the crossing time is
$T_{\alpha}(\xi')$, so the assignment of a time to the event is deterministic for each initial condition.
A trajectory that never crosses $\partial C$ lies in the null set $N_{\alpha}$.}
\label{event1}
\end{figure}

 We can now proceed to the definition of events in classical physics. Intuitively, an event occurs when something new happens. This means that a proposition that was false earlier became true later. Since a single-time proposition is represented by  subset $C$ of the state space $\Gamma$, an event occurs whenever the path of the system first enters $C$---or, equivalently, when it crosses the boundary $\partial C$ of $C$ for the first time---see Fig. (\ref{event1}). 
 
 It is a matter of convention, whether we will treat the transitions $\Gamma - C \rightarrow C$ and  $C \rightarrow \Gamma - C$ as representing two different events or just one. Here, we will choose the latter, so we are led to the following definition.

\begin{definition}
An event is the  crossing of a surface of codimension one in the state space $\Gamma$. 
\end{definition}
 
A surface  of codimension one is locally determined by the vanishing of one  functions on $\Gamma$, hence, we can represent an event $\alpha$ by a function $f_{\alpha}: \Gamma \rightarrow \R$. We can choose the sign   so that $f_{\alpha}$  is positive in  $C$ and negative in $\Gamma - C$.

\subsubsection*{Examples}
\begin{enumerate}
\item For a particle at a line, with state space $\Gamma = \R^2$, an event $\alpha$ may correspond to the particle crossing  $x = 0$. Then $f_{\alpha}(x, p) = x$.

\item For $N$ particles at a line, we can define $n$ distinct events $\alpha_n$, with $f_{\alpha_n}(x_1, \ldots, x_N, p_1, \ldots, p_N) = x_n$. The functions $f_{\alpha_n}$ are functionally independent, hence, the events are independent.
    
\item In a horse race, the most interesting events correspond to each horse crossing the finish line---ideally described in terms of the geometry for the track.

\end{enumerate}
 
Let $\sigma_t$ be the one-parameter family of canonical transformations generated by the Hamiltonian. For any event $\alpha$, we define its null set  $N_{\alpha}$, such that the equation $f_{\alpha}[\sigma_t(\xi)] = 0$ has no solution for any $t \geq 0$. If the system starts from $\xi \in N_{\alpha}$, it will never cross the surface $f_{\alpha} = 0$. 

Then, we can define the $T_{\alpha}$ of the event $\alpha$, as a function $T_{\alpha}: \Gamma - N_{\alpha} \rightarrow \R^+ $, such that $T_{\alpha}(\xi)$ is the smallest positive value of $t$ that solves the equation $f_{\alpha}[\sigma_t(\xi)] = 0$. This means that $T_{\alpha}(\xi)$ is the time it takes a trajectory that starts at $\xi$ to cross the surface $f_{\alpha} = 0$ for the first time.  We therefore define the time of an event as a physical observable taking values in Newtonian time. We have to keep in mind, however, that $T_{\alpha}$ takes values in $\R^+ \cup \{N\}$, where $\{N\}$ corresponds to the event never taking place. This is a generic situation for observables conditioned upon events---an event may simply not happen.

If we have  set $E$ of independent events, then we can define one time function $T_{\alpha}$ for each of them. Hence, for each initial state $\xi$, there is:
a unique clock $T_{\xi}$ on $E$, defined by
\bey
T_{\xi}(\alpha, \beta) = \left\{ \begin{array}{cc} T_{\beta}(\xi) - T_{\alpha}(\xi), & \xi \notin N_{\alpha} \cup N_{\beta}\\
\infty, &  \xi \notin N_{\alpha}, \xi \in N(\beta) \\ -\infty, &  \xi \in N_{\alpha}, \xi \notin N(\beta) \\
N(\mbox{indefinite}), & \xi \in N_{\alpha} \cap N_{\beta} \end{array}\right.
\eey
From this function, we define a causal order  in $E$:
\bey
\alpha \po \beta \;\; &\mbox{if}&\; \;  T_{\xi}(\alpha, \beta) < 0 \;\; (\mbox{including}\; - \infty)\nonumber \\
\beta \po \alpha \;\; &\mbox{if}&\; \;  T_{\xi}(\alpha, \beta) > 0 \;\; (\mbox{including}\; + \infty)\nonumber \\
\alpha \sim \beta \; \; &\mbox{if}& \;\; T_{\xi}(\alpha, \beta) = 0 \nonumber \\
\alpha | \beta \; \; &\mbox{if}& \;\; T_{\xi}(\alpha, \beta) = N.
\eey

 \begin{figure}[tb] 
\includegraphics[height=7cm]{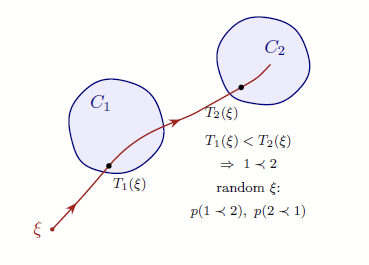} \caption{ Two events, defined by two
surfaces $\partial C_1$ and $\partial C_2$  in the same state space. A single trajectory crosses them at times
$T_1(\xi)$ and $T_2(\xi)$, which determine which event happened first, i.e. the causal order.}
\label{event2}
\end{figure}

Since Hamiltonian mechanics is deterministic, both the clock function and the causal order are fixed by the initial conditions $\xi(0)$. However, suppose that the initial condition  is not known exactly, and the system is described by a probability distribution. Then, we must construct   joint probability distributions for the times of events
\bey
p(t_1, t_2, \ldots, t_n) = \int d\xi \rho(\xi) \delta[T_1(\xi) - t_1] \delta[T_2(\xi) - t_2] \ldots \delta[T_n(\xi) - t_n] . 
\eey
These probability densities are not normalized to unity. For proper normalization, we have to include the probability densities for no events, which corresponds to the null sets $N_{\alpha}$.

For example, for $n = 2$, we have the probability densities $p(t_1, t_2)$ as above, together with the probability densities
\bey
p(N_1, t_2) = \int d \xi \chi_{N_1}(\xi) \delta[T_2(\xi) - t_2] \\
p(t_1, N_2) = \int d \xi \chi_{N_2}(\xi) \delta[T_1(\xi) - t_1]\\
p(N_1, N_2) = \int d\xi \chi_{N_1}(\xi)\chi_{N_2}(\xi).
\eey
Then, we obtain the associated probabilities  
\bey
p(1\po 2) &=& \int_0^{\infty} dt_1 \int_0^{t_1} dt_2 p(t_1, t_2) + \int_0^{\infty} dt_1 p(t_1, N_2) \nonumber \\
p(2 \po 1) &=& \int_0^{\infty} dt_2 \int_0^{t_2} dt_1 p(t_1, t_2) + \int_0^{\infty} dt_2 p(N_1, t_2) \nonumber \\
p(1 | 2) &=& p(N_1, N_2) \nonumber \\
p(1 \sim 2) &=& \int_0^{\infty} dt p(t, t). \label{problat}
\eey
These probabilities are defined on the space $CO(E)$ of causal orders on $E$.

 Causal orders of events can easily be probabilistic within classical physics, even if the spacetime causal structure is absolute and unchanging. This is because the causal order of events depends on the embedding of the events in spacetime---here represented by the functions $T_{\alpha}$---which are physical observables.
 
In a certain sense, the assignment of probabilities to causal orderings of events is among the oldest uses of probability. The odds set by bookmakers in horse or chariot racing provide a direct example.

\begin{exercise}
Let the state space $\Gamma = \{ x_1, x_2, p_1, p_2\}$ describe a system of   two free  particles of mass $m$ at a line. Event 1 corresponds to particle 1 crossing the line $x_1 = 0$ and event 2 corresponds to particle 2 crossing the line $x_2 = 0$. Evaluate the probabilities (\ref{problat}) assuming an initial state 
\bey
\rho(x_1, x_2, p_1, p_2) = \delta(x_1 - x_0) \delta(x_2 - x_0) f(p_1) f(p_2), \nonumber 
\eey
 where $x_0 < 0$. 
\end{exercise}

\subsubsection{Incorporating randomness}

The analysis above extends with little modification to classical stochastic systems. 
Such systems involve random dynamics, and probabilities are expressed in terms of a 
probability measure $\mu$ on the space of histories $\Pi$. The expectation value of 
any history observable $A$ is then given by
\bey
\langle A\rangle = \int d\mu[\xi(\cdot)] A[\xi(\cdot)]] \nonumber 
\eey

An event $\alpha$ is again defined by the first intersection of a path with a surface, 
and it can be represented by a function $f_{\alpha}$ on $\Gamma$. We can therefore 
define a null set $N_{\alpha}$ and a time functional $T_{\alpha}$, now with respect 
to the space of histories $\Pi$ rather than $\Gamma$.

Specifically, we define $N_{\alpha}$ as the subset of $\Pi$ consisting of paths 
$\xi(\cdot)$ for which the equation
$f_{\alpha}(\xi(t)) = 0$
has no solution for any $t \in [0,T]$. For any path 
$\xi(\cdot) \in \Pi \setminus N_{\alpha}$, we define the time functional 
$T_{\alpha}[\xi(\cdot)]$ as the smallest value of $t$ such that
$f_{\alpha}(\xi(t)) = 0$.

The remainder of the construction is unchanged. In particular, the joint probability 
density for $n$ events is defined by
\begin{equation}
p(t_1,\ldots,t_n)
=
\int d\mu[\xi(\cdot)]\,
\delta\!\left(t_1-T_1[\xi(\cdot)]\right)
\cdots
\delta\!\left(t_n-T_n[\xi(\cdot)]\right). \nonumber
\end{equation}

 \section{Relativistic spacetime}
 
With special relativity, Einstein introduced an important revision 
of the Newtonian conception of time. The fact that the speed of light is the same for all observers is not compatible with absolute time. Nonetheless, the revisions did not affect Newton's core idea that the causal and temporal structure of events is generated by an absolute structure in spacetime. This absolute structure is the spacetime metric $g$. 

In what follows, I will describe the   causal structure for a general background spacetime   $(M,g)$, specializing to the Minkowski case where necessary.

\subsubsection*{Causal Structure}

Let $(\M,g)$ be a spacetime, i.e.\ a smooth manifold $\M$ equipped with a 
Lorentzian metric $g$ of signature $(-,+,+,+)$.

A tangent vector $v \in T_x\M$ is called

\begin{itemize}
\item \textbf{timelike} if $g(v,v) < 0$,
\item \textbf{null} (or lightlike) if $g(v,v) = 0$ and $v \neq 0$,
\item \textbf{spacelike} if $g(v,v) > 0$,
\item \textbf{causal} if it is timelike or null.
\end{itemize}

The set of null vectors at a point $x$ defines the {\em light cone} at $x$. A causal vector $v$ is future-directed if $v^0 > 0$ and past-directed if $v^0 < 0$.

A smooth curve $\lambda : [0, 1] \to \M$ is called timelike if $\dot\lambda(s)$ is timelike for all $s$; null if $\dot\lambda(s)$ is null for all $s$; and causal if $\dot\lambda(s)$ is causal for all $s$.

%------------------------------------------------

For any point $x \in \M$, we define

\begin{itemize}
\item  the chronological future of $x$: $$I^+(x) = \{ q \in \M \mid \text{there exists a future-directed timelike curve from } p \text{ to } x, \}$$  
\item the causal future of $p$: $$J^+(x) = \{ q \in \M \mid \text{there exists a future-directed causal curve from } x \text{ to } q \}.$$
\end{itemize}
We similarly define the past sets $I^-(p)$ and $J^-(x)$. 

We also define the causal future/past of a subset $C$ of $\M$ as $J^{\pm}(C) = \cup_{x\in C}J^+(x)$, and similarly for the chronological past/future.

We can therefore define the following causal relations on $\M$.

\begin{itemize}
\item $x \ll y$ if $y \in I^+(x)$ (chronological order),
\item $x \po y$ if $y\in J^+(p)$ (causal order)
\item $x | y$ if $y \notin J^+(x) \cup J^-(x)$. (spacelike separation)
\end{itemize}
Both chronological and causal orders define a partial ordering on $\M$.  The causal  relations remain unchanged if the metric undergoes a conformal rescaling $g \rightarrow \Omega^2 g$.

We can also define extend these relations to general subsets of $\M$:
\begin{itemize}
\item $C_1 \po C_2$:  for every $x \in C_1$ and every $y \in C_2$, $x \po y$.
\item $C_1 | C_2$: for every $x \in C_1$ and every $y \in C_2$, $x | y$
\end{itemize}
Unlike points that may are either causally connected or spacelike separated, spacetime regions may have no clear causal relation.

No notion of simultaneity can be defined on $\M$ solely through the metric $g$. Some additional structure is required.
 
\subsubsection*{Local clocks}
The proper times associated to the metric define  a clock functional $\tau(\lambda)$ on  timelike curves 
\bey
\tau(\lambda) = \int_{\lambda} ds \sqrt{-g_{\mu \nu} \dot{\lambda}^{\mu} \dot{\lambda}^{\nu}}.
\eey
 
For null curves, the integrand vanishes identically, and hence the proper time
between any two points on the curve is zero. Therefore, null curves do not admit
a non-trivial intrinsic notion of duration.

Although null geodesics admit affine parameters, these are not unique and are
defined only up to linear transformations $s \to a s + b$. 
Consequently, affine parameters cannot be interpreted as durations.

 \subsubsection{Global time}
 A global time function on the spacetime is a scalar function $t: \M \rightarrow \R$, that increases along all future-directed causal curves.  We say that the time function $t$  is  a  {\em temporal function} if  
   $t^{\mu} = g^{\mu \nu} \nabla_{\nu}t$ is a future-directed timelike vector everywhere. The integral curves of $t^{\mu}$ define a global family of observers. The metric does not by itself
contain any information singling out a preferred time coordinate; a global time function
is extra structure.

   Not all conceivable spacetimes admit global time functions: spacetimes with closed timelike curves do not. 
 A spacetime that admits at least one time function is called {\em stably causal}. But if one time function  exists, there is an infinity of them: for any strictly monotonic increasing $f: \R \rightarrow \R$, $f(t)$ is a time function if $t$ is. 
 
 Given a time function $t$, we can define a notion of simultaneity relatively to it. Two   points $x, y \in M$ are simultaneous if $t(x) = t(y)$. All points $x$ such that $t(x) = t$ define a spacelike hypersurface of simultaneity, denoted by $\Sigma_t$.

 There are several different prescriptions to define global time functions: 
\begin{enumerate}

\item \textit{Geroch Time.}
Let $\rho$ be a positive integrable 
function on the spacetime manifold$\M$. The Geroch time function is defined by
\begin{equation}
t(x) = \int_{J^-(x)} \, \sqrt{-g} \; d^4x' \; \rho(x').
\end{equation}
 
Since $J^-(x)$ grown monotonically along future-directed causal curves, 
the function $t$ is strictly increasing along such curves.  However, its gradient may fail to be everywhere timelike.
  
In stably causal spacetimes one can smooth the Geroch time to make it into a temporal function.

\item \textit{Foliation Time.} The most common way to define a global temporal function through spacelike foliation. 

\begin{definition}
A spacelike foliation of spacetime is a family of non-intersecting spacelike 
hypersurfaces $\{\Sigma_t\}_{t\in\mathbb{R}}$ such that $\M = \bigcup_{t\in\mathbb{R}} \Sigma_t$ ,
and every point $x\in \M$ lies on exactly one hypersurface $\Sigma_t$.
\label{spfol1}
\end{definition}

Given a foliation, one defines a global time function by 
\begin{equation}
t(x) = t \qquad \text{if } x\in \Sigma_t . \nonumber
\end{equation}

Since causal curves intersect each hypersurface at most once, this function 
is strictly increasing along future-directed causal curves and hence defines 
a global time function.
Different foliations give rise to different time functions.

Usually we restrict to {\em Cauchy foliations}, that is, foliations by Cauchy surfaces.
\begin{definition}
A Cauchy surface is a subset of the spacetime that is intersected only once by each inextendible causal curve.
\end{definition}
A curve is inextendible if it cannot be extended further (it runs from “past infinity” to “future infinity”, or to singularities).

Spacetimes that possess Cauchy surfaces are said to be {\em globally hyperbolic}. It can be proven that their topology is $\R\times \Sigma$. A stably causal spacetime may fail to be Cauchy because some causal curves may become inaccessible---for example, by entering a black hole---or if it admits a timelike boundary (anti de Sitter spacetime).

\item \textit{Killing Time.} Suppose spacetime $(\M,g)$ admits a timelike Killing vector field $K^\mu$, i.e., $\mathcal{L}_K g_{\mu\nu} = 0$.
A Killing time function $t$ is defined by requiring that $K^\mu$ generates
translations in $t$, $K^\mu \nabla_\mu t = 1$ .
 
The integral curves of $K^\mu$ correspond to worldlines of observers at
constant spatial coordinates, and the hypersurfaces $t=\text{const}$
are invariant under the isometry generated by $K^\mu$.

If $K^\mu$ satisfies the integrability condition $K_{[\mu}\nabla_\nu K_{\rho]} = 0$, then it is 
is hypersurface-orthogonal, and it generates a foliation of spacelike hypersurfaces orthogonal to $K^\mu$.
Then, the spacetime is said to be static.
\end{enumerate}
 
  The most important example of Killing time arises in Minkowski spacetime, where time
translations are generated by the Poincaré symmetry group.

\begin{definition}
The Poincaré group is the group of isometries of Minkowski spacetime. 
It consists of affine transformations
\begin{equation}
x^\mu \rightarrow \Lambda^\mu{}_\nu x^\nu + a^\mu ,
\end{equation}
where $\Lambda^\mu{}_\nu$ is a Lorentz transformation satisfying
\[
\Lambda^\mu{}_\alpha \Lambda^\nu{}_\beta \eta_{\mu\nu}
= \eta_{\alpha\beta},
\]
and $a^\mu$ is a constant translation vector.
\end{definition}

The Poincaré group is therefore the semi-direct product
\[
\mathcal{P} = \mathbb{R}^4 \rtimes O(1,3).
\]

\begin{exercise}
Show that the most general Killing vector on Minkowski spacetime is $K^{\mu} = a^{\mu} + \omega^{\mu}{}_{\nu}x^{\nu}$, where $\omega_{\mu \nu} = - \omega_{\nu \mu}$. Identify the Killing vectors that are everywhere timelike, and compute their time functions.
\end{exercise}
The generators of the Poincar\'e are six antisymmetric matrices $M_{\mu \nu}$ for spacetime rotations, and the four vector $P_{\mu}$ for spacetime translations. The corresponding Lie algebra is,
\begin{align}
[M_{\mu \nu}, M_{\rho\sigma}] &= \eta_{\mu \rho} M _{\nu \sigma} +  \eta_{\nu \sigma}M_{\mu \rho} -\eta_{\nu \rho} M _{\mu \sigma}  -  \eta_{\mu \sigma} M _{\nu \rho}, \\
[M_{\mu \nu}, P_{\rho}] &= \eta_{\mu \rho} P_{\nu} - \eta_{\nu \rho} P_{\mu}, \\ 
[P_{\mu}, P_{\nu}] &= 0.
\end{align}

Using spatial indices, we can express the Lie algebra of the Poincar\'e  group in terms of
 three generators $K_i = M_{0i}$ of boosts,
 three generators $J_i =    \frac{1}{2} \sum_{jk} \epsilon_{ijk} M_{jk}$ of spatial rotations,
 three generators  $P_i$ of space translations, and
  one generator $H := P_0$ of time translations.
The corresponding commutation relations are given in  the following table.

\bigskip
 
\hspace{2cm} \begin{tabular}{  c     @{\vline}  c c c c  }
{}  \hspace{0.3cm} & $J_j$ &  $K_j$ & $P_j$ & $H$   \\[0.1cm]
\hhline{-|----}
$J_i$  \hspace{0.3cm}  & $\sum_k \epsilon_{ijk} J_k$ & $\sum_k \epsilon_{ijk} K_k$ & $\sum_k \epsilon_{ijk} P_k$ & $0$\\[0.2cm]
$K_i$  \hspace{0.3cm} & $\sum_k \epsilon_{ijk} J_k$ &$\sum_k \epsilon_{ijk} J_k$& $\delta_{ij}H$& $P_i$ \\[0.2cm]
$P_i$ \hspace{0.3cm} &$\sum_k \epsilon_{ijk} P_k$ & $-\delta_{ij}H$& $0$&$0$ \\[0.2cm]
$H$ \hspace{0.3cm}  &$0$ &$-P_j$&$0$&$0$
 \end{tabular}

\section{  Relativistic dynamics}
The essence of Newton’s construction does not change in special relativity: we still
have an absolute, non-dynamical spacetime that exists “without reference to anything
external.” Newton’s own words for absolute time apply equally to Minkowski spacetime.
What changes is only the form of the causal structure—a different way of assigning past
and future. This is precisely why we can do so much quantum physics in Minkowski
spacetime, assuming the background structure, and why we are at a loss when there is no
fixed background, as in quantum gravity.

%Special relativity is a reformation of Newton's view of absolute space and time. It still preserves spacetime as an absolute structure, only it changes the form of its causal %structure. We can paraphrase Newton in a way that remains accurate in special relativity. 
%\begin{quote}
%{\small Absolute, true, and mathematical spacetime, in and of itself and of its own nature,
%exists uniformly without reference to anything external.}
%\end{quote}
Whenever we have a proper definition of events, any map ${\cal X}$ embedding the space of events $E$ into spacetime   defines a causal ordering and clock variables on $E$. There is no change here from non-relativistic physics, events can still be defined in terms of hypersurfaces in the state space. We must therefore identify appropriate  state spaces for relativistic systems. 

However, interactions 
cannot in general be formulated in terms of a finite number of particle degrees 
of freedom while preserving locality and Poincaré invariance. Relativistic 
interacting systems are naturally described in terms of fields, whose degrees 
of freedom are associated with spacetime points rather than with individual 
particles. Consequently, the appropriate state space for relativistic systems 
is typically the infinite-dimensional space of field configurations.  

By locality we mean that the dynamics at a spacetime point depends only on the 
values of the fields and a finite number of their derivatives at that same 
point.   Equivalently, the equations 
of motion are differential equations, and disturbances propagate within the 
light cone.

Furthermore, when the spacetime is Minkowski,  the Poincar\'e symmetry is fundamental. It must be built in into both the kinematic and the dynamics of the theory. There are two prescriptions of doing so.

\begin{enumerate}
\item Start with an action on configuration space paths that is invariant under transformations by the Poincar\'e group. The action leads to Poincar\'e-covariant Euler-Lagrange equations.
    
    When describing fields, locality implies that the Lagrangian action is local functional of the fields, that is, it can be written as 
    \bey
    S[\phi] = \int d^4 x {\cal L}(\phi, \partial \phi), \label{covct}
    \eey
    in terms of the Lagrangian density ${\cal L}$ which is a local function of the fields and their first derivatives. Variation of the action yields the Euler-Lagrange equations
    \bey
    \partial_{\mu} \left(\frac{\partial {\cal L}}{\partial (\partial_{\mu}\phi_a)}\right) - \frac{\partial {\cal L}}{\partial \phi_a} = 0.
    \eey
    
\item Start with a phase space $\Gamma$ that carries a symplectic action of the Poincar\'e group. This means that there exist functions $H, J_i, K_i, P_i$ on $\Gamma$ with Poisson brackets that reproduce the Lie algebra of the Poincar\'e group.
    
    The symplectic actions of the Poincaré group were first classified using
Kirillov's coadjoint orbit method \cite{Kiril}, and applied explicitly to relativistic
systems by Souriau \cite{Souriau}. The resulting classification reproduces the
types of relativistic particles, familiar from quantum theory \cite{Wigner}, as elementary components of any Hamiltonian system: massive particles with spin,
massless particles with helicity, tachyonic orbits, and degenerate cases.
\end{enumerate}

\subsection{Foliation dependence}
The connection between the Lagrangian and the Hamiltonian description requires a relation between velocities and momenta. Velocity explicitly depends on  choice of time function, usually introduced through  foliation. Thus, the Lagrangian-Hamiltonian correspondence is foliation dependent.

We will see this by analyzing the case of a real scalar field $\phi$ in Minkowski spacetime $\M$. Let $P(\M)$ be the space of scalar functions $\phi : \M \rightarrow \R$. The action is a functional $S:P \rightarrow \R$ defined as $S = \int d^4 x {\cal L}[\phi(x), \partial \phi(x)]$, in terms of the Lagrangian density 
\bey
{\cal L} = -\frac{1}{2}\eta^{\mu \nu} \partial_{\mu}\phi \partial_{\nu}\phi - \frac{1}{2} m^2 \phi^2.
\eey
It is straightforward to verify that ${\cal L}$ in invariant under the Poincrar\'e transformations 
\bey
\phi(x) \rightarrow \phi(\Lambda^{-1} x - a).
\eey
To introduce the conjugate momentum, we select a Killing time $t$, with an associated timelike Killing vector $n = \partial/\partial t$. This defines a foliation of $M$ by Cauchy hypersurfaces $\Sigma_t$  with Euclidean spatial coordinates $\mathbf{x}$ and a (flat) three-metric $\delta_{ij}$. Hence, $\dot{\phi} =n^{\mu}\partial_{\mu}\phi$. 

Then, we select a reference time $t_0$, and define the Lagrangian at $\Sigma_{t_0}$, $L = \int_{\Sigma_{t_0}} d^3 x {\cal L}$, from which we derive the conjugate momentum
\bey
\pi(\mathbf{x}) = \frac{\delta L}{\delta \dot{\phi}(\mathbf{x})} = \dot{\phi}(\mathbf{x})
\eey
The state space $\Gamma$ then consists then by the pairs $(\phi(\mathbf{x}), \pi(\mathbf{x}))$ and it can be identified with $T^*P(\Sigma_{t_0})$, the cotangent bundle over the space of scalar functions on $\Sigma_{t_0}$.
The symplectic form is
\bey
\omega = \int_{\Sigma_{t_0}} \delta \pi(\mathbf{x}) \wedge \delta \phi(\mathbf{x});
\eey
the fields $\phi(\mathbf{x})$ and $\pi(\mathbf{x})$ defining Darboux coordinates.
 The Hamiltonian is
\bey
H = \frac{1}{2} \int_{\Sigma_{t_0}} d^3 x \left(\pi^2 +(\nabla \phi)^2 + m^2 \phi^2 \right).
\eey
We see that the whole state space depends not only on the time function but also on the choice of a reference hypersurface $\Sigma_{t_0}$. This dependence causes no problems at the classical level, because it does not affect the equations of motion, but it is mildly puzzling at the quantum level, and alarming when we attempt to quantize gravity. 

\begin{exercise}
Identify the ten generators of the Poincar\'e group on the state space $\Gamma$ of the scalar field. 
\end{exercise}
\subsection{Covariant state space}
We can characterize the state space in a way that does not depend on the choice of a foliation. Consider the action (\ref{covct}) defined on the space $P(\M)$ of all field configurations. Its variation 
 can be written as
\begin{equation}
\delta S = \int_{\M} d^dx \left( E_i(\phi)\, \delta \phi^i + \partial_\mu \theta^\mu(\phi, \delta \phi) \right),
\end{equation}
where $E_i(\phi)=0$ are the Euler--Lagrange equations, and  
\bey
\theta^\mu(\delta \phi) = \frac{\partial \mathcal{L}}{\partial (\partial_\mu \phi_a)} \, \delta \phi_a.
\eey
 
The covariant phase space $\Gamma_c$ is defined as the space of all solutions to the equations of motion,
\begin{equation}
\Gamma_c = \{ \phi \in P(\M) \, | \, E_i(\phi)=0 \}.
\end{equation}
Tangent vectors $\delta \phi_a$ on $\Gamma_c$ correspond to solutions of the equations of motion.

We define the two-form
\bey
\omega(\delta_1\phi, \delta_2\phi) = \int_{\Sigma}d^3x n_{\mu}\left[\delta_1\theta^{\mu}(\delta_2\phi) - \delta_2\theta^{\mu}(\delta_1\phi)\right], \label{covps}
\eey
where $n_{\mu}$ is the unit normal to the spacelike Cauchy surface $\Sigma$. 
The two-form $\omega$ satisfies by construction $d\omega = 0$, and it is independent of the choice of $\Sigma$. In absence of constraints, it is invertible, hence, it defines a symplectic form on $\Gamma_c$.
\begin{exercise}
Construct the two-form (\ref{covps}) for a massive scalar field, and confirm that it does not depend on the choice of $\Sigma$.
\end{exercise}

The covariant state space  provides a formulation of classical field theory that does not depend on a choice of time or foliation, and treats solutions as fundamental objects rather than initial data. However, since its points correspond to histories, it cannot be used to define single-time observables, and the notion of time evolution becomes opaque.

\subsection{Symplectic histories description}

The symplectic histories formulation has the advantages of the covariant state space method with none of its drawbacks \cite{Sav03}. We start from the space $\Pi = T^*P(\M)$ of state space histories, the cotangent bundle of $P(\M)$. 
Its elements are pairs of fields $(\phi(x), \pi(x))$, and it carries no dependence on a time function. For any choice of timelike Killing field $n$, we define the action
\bey
S_n[\phi(\cdot), \pi(\cdot)] = \int d^4 x \left[\pi n^{\mu}\partial_{\mu} \phi - \frac{1}{2}\left(\pi^2 + (\eta_{\mu \nu} + n^{\mu}n^{\nu})\partial_{\mu}\phi \partial_{\nu}\phi + m^2 \phi^2\right)\right]. \label{sn}
\eey
Then, we vary the action over $\pi(x)$, to obtain $ \pi = n^{\mu}\partial_{\mu} \phi$, and using this to eliminate $\pi$ we retrieve the standard Lagrangian action. In this picture, the space of Lagrangian and Hamiltonian histories are fully covariant, and their dynamics are related through the foliation dependent action (\ref{sn}). The crucial point here is that the phase space action is more fundamental, and the covariant Lagrangian action is derived. 

\subsection{The gauge problem}
A second problem in the Lagrangian-Hamiltonian correspondence is that 
 the state spaces of relativistic systems may not be expressible in terms of local fields.
   This creates a consistency problem that again becomes very acute in the context of quantum gravity.
  
  The standard way out is that we work on a state space with variables additional to the physical ones. In this state space, a canonical form can be defined, and so can a unique Lagrangian. However, the additional variables must be eventually eliminated in order to recover the fundamental description. 

The most important manifestation of this problem is in the Yang-Mills theory, 
which admits a local Lagrangian formulation in terms of the gauge
potential $A_\mu^a(x)$,
\[
L = -\frac14 F_{\mu\nu}^a F^{\mu\nu a},
\qquad
F_{\mu\nu}^a =
\partial_\mu A_\nu^a
-\partial_\nu A_\mu^a
+ f^{abc} A_\mu^b A_\nu^c,
\]
where $f^{abc}$ are the structure constants of the associated group $G$.

However, the variables $A_\mu^a$ are not physical, since the theory is invariant
under gauge transformations $g: M \rightarrow G$
\[
A_\mu \rightarrow g^{-1} A_\mu g + g^{-1} \partial_\mu g.
\]
Thus the Lagrangian formulation involves redundant (dummy) variables. 

To identify the true variables describing the system, we need a detailed analysis of constraints. The key point is that these true variables are obtained from  solution of a non-linear differential equation, and they are non-local variables of the  gauge potential. 
This results into a state space $\Gamma_{phys}$ in which there are no Darboux coordinates expressed as local fields, and the Hamiltonian is non-local. 
 
 Thus, Yang--Mills theory provides an example where
  a local Lagrangian exists in redundant variables, but the fundamental dynamics is non-local. We will come back to this issue    in Chapter 5, where we will explain its implications to quantisation of gravity.

\chapter{Time in quantum physics}

\section{The principles of quantum theory: probabilistic structure}
It is convenient to express quantum theory in terms of a number of fundamental principles that take the form of axioms. A common characterization involves six principles: four determine the probabilistic structure and two involve aspects that are essentially temporal.

The first four principles are the following. 

\begin{enumerate}
\item  \textbf{Quantum states:} Any physical system is described by a complex Hilbert space ${\cal H}$. A state of the system at a given time is represented by a density operator $\hat{\rho}$ on ${\cal H}$.
\item \textbf{Observables:} Physical quantities are represented by self-adjoint operators $\hat{A}$ on ${\cal H}$. Measurement outcomes correspond to subsets of the spectrum of $\hat{A}$.
\item \textbf{Probabilities:} For a system prepared in the state $\hat{\rho}$, a measurement of the observable associated with the operator $\hat{A}$ yields outcomes in a subset $C$ of the spectrum of $\hat{A}$ with probability
\begin{eqnarray}
\mathrm{Prob}(C) = \mathrm{Tr}\!\left(\hat{\rho} \hat{P}_C\right), \label{probaxiom}
\label{prob0}
\end{eqnarray}
where $\hat{P}_C=\chi_C(\hat{A})$ is the spectral projector associated with the set $C$.
\item \textbf{Combination of subsystems:}  A $n$-partite system is described by the tensor product ${\cal H}_1 \otimes {\cal H}_2 \otimes \ldots \otimes {\cal H}_n$ of the Hilbert spaces  ${\cal H} _i$, $i =1, 2, \ldots, n$  of its subsystems.
\end{enumerate}

\subsubsection*{Remarks}
\begin{enumerate}

\item Quantum states are usually interpreted as encoding the way a system has been prepared, either in the laboratory or naturally, prior to measurement. All information about the preparation is contained in the density operator, from which probabilities for measurement outcomes follow through Eq.~(\ref{prob0}). This {\em minimal interpretation} is logically consistent and suffices to account for the predictive success of quantum theory.

However, the quantum state is not an {\em objective property} of an {\em individual} system. By an objective property we mean one that can be determined by measurements without knowledge of the system’s past. In classical physics, states are objective in this sense: the state of a particle is specified by position $x$ and momentum $p$, which can in principle be measured to identify the state. In quantum theory this is impossible because quantum states are not mutually exclusive. If a measurement of $\hat A$ yields the value $a$ associated with eigenvector $|a\rangle$, the pre-measurement state need not have been $|a\rangle$; the outcome is compatible with any state $|\psi\rangle$ with $\langle\psi|a\rangle\neq0$. No measurement on an individual system determines a unique quantum state, hence the state cannot be an objective property\footnote{This holds for a measurement on an individual system. Given a large statistical ensemble, and measurements of  different observables, the quantum state is reconstructed through tomography.}. 

The quantum state must therefore be regarded as partly referring to our knowledge about the system. This departs from the Newtonian tradition, where the basic elements of a theory correspond to objective features of the world. Many attempts have been made to restore objectivity, but proposed interpretations face significant conceptual or technical difficulties.  Consequently, no interpretation beyond the minimal one has universal acceptance. How even this minimal interpretation should be applied in quantum gravity—particularly in cosmology—remains unclear.

\item Mixed states are essential because physical preparations often involve statistical uncertainty: a system may be prepared in state $|\psi_i\rangle$ with probability $p_i$, yielding the density operator
\begin{equation}
\hat{\rho}=\sum_i p_i |\psi_i\rangle\langle \psi_i|.
\end{equation}
Pure states arise as the special case in which one probability equals unity, $\hat{\rho}=|\psi\rangle\langle\psi|$. In general, however, the decomposition of a mixed state into pure states is not unique: the same $\hat{\rho}$ can be decomposed in infinitely many ways.

Since all measurement probabilities depend only on $\hat{\rho}$ through $\mathrm{Tr}(\hat{\rho}\hat P)$, these different ensembles are operationally indistinguishable. This non-uniqueness implies that a mixed state does not represent ignorance about an underlying “true” pure state of an individual system; rather, the density operator itself is the complete physical description of the preparation.

\item The   use of complex  numbers at the fundamental level of the theory  does not originate from the $i$ in Schr\"odinger's equation. It is necessary in order to account for the observed properties of interference. 
 In a real Hilbert space a superposition of two states has the form $|\psi\rangle = a|1\rangle + b|2\rangle$, where $a,b \in \mathbb{R}$.
Hence,  the only relative “phase” is the sign of $b/a$. The probability for an outcome associated with $|1\rangle$ depends on quadratic combinations such as $ab$, yielding only constructive or destructive interference.

However, interference experiments exhibit a continuously tunable phase shift. In complex quantum theory one may define states $|\psi\rangle = a|1\rangle + b e^{i\phi}|2\rangle$ 
and the corresponding probability contains the interference term
\begin{equation}
P  
= |a|^2 + |b|^2 + 2|a||b|\cos\phi .
\end{equation}
The continuous dependence on the phase $\phi$ does not naturally occur in a   real Hilbert space. To recover the observed interference pattern one---involving a $U(1)$ phase---we must introduce an additional structure equivalent to multiplication by $i$, rendering the theory effectively complex.  

\item As in classical physics, the notion of a state refers to a single moment of time. As such, it presupposes some form of simultaneity. In a relativistic setting, the Hilbert space of quantum states is usually assigned to a Cauchy hypersurface.

\item There is a more general definition of observable, in terms of  Positive-Operator-Valued measures (POVMs). A POVM is roughly an assignment of a positive operator $\hat{E}_x$ to each point in a set of alternatives, such that $\int_{\Gamma} dx \hat{E}_x = \hat{I}$. Through POVMs we can construct probabilities for alternatives that cannot be expressed in terms of self-adjoint operators, including regions of classical state space and---as we will see---time.

\end{enumerate}

In what follows, we will mostly work with pure states, tacitly assuming that the results straightforwardly apply to the mixed case. We will use mixed states only when their mixedness is physically important, for example, in discussing non-unitary dynamics.

\section{The principles of quantum theory: time evolution}

We can describe time evolution in either the Schr\"odinger or the Heisenberg picture. Hence, the time evolution principle can take two forms, both expressed in terms of the Hamiltonian $\hat{H}$ of the system, and the associated group of unitary operators.

\bigskip

\noindent \textbf{ Time evolution, Schr\"odinger form:} In a closed physical system and in absence of measurements, the quantum state evolves under the action of a family of unitary operators
 $e^{-i \hat{H} t}$ as $|\psi(t) \rangle = e^{-i \hat{H} t}|\psi(0) \rangle$, where  $t$ is the time.  Observables do not evolve. 
 
 \medskip
 
 \noindent \textbf{ Time evolution, Heisenberg form:} In a closed physical system and in absence of measurements, observables evolve  as
  $\hat{A}(t) = e^{i\hat{H}t}\hat{A}(0)e^{-i\hat{H}t}$. States do not evolve. 
 
 \bigskip
 
We can write the evolution equations in differential form as:
\begin{itemize}
\item Schr\"odinger's equation: $i\dfrac{d}{d t} |\psi(t)\rangle = \hat{H}|\psi(t)\rangle$,
\item Heisenberg's equation: $i \dfrac{d \hat{A}(t)}{d t} = [\hat{A}(t), \hat{H}]$.
\end{itemize}
 
The two pictures are completely equivalent, as far as physical predictions are concerned. We could   equally well use a modification of the probability principle (\ref{probaxiom}), asserting that if a measurement is carried out at time $t$ after the preparation, the associated probabilities are
     
 $$   \mathrm{Prob}(C, t) = \mathrm{Tr}\!\left(e^{-i \hat{H} t}\hat{\rho} e^{i \hat{H} t}\hat{P}_C\right).$$
 It is a matter of convenience whether one groups the action of the unitary operator with the projector  or with the state.
 
 \subsubsection*{Remarks}
 \begin{enumerate}
 
 \item The time $t$ in the evolution is determined by the spacetime causal structure, which is assumed to be absolute. It is not an observable. In practice, the time $t$ is recorded by a classical clock external to the system. 
     
     Bohr had said that time is part of the classical background, necessary to interpret any experiment:
     \begin{quote}
   {\small   The description of the experimental arrangement and the recording of observations must be given in plain language, suitably supplemented by the terminology of classical physics. This is a simple logical demand, since by the word ‘experiment’ we can only mean a procedure regarding which we are able to communicate to others what we have done and what we have learned. In particular, the space-time coordination of the observations belongs to the experimental arrangement.}\cite{Bohr}
     \end{quote}
 Bohr's conception appears to contradict the idea that gravity must be quantized. 

\item Dirac had insisted that the time $t$ in Schr\"odinger's equation is not a dynamical variable but a number. He motivated it by the observation that if there existed a self-adjoint time operator $\hat{T}$, with the Hamiltonian $\hat{H}$ generating time-translations, then we would have
     $$e^{i\hat{H}t}\hat{T}e^{-i\hat{H}t} = \hat{T} + t \hat{I}.$$
This would imply that 
     $[\hat{T}, \hat{H}] = i\hat{I}$. But then the spectrum of the Hamiltonian would have to be the full  real line. It would not be bounded from below, hence, it would be unphysical.
    This argument for the impossibility of a time operator  is now known as {\em Pauli's theorem}.
    
    \begin{exercise}
    For a free particle with mass $m$, the Hilbert space is $L^2(\R, dx)$ and the associated Hamiltonian $\hat{H} = \frac{\hat{p}^2}{2m}$ has positive spectrum. The operator $\hat{T} = \frac{m}{2}(\hat{x}\hat{p}^{-1} + \hat{p}^{-1}\hat{x})$ satisfies $[\hat{T}, \hat{H}] = i\hat{I}$. Doesn't this contradict Pauli's theorem?
    \end{exercise}

\item  The time evolution principle presupposes the existence of a time function, and of a time-independent Hamiltonian $\hat{H}$. It can be relaxed to include time-dependent Hamiltonians, and even non-unitary dynamics (see below), but it cannot work in absence of a time-function. The problem is that the notion of a quantum state is intrinsically linked with the notion of an instant of time, or, equivalently, a hypersurface of simultaneity. In absence of a time function, such hypersurfaces exist only locally, and it is not obvious that the notion of a quantum state is appropriate.
    
    Indeed, the Hilbert space ${\cal H}$ of the system ought to carry an index $t$ labelling the hypersurface of   simultaneity. However, if all surfaces are geometrically identical, we can identify different Hilbert space ${\cal H}_{t_1}, {\cal H}_{t_2}, \ldots$ through a natural isometry map.

\item The absence of a time operator implies that the Heisenberg time-energy uncertainty relation $\Delta E \Delta t \geq \frac{1}{2}$ cannot be expressed in terms of operator commutators. There are several alternative ways to derive uncertainty relations for time and energy, but these are all context-dependent, each reflecting a different physical or operational notion of time \cite{Busch}.

 \end{enumerate}
 
 \subsubsection*{Relativistic systems}
For relativistic systems in Minkowski spacetime, the time evolution principle must be substituted by a stronger condition. The whole Poincar\'e group must be represented on the Hilbert space by unitary transformations, not only time translations. This requirement places a strong restriction on the structure of relativistic quantum theories. 

In particular, the Hilbert space ${\cal H}$ of the system must carry the generators of the Poincar\'e group,   self-adjoint operators $\hat{M}_{\mu \nu}$ and $\hat{P}_{\mu}$ that satisfy commutation relations
\begin{align}
[\hat{M}_{\mu \nu}, \hat{M}_{\rho\sigma}] &= i( \eta_{\mu \rho} \hat{M} _{\nu \sigma} +  \eta_{\nu \sigma}\hat{M}_{\mu \rho} -\eta_{\nu \rho} \hat{M} _{\mu \sigma}  -  \eta_{\mu \sigma} \hat{M} _{\nu \rho} ),\\\
 [\hat{M}_{\mu \nu}, \hat{P}_{\rho}] &= i (\eta_{\mu \rho} \hat{P}_{\nu} - \eta_{\nu \rho} \hat{P}_{\mu}) \\\
[\hat{P}_{\mu}, \hat{P}_{\nu}] &= 0.
\end{align}

The unitary representation of the Poincar\'e group also leads to the Wigner classification of relativistic particles \cite{Wigner}. Irreducible unitary representations are labeled by the eigenvalues of the Casimir operators
\begin{equation}
\hat{P}^\mu \hat{P}_\mu \quad \text{and} \quad \hat{W}^\mu \hat{W}_\mu ,
\end{equation}
where $\hat W^\mu=\frac{1}{2}\epsilon^{\mu\nu\rho\sigma} \hat P_\nu \hat M_{\rho\sigma}$ is the Pauli--Lubanski vector. These correspond physically to the invariant mass and spin (or helicity for massless systems). Massive representations are characterized by mass $m>0$ and spin $s$, while massless representations are labeled by helicity. In this way, the assumption of unitary Poincar\'e symmetry determines the kinematical structure of relativistic quantum systems and provides the group-theoretic definition of elementary particles.
 
 \subsubsection*{Open system dynamics}
  The most general evolution law compatible with the probabilistic structure of quantum mechanics is a Completely Positive (CP) map, that is, a map $\Phi$ acting on density operators as
    \bey
    \Phi(\hat{\rho}) = \sum_a \hat{K}_a \hat{\rho}\hat{K}_a^{\dagger},
    \eey   
for general operators $\hat{K}_a$, subject only to the normalization condition $\sum_a \hat{K}^{\dagger}_a\hat{K}_a = \hat{I}$. Assuming time-homogeneity, we can express the most general quantum evolution law, as 
\bey
\hat{\rho}(t) = \Phi_t[\hat{\rho}(0)],
\eey
where $\Phi_t$ satisfies the semigroup property $\Phi_t[\Phi_{t'}[\hat{\rho}]] = \Phi_{t+t'}[\hat{\rho}]$. The associated evolution law is given by the Gorini-Kossakowski-Lindblad-Sudarshan (GKLS) equation
\bey
\frac{d\hat{\rho}(t)}{d t} = - i [\hat{H}, \hat{\rho}(t)] - \frac{1}{2} \sum_{\alpha}\left( \hat{L}_{\alpha}^{\dagger} \hat{L}_{\alpha} \hat{\rho}(t) +   \hat{\rho}(t) \hat{L}_{\alpha}^{ \dagger} \hat{L}_{\alpha} - 2    \hat{L}_{\alpha}\hat{\rho}(t) \hat{L}_{\alpha}^{ \dagger} \right), \label{LKme}
\eey
where $\hat{H}$ is the Hamiltonian, and the operators $\hat{L}_{\alpha}$ are known as  {\em Lindblad generators}. An analogous equation can be derived for Heisenberg-type evolution.

Then, why don't we take the GKLS equation as the fundamental equation for a closed system? There are three reasons.
\begin{itemize}
\item The GKLS equation takes pure states into mixed states, so it implies the loss of information. There is a persistent intuition that a closed system, at its fundamental level of description, should not lose information. However, the only CP maps that preserve purity, and do not lead to trivial dynamics, correspond to unitary transformations.
\item The Poincar\'e group cannot be represented by CP maps unless these reduce to unitary transformations. In particular, CP maps cannot handle the reversibility of time-translations. By restricting to future-directed time translations, we   can define a Poincar\'e semigroup that is represented by CP maps. However, this object does not carry the full symmetry of Minkowski spacetime, and there is little justification of its introduction. Furthermore, we lose important mathematical results, such as   Wigner's classification of particles.
\item No probes of elementary particles have yet uncovered the need to introduce non-unitary dynamics at the fundamental level. The Standard Model is fully unitary.
\end{itemize}
Having said that, it remains plausible that quantum gravity might involve some form of non-unitarity. Hawking famously suggested that this must be the case, as he argued that a unitary quantum gravity theory cannot possibly account for   black hole formation and evaporation, a process that appears to involve  tremendous loss of information \cite{Hawk76}. The issue of unitarity in black hole collapse has generated a tremendous debate lasting now fifty years. 

A key point is that the semiclassical spacetimes invoked in scenarios of complete black-hole evaporation need not admit global Cauchy surfaces. The standard framework of quantum evolution, in which a state specified on one Cauchy surface evolves unitarily to another, may therefore be inapplicable. The apparent loss of information may reflect not a failure of unitarity within this framework, but a failure of the spacetime to support the global state-evolution structure that unitarity presupposes.

\begin{exercise}
Show that a CP map that preserves purity for all states is either a  unitary transformation, $\hat{\rho} \rightarrow \hat{U} \hat{\rho}\hat{U}^{\dagger}$, or a reset channel, $\hat{\rho} \rightarrow |u\rangle \langle u|$, for a fixed vector $|u\rangle$. (Obviously, reset channels are not appropriate for  dynamics, as they collapse  the system into an effective one-dimensional Hilbert space.)

\end{exercise}

\subsubsection*{Path integrals}

Feynman's formulation of quantum dynamics expresses the evolution operator as a sum over paths. For a particle of mass $m$ in one dimension with Hamiltonian
\begin{equation}
\hat H=\frac{\hat p^2}{2m}+V(\hat x),
\end{equation}
the propagator
\begin{equation}
G_t(x_f,x_i)=\langle x_f|e^{-i\hat H t}|x_i\rangle
\end{equation}
can be constructed by dividing the time interval into small steps and inserting resolutions of the identity in the position basis. Evaluating the short-time matrix elements and taking the continuum limit leads formally to
\begin{equation}
G_t(x_f,x_i)= 
\int_{x(0)=x_i}^{x(t)=x_f}
\mathcal D x(\cdot)\, D p(\cdot)\,
e^{i S[x(\cdot), p(\cdot) ]},
\end{equation}
where $S[x(\cdot), p(\cdot) ] = \int_0^t ds\left[p\dot{x} - \frac{p^2}{2m} - V(x)\right]$ is the state space action.  The integration is taken over all paths connecting the initial and final points of the configuration space---momentum is not constrained at the endpoints.

The integral over momenta is Gaussian, and it can be performed exactly. This yields a path integral solely over configuration space histories,
\bey
G_t(x_f,x_i)= 
\int_{x(0)=x_i}^{x(t)=x_f}
\mathcal D x(\cdot)\, 
e^{i S[x(\cdot)]},
\eey
where now
\begin{equation}
S[x(\cdot)]
=\int_0^t ds \left[\frac{1}{2}m\dot x^2 - V(x)\right]
\end{equation}
is the classical Lagrangian action.

This expression, known as the Feynman path integral, represents quantum dynamics as an interference of contributions from all classical histories, each weighted by the phase $e^{iS}$. It provides a powerful computational tool and a convenient quantisation procedure, and it can be generalized to more complicated systems, including quantum fields.

\medskip

For applications to field theory, the   path integral acquires a broader   role. In particular, it provides a direct framework for the construction of the $S$-matrix, when the time interval is taken to be infinite,
and the field configurations satisfy appropriate asymptotic conditions as $t \to \pm \infty$. 

A typical Hamiltonian has the form
\begin{equation}
H = H_0 + g H_{\text{int}},
\end{equation}
where $H_0$ is quadratic in the fields and $H_{\text{int}}$ contains higher-order terms, with $g$ a small parameter. Expanding in powers of $g$, one obtains expressions involving integrals of the form
\begin{equation}
\int D q(\cdot)\, D p(\cdot) \exp \left\{ \frac{i}{\hbar} \int \left[ \sum_i p_i \dot{x}^i 
- \sum_{i,k} \left( A^{ik} p_i p_k + B_i^{\;k} q^i p_k + C_{ik} q^i q^k \right) \right] dt \right\}
Q[q(t), p(t)] 
\end{equation}
where $A^{ik}$, $B_i^{\;k}$, and $C_{ik}$ are constant matrices, and $Q$ is a polynomial in the dynamical variables. These integrals are Gaussian and can be evaluated explicitly. The resulting perturbative expansion is naturally organized in terms of Feynman diagrams. 
When $H_{\text{int}}$ depends only on $q_i$, the integration over momenta can be carried out, expressing the Feynman diagrams in terms of a configuration space path integral.

This structure is captured more systematically by introducing the generating functional
\begin{equation}
Z[J]=\int \mathcal D\phi \;
\exp\left\{i\left(S[\phi]+\int d^4x\,J(x)\phi(x)\right)\right\}, \label{zj}
\end{equation}
where the fields are assumed to vanish at spacelike and timelike infinity, and $J(x)$ is an external source. Functional derivatives of $Z[J]$ generate time-ordered correlation functions,
\begin{equation}
\langle 0|T\{\phi(x_1)\cdots \phi(x_n)\}|0\rangle
=
\left. \frac{1}{i^n}
\frac{\delta^n Z[J]}{\delta J(x_1)\cdots \delta J(x_n)}
\right|_{J=0}.
\end{equation}

These correlation functions encode the observable content of the theory and, through the LSZ reduction formula, determine the $S$-matrix describing scattering processes between asymptotic states. Thus, the path integral provides not only a representation of propagation, but also a generating functional for all scattering amplitudes of the theory.

Despite its usefulness, the path-integral formulation does not replace the Hilbert-space formulation of quantum theory. It provides a representation of the evolution operator and a generating functional for the S-matrix, while the Hilbert-space structure remains essential for defining observables, analysing quantum states, and describing measurements.
 
\section{The principles of quantum theory: state update}
Every probabilistic theory needs a rule about the incorporation of new information into  its models. Correspondingly, quantum theory requires a rule for updating the quantum state upon the results of measurements. In other words, this rule applies when I extracted some information
from the system, and  upgrade my future predictions by incorporating this information into my model.

For the simple case of measuring  an operator  $\hat{A} =\sum_na_n \hat{P}_n$ with discrete spectrum, the rule is the following.

\bigskip 

\noindent \textbf{State update rule:} If the measurement of $\hat{A}$ on a system prepared on a state $|\psi\rangle$ gives outcome $a_n$, the system after measurement is described by the state vector
\begin{eqnarray}
| \psi; a_n\rangle = \frac{\hat{P}_n |\psi \rangle }{\sqrt{\langle \psi|\hat{P}_n|\psi \rangle}}. \label{stupd}
\end{eqnarray}

 \medskip

Suppose now we carry a measurement of  $\hat{B} = \sum_m b_m \hat{Q}_m$  after the measurement of $\hat{A}$ in an ensemble described by
the state  $|\psi\rangle$.
The {\em conditional probability} of measuring  
  $b_m$ in the second measurement, provided that   $a_n$ was measured in the first is
  \begin{eqnarray}
\mbox{Prob}(b_m|a_n) =  \langle \psi; a_n |\hat{Q}_m | \psi; a_n \rangle = \frac{\langle \psi| \hat{P}_n\hat{Q}_m \hat{P}_n|\psi\rangle }{\langle \psi|\hat{P}_n|\hat{\psi}\rangle}. \label{condprob}
\end{eqnarray}

The {\em joint probability} of finding  $a_n$ in the first measurement and then finding $b_m$ in the second measurement is
\begin{eqnarray}
\mbox{Prob}(a_n, b_m) = \mbox{Prob}(b_m|a_n) \cdot \mbox{Prob}(a_n) =  \langle \psi| \hat{P}_n\hat{Q}_m \hat{P}_n|\psi\rangle. \label{jointprob}
\end{eqnarray}

Note that we can take Eq. (\ref{jointprob}) for joint probabilities as the fundamental principle for state update, rather than Eq. (\ref{stupd}). This is more satisfying conceptually, because Eq. (\ref{stupd}) appears to suggest that the  change in the quantum state as an actual physical process. 

The order of the measurements affects joint probabilities:
\begin{eqnarray}
\mbox{Prob}(a_n, b_m) \neq \mbox{Prob}(b_m, a_n)\label{jopruneq}
\end{eqnarray}
unless $[\hat{P}_n, \hat{Q}_m] = 0$ for all $n$ and $m$, or, equivalently, if $[\hat{A}, \hat{B}] = 0$. In that case,
   \begin{eqnarray}
\mbox{Prob}(a_n, b_m)   =  \langle \psi| \hat{P}_n\hat{Q}_m |\psi\rangle. \label{joprcom}
\end{eqnarray}
The corresponding joint probabilities for classical systems do not depend on the order of the two measurements. The difference is not a small error of a few percent: changing the order of measurements gives dramatically different results. The quantum state update rule introduces a direction of time. To quote Landau and Lifshitz \cite{LL}:

\begin{quote}
{\small The measuring process in quantum mechanics has a ``two-
faced" character: it plays different parts with respect to the past and future
of the electron. With respect to the past, it ``verifies" the probabilities of the
various possible results predicted from the state brought about by the
previous measurement. With respect to the future, it brings about a new state. Thus the very nature of the process of measurement involves
a far-reaching principle of irreversibility.

... The basic equations of quantum mechanics are in
themselves symmetrical with respect to a change in the sign of the time; here
quantum mechanics does not differ from classical mechanics. The irreversibility of the process of measurement, however, causes the two directions
of time to be physically non-equivalent, i.e. creates a difference between the  
future and the past.}
\end{quote}
 The ordering of quantum measurements must be compatible with the causal structure of spacetime. In a multi-measurement set-up involving self-adjoint operators $\hat{A}_1, \hat{A}_2, \ldots, \hat{A}_n$, there should exist   a map ${\cal X}$ mapping each operator on a subset ${\cal X}(\hat{A}_i)$ of spacetime. The spacetime causal order $\po$ should then be reflected on the order of measurements: 
 \begin{itemize}
 \item If ${\cal X}(\hat{A}_i) \po {\cal X}(\hat{A}_j)$, then the spectral projectors of $\hat{A}_i$ must act on $|\psi\rangle$ prior to the spectral projectors of $\hat{A}_j$.
 \item If ${\cal X}(\hat{A}_i) | {\cal X}(\hat{A}_j)$ the order of the corresponding projectors is irrelevant. This is only possible if $[\hat{A}_i, \hat{A}_j] = 0$. Hence, operators corresponding to measurements at spacelike separated regions must commute. 
 \end{itemize}
However, the rule (\ref{stupd}) is rather inflexible regarding the nature of the map ${\cal X}$. It amounts to a global and instantaneous change of the quantum state, and since the state is defined at a Cauchy surface, ${\cal X}(\hat{A})$ can only be a Cauchy surface corresponding to an instant of time. This means that we must generalize Eq. (\ref{jointprob}) to take into account time evolution through a Hamiltonian $t$. 
\begin{eqnarray}
\mbox{Prob}(a_n, t_1; b_m, t_2) = Tr(\hat{Q}_m e^{-i\hat{H}(t_2-t_1)}\hat{P}_n   e^{-i\hat{H} t_1} \hat{\rho} e^{i\hat{H} t_1}\hat{P}_n e^{i\hat{H}(t_2-t_1)}). \label{joint2}
\end{eqnarray}

 However, the identification of ${\cal X}(\hat{A})$ with a Cauchy surface creates problems in relativistic setups. Certainly, all measurements occur in a finite spacetime region, and the use of Cauchy surfaces is an idealization. The difficulty arises because we cannot order Cauchy surfaces invariantly.

\begin{exercise}
Let $\Sigma_1$ and $\Sigma_2$ be two Cauchy surfaces. Then, none of the following statements is true: $\Sigma_1 \po \Sigma_2$, $\Sigma_2 \po \Sigma_1$, and $\Sigma_1 | \Sigma_2$. 
\end{exercise}
 
Hence, there is no covariant way to assert that one measurement occurred before the other. We cannot employ our spacetime intuitions anymore: we would be willing to say that if all procedures in the measurement of $\hat{A}$ have concluded, and afterwards (with respect to the spacetime causal structure) we start measuring $\hat{B}$, then we expect that ${\cal X}(\hat{A}) \po {\cal X}(\hat{B})$. But we cannot say this because ${\cal X}(\hat{A})$ and  ${\cal X}(B)$ are Cauchy surfaces. The same holds for Bell-type experiments 

If we insist on using our spacetime common sense that measurements occur in finite spacetime regions, we invariably come across paradoxes \cite{Sorkin}. Since our spacetime common sense is scientifically precious---and extremely successful---it is a better option to take Eq. (\ref{joint2}) as a special case, with a restricted domain of validity, and to seek a more general rule for joint probabilities in multiple-measurement set-ups. This motivates much of the recent work in formulating measurements in quantum field theory \cite{QTP1, OkOz, QTP3, FeVe,  GGM22, PTM, PRA24, QTP4, FeVe2}. The ideal end result is a general rule for the probability for the measurement of $n$ different observables $\hat{A}_1, \hat{A}_2, \ldots, \hat{A}_n$ at different spacetime regions ${\cal X} (\hat{A}_1), {\cal X}(\hat{A}_2), \ldots, {\cal X}(\hat{A}_n)$, in a way that the causal relation between the regions will be reflected in the structure of the probabilities. We will come back to this topic when we discuss the notion of quantum events.

\subsubsection*{Quantum post-selection}
 When multiple measurements are performed on a quantum system, one may define \emph{post-selected} probabilities, in which only those experimental runs are retained for which a later measurement yields a specified outcome, while all others are discarded.

Suppose that we first measure the observable $\hat{A} = \sum_n a_n \hat{P}_n$ 
and subsequently the observable
$\hat{B} = \sum_m b_m \hat{Q}_m$.
We restrict attention to those runs in which the second measurement yields the outcome $b_1$. The relevant quantity is then the conditional probability $\mathrm{Prob}(a_n \mid b_1)$,
namely, the probability of obtaining $a_n$ in the first measurement given that the second measurement yields $b_1$. This differs from the usual conditional probability, because the conditioning refers to a later measurement outcome.

For an initial state $|\psi\rangle\langle \psi|$ and  a rank-one projector $\hat{Q}_1 = |\phi\rangle\langle\phi|$, we obtain
\begin{equation}
\mathrm{Prob}(a_n \mid b_1)
= \frac{\mathrm{Prob}(a_n, b_1)}{\sum_n \mathrm{Prob}(a_n, b_1)}
= \frac{|\langle \psi | \hat{P}_n | \phi \rangle|^2}{\sum_n |\langle \psi | \hat{P}_n | \phi \rangle|^2}
\label{postsel}
\end{equation}
The denominator in Eq.~(\ref{postsel}) is the probability of obtaining $b_1$ in the second measurement, irrespective of the outcome of the first.
 
In Eq.~(\ref{postsel}), there is a symmetry between the states $|\psi\rangle$ and $|\phi\rangle$: it is no longer possible to distinguish which represents the initial state and which the final. Post-selected measurements thus suggest a description in terms of two states, the initial state $|\psi\rangle$ defining the pre-selection, and the final state $|\phi\rangle$ defining the post-selection.

One may therefore take Eq.~(\ref{postsel}) as a fundamental principle of quantum theory, replacing the usual probability postulate by one that depends on both initial and final conditions \cite{ABL}. In most experimental situations, no post-selection is performed, and summing over all possible final states recovers the standard probability rule. Hence, within a minimal interpretation focused solely on measurement outcomes, this modification makes little practical difference.

However, the conceptual implications are profound for quantum cosmology. Equation~(\ref{postsel}) places final conditions on the same footing as initial ones in determining quantum probabilities. This opens the possibility of a scientific description not only of the initial conditions of the Universe—long a central concern in physics—but also of its final conditions, a rather distinct perspective that   cannot be discounted {\em a priori}. It turns out that it can also make observable predictions at the cosmological level \cite{Ana25}.
 
\section{Quantum histories}
Classical physics admits both an evolutionary and a histories description. In the former, the basic object is a microstate evolving in time, as in Hamiltonian mechanics. In the latter, the basic object is the set of possible histories (trajectories), and a principle that selects the physically realized ones (the action principle). 

Quantum theory is standardly formulated as a Hamiltonian theory, that is, in an evolutionary form. Path integrals, when used merely as a representation of the propagator, do not define a histories description.

A histories formulation of quantum theory exists, but it has several complications, as it touches upon many issues of quantum foundations.
The starting  point  is a
    {\em history}, that is, a sequence of  properties of a physical
system at successive instants of time.  An $N$-time history
$\alpha$
corresponds to a sequence $\hat{P}^{(1)}_{a_1}, \hat{P}_{a_2}^{(2)}, \ldots \hat{P}^{(N)}_{a_N}$ of projectors
each corresponding to an outcome $a_i$ in the measurement of an observable $\hat{A}^{(i)}$ at time $t_i$, where $i = 1, 2, \ldots, N$. The associated joint probabilities are given by the extension of Eq. (\ref{joint2}) to $N$ measurements. To express them in a compact form, we define the 
 history operators \begin{eqnarray}
\hat{C}_{\alpha} =  \hat{P}^{(N)}_{a_N}(t_N) \ldots \hat{P}^{(2)}_{a_2}(t_2) \hat{P}^{(1)}_{a_1}(t_1), \label{classoper}
\end{eqnarray}
where $\hat{P}^{(i)}_{a_i}(t_i) : = e^{i\hat{H}(t_i-t_0)}\hat{P}^{(i)}_{a_i}e^{-i\hat{H}(t_i - t_0)}$ is the Heisenberg-picture evolution of $\hat{P}^{(i)}_{a_i}$. 
Then, the probability of $\alpha$ is
\begin{equation}
p(\alpha) = \langle \psi| \hat{C}_{\alpha}^{\dagger}\hat{C}_{\alpha} |\psi\rangle. \label{histprob2}
\end{equation}
Eq. (\ref{histprob2}) was proposed by Wigner and collaborators as the essential content of quantum theory \cite{Houtapel}, as it provides the probabilities of any conceivable quantum experiment. They noted, however, that the history operators should probably be modified in relativistic QFT, as the definition (\ref{classoper}) presupposes state updates over an entire Cauchy surface.

The decoherent histories approach, developed by Griffiths, Omn\'es, Gell-Mann and Hartle  \cite{Gri, Omn1, Omn2, GeHa1, hartlelo},  starts from the re-interpretation of histories as a sequence of properties of a physical system rather than as a sequence of measurement outcomes. The benefit   is that histories-as-properties  have a powerful logical structure. We can make propositions about histories and relate these propositions by logical operations like AND ($\wedge$), OR ($\vee$) , NOT ($\neg$), and so on. The set of history propositions forms a lattice, and it includes the trivially true proposition $I$ and the trivially false proposition $\emptyset$. 

The price  is that Eq. (\ref{histprob2}) does not satisfy the Kolmogorov additivity condition
\bey
p(\alpha \vee \beta ) = p(\alpha) + p(\beta) \label{eqkol}
\eey
 for any pair of disjoint histories $\alpha$ and $\beta$.  

There is a partial resolution: we can define probability measures when restricting to specific sets of histories.  To this end, we first define the {\em decoherence functional} $d$, a complex-valued function of pairs of histories, as \index{decoherence functional}
\begin{equation}
d(\alpha, \beta) = Tr \left( \hat{C}_{\alpha} \hat{\rho}_0 \hat{C}_{\beta}^{\dagger}\right). \label{decfun}
\end{equation}
The diagonal elements $d(\alpha, \alpha)$ of the decoherence functional coincide with the probabilities $p(\alpha)$ of Eq. (\ref{histprob2}). 

Let $\Omega$ be an exclusive and exhaustive set of histories, that is, a set of histories  $\alpha_i$ labeled by an index $i$, such that $\alpha_i \wedge \alpha_j = \emptyset$ for $i \neq j$, and $\vee_i \alpha_i = I$. 
 If all histories in $\Omega$ satisfy the decoherence condition
\begin{equation}
 d(\alpha_i, \alpha_j) = 0, \;\;\; \mbox{for} \;\;\; i \neq j, \label{decc}
\end{equation}
then, Eq. (\ref{eqkol}) is satisfied, and we can define a probability measure on $\Omega$. Then, $\Omega$ is called a {\em consistent set} or a {\em framework}. 
In fact, the weaker condition $\mbox{Re} d(\alpha, \beta) = 0, \,\, \mbox{if}\,\, \alpha \neq \beta$, suffices for the definition of a framework, but the condition (\ref{decc}) is more appropriate for discussions of the classical limit. The classical limit is a  set of histories that  can be described approximately by a stochastic process; in this case, we can relax 
\emph{} condition (\ref{decc}) to that of approximate decoherence, $|d(\alpha, \beta)| < \epsilon << 1$ for all $\alpha \neq \beta$.

A framework in the decoherent histories approach plays the role of a sample space, within which one can reason using classical logic and standard probability theory. Histories can be combined with logical connectives such as AND, OR, and NOT, and implication is defined probabilistically, with $\alpha \rightarrow \beta$ whenever $p(\alpha \cap \beta)=p(\alpha)$. Within a given framework, the theory allows one to consistently assign properties to physical systems without reference to measurement. Measurements can then be treated as ordinary physical processes, and logical implication can relate macroscopic pointer values to microscopic quantities. In suitably coarse-grained frameworks, the probability assignments reproduce approximate classical equations of motion, providing a precise sense in which classical behavior emerges from quantum theory.

However, this construction is inherently framework-dependent, and there is in general an infinite number of incompatible frameworks, none of which is singled out by the theory. As anticipated by the Kochen–Specker theorem \cite{KoSp}, combining propositions from different frameworks leads to contradictions, and different frameworks may yield mutually incompatible descriptions of reality. This leads to a strong form of complementarity: there is no single exhaustive description of a system, but rather multiple incompatible ones that cannot be combined. 

While this pluralism is formally consistent, it raises difficulties in practice. 
When describing measurements, there are infinitely many frameworks that
could describe it---but the experimentalist has only one. Since the theory provides no principle for selecting the relevant framework, one must choose it based on prior notions of classicality. In this sense, although the decoherent-histories approach removes measurement from the fundamental probability rules, it reintroduces a special role for measurement through the need to select frameworks that reflect our prior classical understanding of how a measuring apparatus operates.

However, decoherent histories is the basis  for  genuinely spacetime-based generalizations of quantum theory. In such generalizations, we can consider  
  more general types of histories than  sequences of projectors \cite{Ish94}, and more general decoherence functionals than those of Eq. (\ref{decfun}) \cite{IL94, ILS, AnSav26}. Indeed, the space of histories ${\cal V}$ needs not be tied to the standard Hilbert space structure, and the decoherence functional can be defined abstractly as 
  a map $d: {\cal V} \times {\cal V} \rightarrow \C$ that satisfies specific axioms. 
  
  We will return to the decoherent histories in more detail in Chapter 6.

 \section{Events and clocks}
 \subsection{Events}
 The notion of events in quantum theory is rather more complex than the classical one. They can no longer be identified with surfaces in the state space. Such surfaces have no physical meaning in quantum theory, where properties of a system (after measurement) correspond to linear subspaces of the Hilbert space. 
 
A large class of quantum events---and, according to some interpretations, all quantum events---are defined by reference to measurement and associated with the production of a macroscopic record. A measurement, however, may be a complex process extending over a substantial interval of time and involving several stages of interaction between the quantum system and the apparatus. A particle, for example, may travel for a long time through the arms of a large interferometer before eventually being detected. Which stage of this extended process should be identified with the occurrence of the measurement event?

Following a suggestion of Wheeler \cite{Wheeler}, many experiments have been carried
out that involve a delayed choice of the observable that is being measured, i.e.,
the observable is selected long after the particle has entered an interferometer.
The result is unambiguous---see, for example,\cite{Jacques}---and it is also
conceptually simple. A measurement occurs when a macroscopic record appears on
the detector. What happened before the emergence of this record should not be
viewed as part of the measurement per se, but part of the time evolution of the
quantum system. In particular, the state update rule should be applied at the moment that we
obtain information from the quantum system.
 
 If we identify a quantum event with the emergence of a macroscopic record of observation, it is possible to develop a formalism of quantum events leading to a probability distribution for the time of the event's occurrence. This is the Quantum Temporal Probabilities (QTP) approach to measurements \cite{QTP1, QTP3}.
  In this framework, we consider a composite physical system that consists of a microscopic and a macroscopic component. The microscopic component is the quantum system to be measured and the macroscopic component is the measuring device.

 Let ${\cal H}$ be the Hilbert space of the composite system, and $\hat{H}$ the associated Hamiltonian operator. A measurement event is defined 
 with respect to a split of ${\cal H}$ in two subspaces: in ${\cal H}_+$ a definite macroscopic record of detection exists; in ${\cal H}_-$ is does not.
The subspace ${\cal H}_+$ describes the states of the system in which a macroscopic record of detection exists.   We
denote  the projection operator onto ${\cal H}_+$ as $\hat{P}$ and the projector onto ${\cal H}_-$ as $\hat{Q} := 1  - \hat{P}$.  

Our first aim is to construct a history operator $\hat{C}_{B}$ that corresponds to the transition ${\cal H}_- \rightarrow {\cal H}_+$ taking place at the interval $B = [t_1. t_2]$. First, we will choose $B = [t, t + \delta t]$, and   keep only leading-order terms with respect to $\delta t$. At times prior to $t$, the state
lies in ${\cal H}_-$. This is taken into account by evolving an initial state $|\psi_0 \rangle \in {\cal H_-}$ at $t = 0$, with the restricted propagator in ${\cal H}_-$,
 \begin{eqnarray}
 \hat{S}_t =  \lim_{N
\rightarrow \infty} (\hat{Q}e^{-i\hat{H} t/N} \hat{Q})^N. \label{restricted}
\end{eqnarray}
By assumption, the transition occurs at some instant within the time interval $[t, t+\delta t]$, hence, there is no constraint in the propagation from $t$ to $t + \delta t$; the propagation  is implemented by  the unrestricted evolution operator   $e^{-i \hat{H} \delta t} \simeq \hat{I} - i \delta t \hat{H}$. Then, the system has transitioned to ${\cal }_+$ and there is the action of $\hat{P}$. For times greater than $t + \delta t$, there is no constraint, so the amplitude
evolves as $e^{-i \hat{H} (T-t)}$  until some final moment $T$.

The successive operations above yield  a state vector $ e^{-i\hat{H}T} \hat{C}_t \delta t |\psi_0\rangle$
where 
\bey
 \hat{C}_t := e^{i \hat{H}t} \hat{P} \hat{H}\hat{S}_t.
\eey
 Since  the resulting state vector is  proportional to $\delta t$, it defines  a {\em density} with respect to time. Hence, for any time interval $B \subset \R^+$, we can define
 $$\hat{C}_{B} = \int_B dt e^{i \hat{H}t} \hat{P} \hat{H}\hat{S}_t.$$

The resulting probabilities are  
\begin{eqnarray}
\mbox{Prob}(t \in B) = 
  \int_B \,  dt \, \int_B dt' \langle \psi_0|\hat{C}^{\dagger}_{t'} \hat{C}_t|\psi_0\rangle. \label{prob1}
\end{eqnarray}
 These quantities   do not define a probability measure with respect to time, because in general, they do not satisfy the Kolmogorov additivity condition $\mbox{Prob}(B_1 \cup B_2) = \mbox{Prob}(B_1) + \mbox{Prob}(B_2)$ for $B_1 \cap B_2 = \emptyset$. They only do so, if
\begin{eqnarray}
{\cal D}(B_1, B_2):=  2 Re \left[ \int_B \,  dt \, \int_{B'} dt' \langle \psi_0|\hat{C}^{\dagger}_{t'} \hat{C}_t|\psi_0\rangle \right] \label{decond}
 \end{eqnarray}
 vanishes. 
In a macroscopic system (or in a system with a macroscopic component) one expects that Eq. (\ref{decond}) holds with a good degree of approximation, given a sufficient degree of coarse-graining \cite{GeHa2, hartlelo}.
   Thus, a necessary condition for the time of transition to be associated to macroscopic records is the existence of a
    coarse-graining time-scale $\sigma$, such that  $|{\cal D}(B_1, B_2)|$ is negligible
    for  $ [B_1], [B_2] >> \sigma$; $[B]$ stands for the width of the interval.
    
Then, the family of positive operators  $\hat{E}_B = \hat{C}^{\dagger}_B     \hat{C}_B$  determines the probability of occurrence of any event for any initial  state in ${\cal H_-}$. Crucially, the time of the event is a random variable. Note that $\hat{E}_{\R_+} \leq \hat{I}$, so we need to also consider the positive operator $\hat{E}_N = \hat{I} - \hat{E}_{\R_+}$  that no such transition ever occurred.

\subsubsection*{Remarks}
\begin{enumerate}
\item We can split the half-line, into intervals $B_n= [t_{n}, t_{n+1}]$, where $t_n = n \Delta$, where $\Delta >> \sigma$. We can then define a time operator
\bey
\hat{T} = \sum_{n=0}^{\infty} t_n \hat{E}_{B_n}.
\eey
 
This vanishes trivially on ${\cal H}_+$, but is ill-defined on the subspace where $\hat{E}_N$ vanishes.

\item A  system  localized in a spatial region $K_t$ corresponds to a world-tube $\cup_t K_t$ in spacetime. Hence, each positive operator $\hat{E}_B$ corresponds to a spacetime region $\cup_{t\in B}K_t$, that is, a think slice of the world-tube. In this sense, the collection of all $\hat{E}_B$ for an event $\alpha$ is the direct analogue of the classical map ${\cal X}(\alpha)$ that embeds the event into a spacetime region. 
    
   Correspondingly, we can use the maps $\hat{E}_B$ to define causal relations between different events. If the events occur in different systems, each described by Hilbert spaces ${\cal H}_{\alpha}$, $\alpha = 1, 2, \ldots, n$, we can associate one family of positive operators $\hat{E}^{\alpha}_B$ on ${\cal H}_{\alpha}$ to each event.  

\begin{exercise}
Take $n=2$, and assume that both operators share the same world-tube. Construct positive operators $\hat{E}_{\alpha_1 \po \alpha_2}$ and $\hat{E}_{\alpha_1 | \alpha_2}$ associated to $\alpha_1 \po \alpha_2$ and $\alpha_1 | \alpha_2$, respectively.
\end{exercise}
We see that the partial ordering is a random variable, expressed by a POVM, defined through the absolute causal structure of Minkowski spacetime. 
In particular this means that we can construct states of indefinite causal order, for examples superpositions of the form $|\psi_1 \rangle + |\psi_2\rangle$, where $|\psi_1 \rangle$ is an eigenvector of $\hat{E}_{\alpha_1 \po \alpha_2}$ and $|\psi_2\rangle$ of $\hat{E}_{\alpha_2 \po \alpha_1}$.

\item Using explicit models for the measurement apparatus, we can construct quantum probabilities for any set-up that involves temporal observables. 
\begin{exercise}
Construct an elementary measurement model. Let the measured system is described by the Hilbert space ${\cal H}_s$ and the macroscopic apparatus by ${\cal H}_{a}$.  The latter is split as ${\cal H}_{a+} \oplus{\cal H}_-$. The Hamiltonian splits as $\hat{H}_s + \hat{H}_a + \hat{V}$, where we take $\hat{V} = \hat{K}\otimes \hat{J}$. It is convenient to assume that ${\cal H}_{a-}$ consists only of the ground state $|\Omega\rangle$ of the apparatus, where $\langle \Omega|\hat{J}|\Omega\rangle$. Show that to leading order in the interaction
$$ \hat{C}_B = \hat{K}(t) \otimes \hat{J}(t)|\Omega\rangle \langle \Omega|, $$
in terms of interaction picture operators $\hat{A}(t)$ and $\hat{J}(t)$.
This model can be augmented to account for records of any microscopic observable in addition to the detection time (energy, momentum, spin, internal degrees of freedom).
\end{exercise}
The application of this formalism to the time-of arrival problem is instructive. Rather than asking ``where is the particle now" as in standard measurement theory, we ask  ``When is the particle here?", a question that is often more appropriate for many experimental set-ups.
    Through QTP,  we can construct the probability $P(L,t) dt$  that a detector located at $x = L$ records a particle at time $t$ for any relativistic particle \cite{QTP3}. Note that in this construction,  $L$ is a constant and $t$ is a random variable. 
 
\item There is an alternative concept of events, motivated by recent studies of indefinite causal ordering in quantum information \cite{OCB12, CDPV13}. In this context,  an event is identified with a local operation on a quantum system, such as a step in a computation. Then, events are associated with quantum channels or operations, and their ordering need not be definite, as exemplified by the quantum switch, where the sequence of operations is controlled by another quantum system. Such indefinite causal structures have even been realized experimentally \cite{exp1, exp2, GR20}. 
    
    However, this notion of event, rooted in quantum information theory, differs from---and may conflict with---the notion of  events presented here, and with the notion of an event in classical general relativity, highlighting the importance of clarifying what is meant by an event in discussions of causality \cite{ViCo, ViRe}. On the other hand, a splitting of the notion of an event in different sub-concepts may be what is needed in order to address one of the key problems of quantum gravity, the problem of agency---see Sec. 4.1. Ref. \cite{AnSav26} proposes that the indefinite causal structure of events of this form can be interpreted as a manifestation of a more general structure, the superposition of dynamics, that is made possible by histories theory.
 
\end{enumerate}

\subsection{Clocks}

Classical clocks admit three complementary characterizations:
\begin{itemize}
\item as periodic or quasi-periodic systems—such as a pendulum or the Earth's rotation—where time is inferred by counting cycles and assigning a duration to each period;

\item as dynamical variables that evolve in a one-to-one and predictable manner with respect to a background time parameter, so that their instantaneous value directly labels time;

\item as sequences of events with a predictable (possibly probabilistic) distribution in time—for example, radiometric dating relies on estimating the statistics of nuclear decay events.
\end{itemize}

All three notions have analogues in quantum theory, but each encounters characteristic limitations. In the first case, quantum clocks based on rotations or phase evolution (e.g., atomic transitions) encode time in the accumulated phase, with a macroscopic signal registered upon completion of a full $2\pi$ cycle. However, this description is intrinsically approximate: phase cannot be measured directly without destroying coherence, and practical readout requires large ensembles, leading to uncertainties from quantum fluctuations, finite coherence times, and measurement noise\footnote{In atomic clocks, the quantum system is not itself the clock. Rather, it stabilizes a classical oscillator through intermittent measurements, and time is identified with the phase of the stabilized oscillator.}.

A simple model of such a quantum phase clock is provided by equally spaced energy levels,
\[
\hat{H} = \omega \sum_{n=0}^{N-1} n \, |n\rangle\langle n|,
\]
and the corresponding phase states
\[
|\theta \rangle = \frac{1}{\sqrt{N}}\sum_{n=0}^{N-1} e^{in\theta} |n\rangle.
\]
Time evolution gives
\[
e^{-i\hat{H}t}|\theta\rangle = |\theta + \omega t\rangle,
\]
so the phase $\theta$ tracks time linearly. Operationally, however, this information can only be extracted statistically from  measurements in a statistical ensemble, not individual systems, since a single measurement disrupts the phase coherence.

In the second case, one seeks an operator $\hat{X}$ whose expectation value tracks time monotonically and bijectively. This ideal requirement would imply a canonical commutation relation with the Hamiltonian and is ruled out, in general, by Pauli's theorem. A weaker alternative is to require
\begin{equation}
\langle \hat{X}(t)\rangle = a + bt, \label{pseudoclock}
\end{equation}
for suitable constants $a$ and $b$. In principle, such a system could function as a clock without requiring an ensemble, provided $\hat{X}$ can be measured with minimal disturbance. However, this comes at the cost of increasing fluctuations: in generic systems, the variance of $\hat{X}(t)$ grows with time, eventually rendering the clock unreliable.

\begin{exercise}
Equation (\ref{pseudoclock}) is satisfied by the position operator of a non-relativistic free particle. Show that the uncertainty $\Delta X(t)$ increases asymptotically with $t$, demonstrating that wave-packet dispersion destroys the clock behavior.
\end{exercise}

The third case is exemplified by systems that generate events with well-defined statistical distributions. A standard example is a population of unstable systems decaying with rate $\Gamma$, where the clock variable is the probability distribution $P(t)$ for detecting decay products at time $t$. Owing to the robustness of the exponential decay law, such clocks can remain accurate over long timescales, and their precision improves with the size of the ensemble. Radiometric dating is a common application of such clocks.

In summary, while classical notions of timekeeping can be emulated in quantum systems, they cannot be realized exactly at the level of individual systems. Reliable quantum clocks instead emerge either as approximate constructions valid over finite intervals, or as statistical devices based on large ensembles and well-controlled probability distributions.

\subsubsection*{Ordering events without clocks}
It is important to distinguish between coarse-grained and fine-grained descriptions of causal ordering. In the examples discussed in Chapter 2, the measurement of causal order is coarse-grained. The apparatus records detection events together with their associated times, and the ordering of events is inferred by comparing these time records. In this sense, coarse-graining means that detailed information about the precise detection process is discarded, and only the clock readings are retained.

Quantum theory, however, allows for a more refined notion of causal ordering. In principle, one may define probabilities directly for the ordering of events, without referring to detection times at all. In such a framework, the apparatus determines ``which event happened first'' without recording the time at which either event occurred. Temporal order is then encoded directly in the structure of the measurement outcomes rather than inferred from clocks.

The notion that causal order can be established operationally without measuring time is already familiar from pre-Newtonian physics. A simple example is a race decided by a string stretched across the finish line. Each runner defines an event by interacting with the string, and the winner is the one who breaks it first. The ordering of events is determined directly through a local physical process, without any reference to synchronized clocks or temporal coordinates. In this case, causal order is operationally more primitive than the measurement of time intervals.

The key idea is to correlate the ordering of detection events with distinct transitions in a quantum system. Consider two particles directed towards a detector that can register either one, but not both simultaneously \cite{AnPl23}. As a concrete realization, we use a three-level system (3LS) with states $|0\rangle, |1\rangle$, and $|2\rangle$. Particle 1 couples only to the transition $0 \rightarrow 1$, while particle 2 couples only to the transition $0 \rightarrow 2$. After the interaction, if the system is found in state $|1\rangle$, we infer that particle 1 has been detected, and similarly for state $|2\rangle$.

Strictly speaking, to conform with the definition of an event as a recorded measurement, one could place a second identical 3LS downstream, so that the particle not absorbed by the first system may also be detected. However, this is unnecessary for determining the causal ordering of events. If particle 1 is recorded in the first 3LS, then particle 2 will either be recorded later or not recorded at all. In both cases, the ordering is the same: the record of particle 1 precedes that of particle 2. Hence, for the purpose of determining the causal order of events a single 3LS suffices.

This setup generalizes naturally to $n$ events. One requires $n$ distinguishable particles (e.g., by their energies) and a sequence of detecting systems with appropriate level structures, so that each detection event corresponds to a distinct order of events---again without measuring time. Thus, there exist ordering observables in
quantum theory that do not reduce to the measurement of time.

\begin{exercise}
To implement the model, we describe the particles by a free scalar field $\hat{\phi}(x)$ of mass $m$, prepared in states with distinct energies. The particles interact with a single 3LS located at $\mathbf{x}=0$. The total Hamiltonian is
$
\hat{H} = \hat{H}_{\phi} + \hat{H}_{3LS} + \hat{H}_{int},
$
where
\begin{equation}
\hat{H}_{\phi} = \int d\mathbf{k} \, \epsilon_{\mathbf{k}} \, \hat{a}^{\dagger}_{\mathbf{k}} \hat{a}_{\mathbf{k}},
\end{equation}
with $d\mathbf{k}= d^3k/(2\pi)^3$, and $\hat{H}_{3LS} = \Omega_1 |1\rangle \langle 1| + \Omega_2 |2\rangle \langle 2|$.
The interaction Hamiltonian is
\begin{equation}
\hat{H}_{int} = \sum_{a=1}^2 \lambda_a \int \frac{d\mathbf{k}}{\sqrt{2\omega_{\mathbf{k}}}}
\left( \hat{a}_{\mathbf{k}} \hat{u}_{a+} + \hat{a}^{\dagger}_{\mathbf{k}} \hat{u}_{a-} \right),
\end{equation}
where $\lambda_a$ are coupling constants associated with the transitions $0 \rightarrow a$, and $
\hat{u}_{a+} = |a\rangle \langle 0|,  \hat{u}_{a-} = |0\rangle \langle a|, \qquad a=1,2.$ Construct the probabilities for the transition $|0\rangle \rightarrow |1\rangle$ and for $|0\rangle \rightarrow |2\rangle$---follow \cite{AnPl23}.
 \end{exercise}
 \section{Time and causality in quantum field theory}

In particle physics, quantum field theory is most commonly applied through the
$S$-matrix. The scattering matrix relates asymptotic ``in'' states, prepared in
the distant past, to asymptotic ``out'' states, detected in the distant future.
For interactions through a Hamiltonian density operator $\hat{\HH}_I(x)$, the S-matrix is given by the Dyson formula
\bey
\hat{S} = \hat{I} + \sum_{n=1}^{\infty}    \frac{(-i)^n}{n!} \int  d^4x_1 \int  d^4x_2 \ldots \int  dx_n   {\cal T} [\hat{\HH}_I(x_1)\hat{\HH}_I(x_2)\ldots \hat{\HH}_I(x_n)]. \label{smat}
\eey
where ${\cal T}$ designates the time ordered product.

The S-matrix framework has been extraordinarily successful, but its operational content
is restricted to asymptotic preparation and detection. Although QFT describes
finite-time dynamics through fields and correlation functions, the $S$-matrix
framework does not by itself provide a general prescription for assigning
probabilities to localized measurements, interventions, and events at finite
times.

This limitation is particularly important for causality. In relativistic QFT,
causality is usually associated with \emph{microcausality}: physical observables
localized in spacelike-separated regions commute,
\[
 [\hat A(x),\hat B(y)]_{\pm}=0
 \qquad \text{if }x\text{ and }y\text{ are spacelike separated},
\]
where the plus sign (anti-commutator) applies when both fields are fermionic. Microcausality expresses
the algebraic compatibility of observables in spacelike-separated regions.
It ensures that the time-ordered products appearing in the S-matrix are compatible with Lorentz
covariance. In particular, microcausality implies that 
\[
 [\hat{\HH}_I(x),\hat{\HH}_I(y)]=0,
 \qquad (x-y)^2<0,
\]
so that  opposite temporal orderings
of two spacelike-separated points, associated to different inertial observers, do not change the time-ordered products in Eq. (\ref{smat}).

%Microcausality guarantees the absence of physical response outside the
%light cone. In linear-response theory, the change in the expectation value of a
%local observable \(\hat A(x)\) produced by a perturbation coupled to
%\(\hat B(y)\) is governed by the retarded commutator
%$$
 %G_{\mathrm R}(x,y)
 %=i\theta(x^0-y^0)\langle[\hat B(y),\hat A(x)]\rangle .
%$$
%Microcausality implies that this expression vanishes at spacelike separation.
%A small localized perturbation can therefore affect an observable only within its
%causal future.  

Microcausality does not by itself guarantee  no faster-than-light transmission of information, especially  set-ups that involve external interventions or measurements.

An operational formulation of relativistic causality is provided by
\emph{non-signalling}. Suppose that $x$ and $y$ label operations performed in
two spacelike-separated regions and that $a$ and $b$ are their respective
outcomes. Non-signalling from the second region to the first requires
\[
 \sum_b P(a,b|x,y)=P(a|x),
\]
independently of $y$. This condition does not require the outcomes $a$ and $b$
to be statistically independent. They may be correlated, as they are in
relativistic Bell experiments; what is excluded is any dependence of the local
outcome statistics on the choice of a remote operation.

To formulate this condition, however, one must already possess a theory of
localized operations and their outcomes. This is precisely where the usual
$S$-matrix formulation is insufficient. As discussed in Sec.~3.5, the notions
of event and state update are considerably more subtle in QFT than in ordinary
quantum mechanics. Even particle localization is problematic: Malament's
theorem shows that, under natural assumptions, there is no relativistically
covariant sharply localized particle-position observable in QFT \cite{Malament}.
 The theorem does not exclude local field observables or localized detector events; rather, it demonstrates that particle localization in QFT cannot be represented by the sharp position observables familiar from nonrelativistic quantum mechanics.
 
Algebraic quantum field theory (AQFT) provides an important conceptual advance beyond
the $S$-matrix framework by formulating QFT directly in terms of observables
localized in spacetime \cite{Haag}. To each spacetime region $\mathcal O$, it
assigns an algebra of observables $\mathcal A(\mathcal O)$. Locality is expressed
by
\bey
[\mathcal A(\mathcal O_1),\mathcal A(\mathcal O_2)]=0
\eey
whenever $\mathcal O_1$ and $\mathcal O_2$ are spacelike separated. This
framework makes the localization structure of QFT precise. If measurements correspond to elements of local algebras, non-signalling follows automatically.

However, local algebras do not by themselves provide an operational theory of measurement. Many experimentally measured quantities—such as total energy, momentum and spin, particle flux, and detection time—do not belong to any bounded-region local algebra. In a time-of-arrival measurement, moreover, the time—or, more generally, the spacetime location—of the detector event is itself a random variable, rather than part of a region fixed in advance. Although relativistic causality is imposed in AQFT at the level of its operator algebra, establishing it for physically relevant variables requires additional structure connecting local interventions with recorded events, probabilities, and state changes.

Problems concerning time, events, causality, and measurement therefore arise even in the absence of gravity. The theories on which quantum gravity is built do not provide a single, unproblematic notion of time that need only be extended to gravitational systems. Rather, they already contain tensions that quantum theories of gravity inherit.

\chapter{Time in general relativity}

\section{The basic structure of GR}
General relativity can be understood as a theory whose configuration space consists of \emph{histories} of fields on a spacetime manifold $\M$. A history is given by a pair $(g, \phi)$, where $g$ is a Lorentzian metric and $\phi$ collectively denotes the matter fields. These fields are defined over the entire manifold and represent complete spacetime configurations rather than instantaneous states. In this sense, the basic objects of the theory are four-dimensional, and the dynamics is formulated in terms of relations between such histories rather than evolution with respect to an external time parameter. Furthermore, since the metric is dynamical both the causal structure and the local-clock structure of spacetime are dynamical, too.

The dynamics is specified by the Einstein--Hilbert action (augmented by matter contributions),
\[
S[g,\Phi] = \frac{1}{2\kappa} \int_{\M} d^4x \, \sqrt{-g}\, R[g] \;+\; S_{\text{matter}}[g,\phi], \label{vvv}
\]
whose stationary points determine the physically allowed histories---$R[g]$ stands for the Ricci scalar and $\kappa = 8 \pi G$.

Variation of this action with respect to the metric yields Einstein’s field equations,
 
\begin{equation}
G_{\mu\nu}  
=
\kappa \, T_{\mu\nu},
\end{equation}
where the Einstein tensor is defined by $G_{\mu\nu}
:=
R_{\mu\nu}
- \frac{1}{2} g_{\mu\nu} R$,
with $R_{\mu\nu}$ the Ricci tensor.
The stress-energy tensor $T_{\mu\nu}$ is defined as the functional derivative of the matter action $S_{\mathrm{m}}$ with respect to the metric,
\begin{equation}
T_{\mu\nu}(X)
:=
-\frac{2}{\sqrt{-g}}
\frac{\delta S_{\mathrm{m}}}{\delta g^{\mu\nu}(X)}.
\end{equation}
Variation of the action (\ref{vvv}) with respect to the matter fields gives their equations of motion. In this formulation, the equations are manifestly covariant and make no reference to any preferred temporal slicing of spacetime, reinforcing the idea that general relativity is fundamentally a four-dimensional, rather than a time-evolution, theory.

A central feature of this framework is invariance under the diffeomorphism group $\DM$, consisting of smooth, invertible maps of the manifold onto itself. Under an active diffeomorphism, the fields $(g,\Phi)$ are transformed by pullback, producing a new configuration that is mathematically distinct but physically equivalent. Thus, two configurations related by a diffeomorphism represent the same physical situation. This symmetry implies that spacetime points have no intrinsic identity independent of the fields, and that physically meaningful quantities must be invariant under $\DM$\footnote{The significance of diffeomorphism invariance is captured by the Einstein \emph{hole argument}. Consider a region $\mathcal{H}\subset\mathcal{M}$ (the ``hole'') in which no matter is present, and suppose $(g,\Phi)$ is a solution of the field equations. Let $\varphi\in\mathrm{Diff}(\mathcal{M})$ be a diffeomorphism that is the identity outside $\mathcal{H}$ but nontrivial inside it. Then $(\varphi^* g,\varphi^*\Phi)$ is also a solution, and it agrees with $(g,\Phi)$ everywhere outside the hole while differing inside. If spacetime points had an independent physical identity, this would lead to a form of indeterminism: the data outside $\mathcal{H}$ would not determine the fields inside it. The standard resolution is to regard diffeomorphically related configurations as physically identical. In this view, only $\mathrm{Diff}(\mathcal{M})$-invariant  quantities are genuine observables, and the apparent indeterminism disappears.}.  For example, if $\phi$ is a scalar field on $\mathcal{M}$, the value $\phi(x)$ at a specific point $x$ is not an invariant quantity, since the point itself can be moved by a diffeomorphism.  Physical observables must be invariant under diffeomorphisms, and are therefore typically nonlocal. A simple example is given by spacetime integrals such as
\[
J[g] = \int_{\M} d^4x \, \sqrt{-g} \, R^{\alpha\beta\gamma\delta}\, R_{\alpha\beta\gamma\delta},
\]
which are invariant by construction. Physically localized observables are also of this form, localization defined with reference to a material object. 
Thus, in general relativity, observables are not associated with field values at points, but with  integrated quantities defined over the whole spacetime.
 
   However, this raises two important difficulties:
   \begin{itemize}
\item The causal structure of spacetime is defined through relations between spacetime points. Relations between bare manifold points are not invariant under $\DM$, because the points themselves have no diffeomorphism-invariant physical identity. In what sense, then, is causal structure observable, and how are the events entering a causal relation to be identified physically?

\item $\DM$ invariance implies in particular that physical predictions are  independent of any particular choice of coordinates. In this sense, time is just one coordinate on the spacetime manifold $\mathcal{M}$ and does not have a fundamental status. But  if time is merely a coordinate with no intrinsic physical meaning, how does the notion of change emerge from the theory? 

\end{itemize}

These  difficulties are easily overcome in the classical theory. Once the field equations are solved for given initial data, we obtain a unique   Lorentzian metric $g$ that provides a definite causal structure on $\mathcal{M}$. This allows us to define notions such as causality and spacelike separation, even though these notions are not themselves diffeomorphism-invariant. The causal structure is essentially fixed by the initial conditions, because the evolution equations are deterministic.

Obviously, this solution does not work in the presence of fluctuations, either quantum or classical. If the metric is allowed to fluctuate dynamically, causal relations become history dependent, and the two difficulties above reappear in full force. A fluctuating metric does not allow us to even define a  microcausality or   non-signalling condition, because the notion of spacelikeness will also fluctuate.

The decisive point is that the problem of time is, at root, a consequence of breaking
determinism for the evolution of the spacetime geometry. In the deterministic theory the two difficulties above are harmless: fixing
initial data and a reference frame yields a unique metric, and one may treat that solved-
for metric as though it were a fixed background—it is dynamical, but in a deterministic
system the distinction makes no operational difference. Quantizing gravity is therefore
not what creates the problem; any loss of determinism does.
In particular, a purely
classical stochastic theory of gravity already breaks determinism and faces exactly the
same questions about what time and causality are, and how they relate to diffeomorphism
invariance.

  There are only a limited number of strategies for addressing time in quantum gravity:
\begin{enumerate}

\item One possibility is to abandon $\DM$ invariance as a fundamental symmetry. This typically involves postulating a preferred foliation of spacetime—hidden at the level of classical general relativity—relative to which an absolute notion of time is defined. A natural candidate is the foliation associated with approximate homogeneity and isotropy in cosmology, whose origin remains unclear. Such theories are tightly constrained by observations (solar-system tests, binary pulsars, gravitational waves, and high-energy astrophysics), which force them to reproduce general relativity to high accuracy at observable scales. If a preferred foliation exists, it must therefore be effectively invisible in the classical regime, appearing only at the quantum level—an assumption that seems somewhat ad hoc.

\item A second strategy is to restrict attention to spacetimes with fixed asymptotic structure, such as asymptotically flat or asymptotically anti-de~Sitter geometries (AdS). In this case, one can define asymptotic notions of time and space, since physical $\DM$ transformations are typically required to  act trivially at infinity. Time evolution can then be associated with the generators of the asymptotic symmetry group (Poincar\'e or AdS). While this framework is well suited for scattering problems or holographic constructions, it is largely disconnected from physically realistic situations. It relies on asymptotic regions and observers at infinity, which are absent in cosmology and in any laboratory setting. Its notion of time is defined only with respect to such idealized boundaries, rather than in terms of local physical clocks. They therefore displace, rather than  resolve, the problem considered here.

\item Another possibility is to define time and causality relationally, by introducing spacetime coordinates $X^{\mu}[g,\phi]$  as functionals of the metric and matter fields. This yields a $\DM$-invariant characterization of spacetime in terms of physical observables, generalizing the idea that time is measured by clocks (e.g., proper time along a worldline defined by physical events). However, this approach raises significant issues in quantisation: since the coordinates depend on dynamical fields, they may fail to commute, casting doubt on their interpretation as spacetime coordinates and raising tensions with the restrictions on time operators we discussed earlier.

\item A more radical approach is to abandon time and causality as fundamental concepts, and to regard them as emergent in an appropriate semiclassical limit. Although this viewpoint has gained traction, its logical status is uncertain: it is not obvious how robust notions such as temporal ordering or causal structure can emerge unless they are already encoded, at least implicitly, in the underlying theory.

\item Finally, one may attempt to generalize quantum theory itself so that causal structure of general relativity is incorporated at a fundamental level, and then seek its unification with gravity. In this perspective, the goal is not to quantize gravity but to ``general-relativize'' quantum theory. This would require a substantial revision of quantum field theory, while maintaining its empirical success---a formidable challenge.

\end{enumerate}

Almost all existing approaches to quantum gravity can be understood as pursuing one or more of these strategies, often in combination.
  
There is an additional structural conflict between general relativity and quantum theory: the \emph{agency problem}. A distinctive feature of gravity is its universal coupling to all forms of energy and momentum. Consequently, any physical system used to define or probe spacetime---including rods, clocks, and detectors---necessarily contributes to the gravitational field and must, at the fundamental level, be included in the dynamical description. This undermines the standard operational viewpoint, in which measurements are modelled as interventions performed by agents external to the system: in gravity, the measuring device itself gravitates and back-reacts on the geometry. Even at a heuristic level, the notion of a localized, externally controlled operation becomes problematic.

By contrast, standard quantum theory is fundamentally operational. Its basic structure is formulated in terms of preparations, transformations, and measurements, all of which presuppose a distinction between the system and an agent who selects and implements these operations. Reconciling quantum theory with gravity would therefore seem to require a formulation in which measurements are treated as dynamical processes within the system itself rather than as interventions imposed from outside.

Such a move faces significant obstacles. The Bell and Kochen--Specker theorems \cite{KoSp,Bell} show that measurement outcomes cannot, in general, be understood as revealing pre-existing values independent of the measurement arrangement. Any underlying description reproducing quantum predictions must be contextual, nonlocal, or relinquish some other assumption entering these theorems. Incorporating the apparatus into the dynamical system therefore cannot eliminate the measurement context; that context must itself be represented within the system.

The agency conflict is arguably \emph{deeper} than the conflict about time. Were quantum theory our only guide, we could plausibly treat causality and temporality themselves as operational notions. General relativity, however, precludes a fundamentally external agent of the kind presupposed by the operational picture and thereby blocks this route. A theory of quantum gravity must therefore explain how measurement contexts, interventions, and definite events arise within the physical system itself---unless it abandons or modifies the universal coupling of gravity.

\section{The canonical formalism}
The main tool for the analysis of the problem of time in quantum gravity is the canonical (Hamiltonian) formulation of general relativity. Many of the deep conceptual problems in quantum gravity are more transparent in the canonical description. The canonical approach also lends itself to quantisation procedures that highlight the Hilbert space structure of the theory, and thus, is more directly relevant to issues pertaining to quantum interpretations and quantum information. Nonetheless, our discussion of the problem of time will not be limited to canonical quantisation methods.

There is a sharper, technical reason why the Hamiltonian formulation is indispensable: the Einstein equations are not all dynamical. Written out explicitly, several of them
contain no second time derivatives and hold instantaneously on each spacelike slice—they
are constraints in the sense of constrained mechanics, rather than evolution equations.
The meaning of such an instantaneous equation, and the way it removes apparent degrees
of freedom, is far more transparent in the Hamiltonian than in the Lagrangian formulation,
which is why the canonical analysis is the natural setting for the problem of time.
 
 \subsection{Spacelike foliation}

Let $(\mathcal M, g)$ be a spacetime, where $\mathcal M$ is a four-dimensional manifold and $g_{\mu\nu}$ a Lorentzian metric. We assume that $\mathcal M$ has topology $\Sigma \times \R$. We denote spacetime coordinates by $X^\mu$ and coordinates on $\Sigma$ by $x^i$.

A diffeomorphism $\E: \Sigma \times \R   \to \M$ defines a spacelike foliations of $\M$ if the hypersurfaces 
\[
\Sigma_t := \{ X \in \mathcal M \;|\; X = \E(x,t), \; x \in \Sigma \}
\]
are spacelike submanifolds of $\M$.

Given $\E$, we define the vector fields
\[
t^\mu(X) := \left.\frac{\partial \mathcal E^\mu(x,t)}{\partial t}\right|_{\mathcal E^{-1}(X)}, 
\qquad
\mathcal E^\mu_i(X) := \left.\frac{\partial \mathcal E^\mu(x,t)}{\partial x^i}\right|_{\mathcal E^{-1}(X)}.
\]

The vectors $\mathcal E^\mu_i$ are tangent to $\Sigma_t$, while $t^\mu$ is transverse to $\Sigma_t$.

Let $n_\mu$ be the unit normal to $\Sigma_t$, normalized by $n^\mu n_\mu = -1$. Then
\[
n_\mu = -N \nabla_\mu t,
\]
where $N>0$ is the \emph{lapse function}. The vector $t^\mu$ decomposes as
\bey
t^\mu = N n^\mu + N^i \mathcal E^\mu_i,
\eey
where $N^i$ is the \emph{shift vector}.

In adapted coordinates $(t,x^i)$ associated with the foliation, the spacetime metric takes the ADM form
\begin{equation}
ds^2 = -N^2 dt^2 + h_{ij}\big(dx^i + N^i dt\big)\big(dx^j + N^j dt\big).
\end{equation}

\begin{exercise}
Show that  $\nabla_{\mu} n_{\nu} - \nabla_{\nu} n_{\mu} = \eta_{\mu} b_{\nu} - \eta_{\nu} b_{\mu}$, where $b_{\mu} =  \partial_{\mu}N/N$.
\label{frob}
\end{exercise}

\subsection{3+1 decomposition of the metric}

The induced metric on $\Sigma_t$ is defined as the pullback
\[
h_{ij} = \mathcal E^\mu_i \mathcal E^\nu_j g_{\mu\nu}.
\]

We introduce the spacetime projector
\[
h_{\mu\nu} := g_{\mu\nu} + n_\mu n_\nu,
\]
so that $h_{\mu\nu} \mathcal E^\nu_i = \mathcal E_{\mu i}$ and $h_{\mu\nu} n^\nu = 0$.
Indices on $\Sigma$ are raised and lowered using $h_{ij}$ and its inverse $h^{ij}$.

The projection of tensors between spacetime and $\Sigma_t$ is defined by
\[
X^i = \mathcal E^i_\mu X^\mu, \qquad 
\omega_i = \mathcal E^\mu_i \omega_\mu,
\]
and conversely
\[
X^\mu = \mathcal E^\mu_i X^i, \qquad 
\omega_\mu = \mathcal E_\mu^i \omega_i,
\]
with $n_\mu X^\mu = 0$ and $n^\mu \omega_\mu = 0$.

The spatial covariant derivative $\nabla_i$ is defined by projection:
\[
\nabla_i X^j = \mathcal E^\mu_i \mathcal E^j_\nu \nabla_\mu X^\nu,
\]
and similarly for covariant tensors. One verifies that $\nabla_i h_{jk} = 0$, 
so $\nabla_i$ is the Levi-Civita connection of $h_{ij}$.

 In adapted coordinates $(t,x^i)$ associated with the foliation, the spacetime metric takes the ADM form
\begin{equation}
ds^2 = -N^2 dt^2 + h_{ij}\big(dx^i + N^i dt\big)\big(dx^j + N^j dt\big).
\end{equation}

\begin{exercise}
Show that $\sqrt{-g} = N \sqrt{h}$.
\end{exercise}

\subsection{Extrinsic curvature}

The extrinsic curvature of $\Sigma_t$ is defined by
\[
K_{ij} := \mathcal E^\mu_i \mathcal E^\nu_j \nabla_\mu n_\nu.
\]

Using Exercise \ref{frob}, we find that $K_{ij}$ is symmetric:
\[
K_{ij} = K_{ji}.
\]

Its trace is
\[
K = h^{ij} K_{ij} = \nabla_\mu n^\mu.
\]
 
 The extrinsic curvature can then be written in terms of the time derivative of the induced metric as
\begin{equation}
K_{ij} = \frac{1}{2N}\left(\dot{h}_{ij} - \nabla_i N_j - \nabla_j N_i\right), \label{excurv}
\end{equation}
where $\nabla_i$ is the Levi-Civita covariant derivative associated with the spatial metric $h_{ij}$, and
\bey
\dot{h}_{ij}
= \E^\mu_i \mathcal E^\nu_j \, \mathcal L_{t} g_{\mu\nu}.
\eey
The extrinsic curvature captures the rate of change of the metric along the leaves of the foliation.

\begin{exercise}
Prove Eq. (\ref{excurv}).
\end{exercise}

The distinction at work here is intrinsic versus extrinsic curvature. A flat sheet of paper
has zero intrinsic curvature—curves drawn on it obey Euclidean geometry no matter how
the sheet is rolled—yet rolling it endows it with extrinsic curvature, which measures how
the sheet bends within the embedding space.  

\subsection{The 3+1 form of the Lagrangian}
We will express the Einstein-Hilbert action for gravity in terms of the three-metric $h_{ij}$, the extrinsic curvature $K_{ij}$, the lapse $N$ and the shift $N_i$.

The key geometric ingredient is the Gauss--Codazzi decomposition of the four-dimensional Ricci scalar,
\begin{equation}
R = {}^{(3)}R + K_{ij}K^{ij} - K^2 + 2 \nabla_\mu \big( n^\mu K - a^\mu \big),
\label{Rsplit}
\end{equation}
where ${}^{(3)}R$ is the Ricci scalar of the spatial metric $h_{ij}$, $K_{ij}$ is the extrinsic curvature,
\begin{equation}
K_{ij} = \frac{1}{2N}\big(\partial_t h_{ij} - \nabla_i N_j - \nabla_j N_i\big),
\end{equation}
$K=h^{ij}K_{ij}$, $n^\mu$ is the unit normal to the slices, and $a^\mu = n^\nu \nabla_\nu n^\mu$ is its acceleration.

Substituting (\ref{Rsplit}) into the action gives
\begin{equation}
S_{EH}
= \frac{1}{2 \kappa}\int dt \int_\Sigma d^3x \, N\sqrt{h}
\Big(
{}^{(3)}R + K_{ij}K^{ij} - K^2
\Big)
+ S_{bdy},
\end{equation}
where $S_{bdy}$ arises from the total divergence term in (\ref{Rsplit}). If appropriate boundary terms are added or boundary contributions are neglected, the bulk action takes the Arnowitt-Deser-Misner form
\begin{equation}
S_{ ADM}
= \frac{1}{2 \kappa}\int dt \int_\Sigma d^3x \, N\sqrt{h}
\Big(
{}^{(3)}R + K_{ij}K^{ij} - K^2
\Big).
\end{equation}

\begin{exercise}
Prove the Gauss-Godazzi relation. (A detailed proof is in Sec. 3.5 of Ref. \cite{Poisson})
\end{exercise}

\subsection{The Hamiltonian structure of general relativity}
To construct the Hamiltonian, we define the momentum conjugate to $h_{ij}$,
\begin{equation}
\pi^{ij}
:= \frac{\partial \mathcal L}{\partial(\dot{h}_{ij})}
= \frac{\sqrt{h}}{2 \kappa}\big(K^{ij} - h^{ij}K\big).
\end{equation}
The momenta conjugate to the lapse and the shift vanish. 
Solving for $K^{ij}$ in terms of $\pi^{ij}$ and performing the Legendre transform yields
\begin{equation}
S_{ADM}
= \int dt \int_\Sigma d^3x
\left(
\pi^{ij}\dot{h}_{ij}
- N \mathcal H
- N^i \mathcal H_i
\right),
\end{equation}
where 
\bey
\mathcal H
=
\frac{2\kappa }{\sqrt h}
\left(\pi_{ij}\pi^{ij} - \frac12 \pi^2 \right)
-\frac{\sqrt h}{2\kappa}\,{}^{(3)}R, \label{hamc}
\eey
is the Hamiltonian constraint, and 
\bey
\mathcal H_i
 = -2 \nabla_j \pi^j{}_i. \label{momc}
\eey 
is the momentum constraint. 

Variation with respect to $N$ and $N^i$ yields the constraint equations, $\HH = 0$ and $\HH_i$ = 0.

We conclude that the phase space $\Gamma$ of general relativity is the cotangent bundle $T^*\mbox{Riem}(\Sigma)$, where $\mbox{Riem}(\Sigma)$ is the space of Riemannian metrics on the surface $\Sigma$. The conjugate variable to $h_{ij}$ is a (2,0) tensor density $\pi^{ij}$

The phase space is equipped with a symplectic form,
\bey
\omega = \int d^3x \delta \pi^{ij}(x) \wedge \delta h_{ij}(x),
\eey
leading to a Poisson bracket 
\bey
\{h_{ij}(x), \pi^{kl}(x')\} = (\delta_i^k \delta_j^l + \delta_i^l \delta_j^k)\delta^3(x, x').
\eey

The Hamiltonian $H = \int d^3x \left( N \mathcal H
+ N^i \mathcal H_i\right)$ vanishes due to the constraints. 

\begin{exercise}
Show that the Hamiltonian constraint $\HH = 0$ is equivalent to the normal component of Einstein's equation in vacuum $G^{\mu \nu}n_{\mu}n_{\nu} = 0$.
\end{exercise}

The vanishing total Hamiltonian should not be mistaken for physical triviality. The
very same 3+1 system is what is integrated numerically to evolve astrophysical spacetimes—
the colliding-black-hole waveforms whose predictions match the LIGO/Virgo detections to
within a few percent are produced from these equations. The “frozen” zero-Hamiltonian
appearance reflects only that diffeomorphism invariance is incompatible with time evolution as usually conceived. It does not mean that change does not occur.

\subsection{The constraint algebra}
 We define the smeared Hamiltonian and momentum constraints as
\begin{equation}
H[N] := \int_\Sigma d^3x \, N(x)\,\mathcal H(x),
\qquad
H[\vec N] := \int_\Sigma d^3x \, N^i(x)\,\mathcal H_i(x),
\end{equation}
where $N$ is a scalar function on $\Sigma$ and $\vec N$ a vector field on $\Sigma$. They are usually called lapse and shift, respectively, even if they are not necessarily the same objects that appear in the action, but arbitrary fields.

Their Poisson algebra is given by
\begin{align}
\{H[\vec N],H[\vec M]\}
&= H\big[[\vec N,\vec M]\big], \\
\{H[\vec N],H[M]\}
&= H[\mathcal L_{\vec N} M], \\
\{H[N],H[M]\}
&= H[\vec K],
\end{align}
where $\mathcal L_{\vec N} M = N^i \partial_i M$, 
\begin{equation}
[\vec N,\vec M]^i = N^j \partial_j M^i - M^j \partial_j N^i,
\end{equation}
is the Lie bracket of the two vector fields,
and
\begin{equation}
K^i = h^{ij}\big(N \partial_j M - M \partial_j N\big). \label{veck}
\end{equation}

This algebra involves structure functions (through $h^{ij}$) rather than structure constants, and is therefore not a Lie algebra. It is known as the Dirac algebra, and it corresponds to hypersurface deformations.

To understand this interpretation, consider a hypersurface $\Sigma$ embedded in spacetime by $\E^\mu(x)$, with tangent vectors $\mathcal E^\mu_i$  and unit normal $n^\mu$.
A pure normal deformation of the hypersurface is $\delta_N \E^\mu = N n^\mu$,
where $N$ is a scalar on $\Sigma$. Given two such deformations, with lapses $N$ and $M$, their commutator is
\[
[\delta_N,\delta_M]X^\mu
=
\delta_N(Mn^\mu)-\delta_M(Nn^\mu) =
M\,\delta_N n^\mu - N\,\delta_M n^\mu.
\]
To compute $\delta_N n^\mu$, we note that the normalization condition   $n_\mu n^\mu=-1$ implies that its variation is orthogonal to $n^\mu$, hence  of the form $\delta_N n^\mu = A^i \mathcal E^\mu_i$.
To determine $A^i$, we vary the orthogonality relation $n_\mu \mathcal E^\mu_i=0$:
\[
(\delta_N n_\mu)\mathcal E^\mu_i + n_\mu \delta_N \mathcal E^\mu_i = 0.
\]
We have $\delta_N \mathcal E^\mu_i = \partial_i \delta {\cal E}^{\mu} 
=
\partial_i(Nn^\mu)
=
(\partial_i N)n^\mu + N\partial_i n^\mu$.
Contracting with $n_\mu$ gives
\[
n_\mu \delta_N \mathcal E^\mu_i
=
(\partial_i N)\, n_\mu n^\mu + N n_\mu \partial_i n^\mu
=
-\partial_i N,
\]
since $n_\mu n^\mu=-1$ and $\partial_i(n_\mu n^\mu)=0$ implies
\[
n_\mu \partial_i n^\mu = 0.
\]
Hence $(\delta_N n_\mu)\mathcal E^\mu_i = \partial_i N$, or equivalently,
\[
\delta_N n^\mu = (\nabla^i N)\mathcal E^\mu_i.
\]

Substituting into the commutator yields
\[
[\delta_N,\delta_M]\E^\mu
=
\big(M\nabla^i N - N\nabla^i M\big)\mathcal E^\mu_i.
\]
Therefore, the commutator of two normal deformations is a tangential deformation with shift vector $K^i$ given by Eq. (\ref{veck}). 
\begin{exercise}
Calculate the commutators $[\delta_N,\delta_{\vec N}]$ and $[\delta_{\vec N}, \delta_{\vec M}]$, where the tangential surface deformations are given by $\delta_{\vec N}\E^{\mu} = N^i \E_i^{\mu}$. 

\end{exercise}

\section{Constrained systems}
To proceed, we will analyze the structure of Hamiltonian constraint systems, a special case of which is GR. The identification of the constraints in a general mechanical system  from the Lagrangian formulation, following  a careful analysis of the Legendre transformation,   will not be discussed here; see Refs. \cite{Dirac64, HenneauxTeitelboim, Sunder}.

Consider a Hamiltonian system with a state space $\Gamma$ and a Hamiltonian $H$.
The system is {\em constrained} if its evolution only takes place in a subset $C$ of $\Gamma$, the {\em constraint surface}. The constraint surface is typically determined by $n$ independent functions
  $\phi_r : \Gamma \rightarrow \R$, $\alpha = 1, 2, \ldots, n$, such that $\phi_{\alpha} = 0$ for all $\alpha$. Time evolution is compatible with the constraints, in the sense that it cannot take the system out of the constraint surface if the Hamiltonian satisfies $\{H, \phi_\alpha\} = 0 $ on $C$ for all $\alpha$.  

A constrained system is called {\em first-class}, if $\{\phi_\alpha, \phi_\beta \}$ vanishes on the constraint surface $C$. This condition implies that $X_{\phi_\alpha}^i\partial_i\phi_\beta = 0$ on the constraint surface. Hence, the Hamiltonian vector fields generated by the constraints are horizontal: they generate diffeomorphisms that stay entirely within $C$. The constraints of GR are first-class.
 
 The constraint surface does not define a symplectic manifold, because the restriction of the symplectic form on $C$ is degenerate: $\omega_{ab} X_{\phi_{\alpha}}^a Y^b = Y^b\partial_b \phi_{\alpha} = 0$, for all horizontal vector fields. The vector fields $X_{\phi_\alpha}$ specify the degenerate directions.
 
Only gauge invariant functions define physical observables, i.e., the functions that satisfy
\bey
\{F, \phi_a\} = 0  \; \; \mbox{on} \; \; C, \label{gaugeinv}
\eey
for all $a$. This means that $X_F(\phi_a) = 0$ on $C$. The Hamiltonian $H$ is obviously a physical observable.

\begin{exercise}
Show that the Poisson bracket of two observables is an observable.
\end{exercise}
 
The constraints   also generate canonical transformations. Any function $G = \sum_{\alpha} r_{\alpha} \phi_{\alpha}$, where $r_{\alpha}$ are functions on $\Gamma$, generates an one-parameter family of canonical transformations through the integral curves on $C$,
\bey
\frac{d \xi^a(s)}{ds} =  \sum_{\alpha} r_{\alpha} \{\xi^a(s),\phi_{\alpha} \} = -  \sum_{\alpha} r_{\alpha} X_{\phi_\alpha}[\xi^a(s)].
\eey
Observables are constant along $\xi(s)$,
\bey
\frac{dF[\xi(s)]}{ds} = \sum_a \frac{\partial F}{\partial \xi^a}  \frac{d \xi^a(s)}{ds} = - \sum_\alpha r_{\alpha} \sum_{a} \frac{\partial F}{\partial \xi^a} X_{\phi_\alpha}[\xi^a(s)]
\nonumber \\
=- \sum_\alpha r_{\alpha} \sum_{a}  X_{\phi_\alpha}[F[\xi(s)]] = \sum_\alpha r_{\alpha} \{F, \phi_\alpha\}[\xi(s)] = 0. \nonumber
\eey
Two points of $C$ are {\em gauge-equivalent} is there is a curve  generated by a gauge transformation that connects them. The set of all points in $C$ that are gauge-equivalent to a point  $\xi$ defines a {\em gauge  orbit} through $\xi$, denoted by $O_{\xi}$. Thus the constraint surface splits into a collection of orbits, such that each point of $C$ belongs to only one orbit.
The set of all orbits is the {\em reduced state space} $\Gamma_{red}$, i.e., the state space of the true degrees of freedom of the system.  Any function on $C$ that can be used as a coordinate along the gauge orbits is called a {\em pure-gauge} variable. Note that, in general, $\Gamma_{red}$ is not a cotangent bundle, even if $\Gamma$ is.

\begin{figure}[tbp]
\includegraphics[height=6cm]{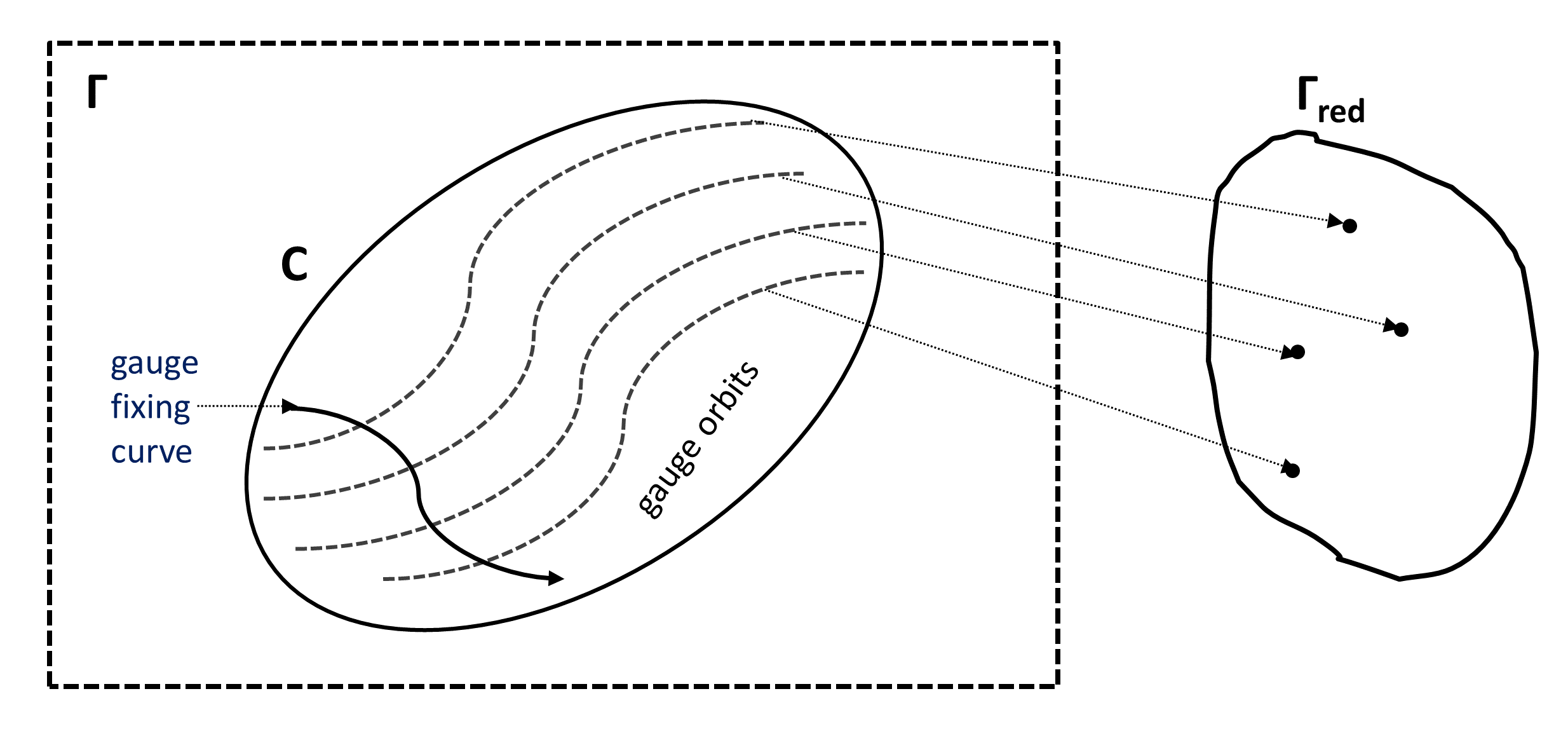} \caption{  Reduction of a first-class constrained system. The constraints restrict from the full space $\Gamma$ to the constraint surface $C$, and they generate gauge orbits. Each gauge orbit defines a point of the reduced state space. We fix the gauge by selecting a curve that intersects each orbit once. Then, we treat the points of intersection as a representative of the orbits.
}
\label{gammareduced}
\end{figure}

Let the dimension of $\Gamma$ be $2N$. In presence of $n$ constraints, the constraint surface has dimension $2N - n$. Each $X_{\phi_a}$ generates a transformation along a different direction in a gauge orbit, so the dimension of a gauge orbit is $n$. It follows that the dimension of the reduced state space is $2N - 2n$. The reduction procedure is sketched in Fig. \ref{gammareduced}.

Each observable $F$ takes a single value in each orbit. Hence, it projects to a function $\tilde{F}$ on $\Gamma_{red}$, defined by
\bey
\tilde{F}(O_{\xi}) := F(\xi).
\eey
Since the Poisson bracket of two observables is also an observable, the reduced state space also accepts a Poisson bracket, defined by
\bey
\{\tilde{F}, \tilde{G}\} = \widetilde{\{F, G\}}.
\eey
Thus, we obtain a well-defined Hamiltonian system on the reduced state space, with time evolution given by the projection $\tilde{H}$ of the Hamiltonian on the original state space $\Gamma$---often referred to as the {\em reduced Hamiltonian}.

\subsubsection*{Gauge fixing}
In general, however, the reduced state space $\Gamma_{red}$ cannot be constructed explicitly as a global manifold. Although it is defined abstractly as the space of gauge orbits, identifying these orbits and endowing their quotient with a smooth manifold structure is highly nontrivial. In practice, one often proceeds by \emph{gauge fixing}: one introduces $n$ independent functions $\chi_\alpha$ on $\Gamma$ such that the combined conditions
\[
\phi_\alpha = 0, \qquad \chi_\alpha = 0
\]
select a unique representative on each gauge orbit, at least locally. The functions $\chi_\alpha$ then serve as coordinate conditions that allow us to parameterize the reduced phase space in terms of the remaining $2N-2n$ degrees of freedom.

We now describe how gauge fixing can be implemented directly at the level of the Hamiltonian action for a general constrained system. We consider an action of the form
\begin{equation}
S[\xi,\lambda]
=
\int dt \left(
\theta_a(\xi)\dot{\xi}^a - H   - \lambda^\alpha \phi_\alpha 
\right),
\end{equation}
where the functions $\phi_\alpha(\xi)=0$ are first-class constraints, and the multipliers $\lambda^\alpha$ are arbitrary functions of time, the variation of which enforces the constraints.

To fix the gauge, we introduce a set of gauge-fixing functions $\chi_\alpha(\xi)=0$, one for each constraint, such that the matrix
\begin{equation}
\Delta_{\alpha\beta} := \{\chi_\alpha,\phi_\beta\}
\end{equation}
is invertible, at least locally on $\Gamma$. We then impose the gauge conditions by adding them to the action with new Lagrange multipliers $\mu^\alpha$,
\begin{equation}
S_{\mathrm{gf}}[\xi,\lambda,\mu]
=
\int dt \left(
\theta_a(\xi)\dot{\xi}^a - H   - \lambda^\alpha \phi_\alpha - \mu^\alpha \chi_\alpha 
\right).
\end{equation}

Variation with respect to $\lambda^\alpha$ and $\mu^\alpha$ enforces $\phi_\alpha = 0$ and $\chi_\alpha = 0$.
 The preservation in time of the gauge conditions determines all the previously arbitrary multipliers $\lambda^\alpha$, thereby eliminating the gauge freedom. The multipliers $\mu^\alpha$ serve only to enforce the gauge conditions and play no dynamical role.
 
Note that if we choose the gauge-fixing conditions so that
\bey
\{\chi_a, \chi_b\} = 0, \label{condgauge}
\eey
 we can always choose Darboux coordinates in which $\chi_a$ are identified with momenta $\bar{p}_a$, where $a = 1, \ldots, n$. Since $\bar{p}_a = 0$, the constraints $\phi_a$ are functions of the conjugate variables $q^a$, plus the remaining Darboux variables $q^*_c, p^{*d}$ on the reduced state space, where $c, d = 1, \ldots, N-n$. 
 We can solve the constraint equations $\phi_a = 0$ for $q^a$, so that $\Gamma_{red}$ can be seen as an embedded submanifold of $\Gamma$, through the conditions $p_a = 0$, and  $q^a = q^a(q^*, p^*)$. 

It should be emphasized  that   gauge fixing is generically only a local construction and may fail globally. The gauge-fixing conditions may not intersect every orbit, or they may intersect some orbits more than once. This leads to ambiguities in the parametrization of $\Gamma_{red}$, reflecting the fact that the reduced state space may have a nontrivial global structure . Thus, while gauge fixing provides a practical method for describing the physical degrees of freedom, it does not in general yield a global and unique characterization of the reduced state space.

The cleanest illustration of this global failure is the Gribov ambiguity.
In Abelian
gauge theory the conditions (Coulomb, Lorenz, axial) are linear, so a single gauge slice
meets each orbit once and reproduces $\Gamma_{red}$ faithfully. In non-abelian Yang–Mills the orbits
curve, and Gribov showed that no local gauge condition can intersect every orbit exactly
once: the same physical configuration is counted multiple times (Gribov copies). This is a
genuine topological obstruction, and it is why gauge-fixed perturbation theory cannot by
itself reach the full reduced space. The same warning applies to gravity: results obtained
after gauge fixing may depend on the gauge chosen and may silently omit part of $\Gamma_{red}$, an
error that becomes acute once one tries to quantize.

Alternatively, one may attempt to eliminate gauge freedom at the level of the action by fixing the Lagrange multipliers, $\lambda^\alpha = f^\alpha(\xi,t)$. This breaks gauge invariance and selects particular representatives along gauge orbits, yielding equations of motion without arbitrary gauge transformations and effectively restricting to a reduced set of trajectories.
However, this procedure is not reliable. Fixing the multipliers does not guarantee that each gauge orbit is represented exactly once: gauge-equivalent solutions may remain, while others may be excluded. In addition, it provides no control over the global structure of the reduced space and obscures the gauge symmetry.

 \begin{exercise} Consider two harmonic oscillators of unit mass and frequency constrained so that their total energy is constant, that is, subject to the constraint
$\phi = \frac{1}{2}(p_1^2 + p_2^2 +x_1^2 +x_2^2) - E = 0$. Determine the reduced state space $\Gamma_{red}$, and express the Hamiltonian $\tilde{H}$ in terms of local coordinates on $\Gamma_{red}$.
\end{exercise}
  
  \begin{exercise}
 Consider   two harmonic oscillators with constant energy difference, i.e., with the constraint
\bey
\phi = \frac{1}{2}(p_1^2 + x_1^2  - p_2^2  - x_2^2) - \delta. \nonumber
\eey
(i) Show that the reduced state space is a cylinder $\R\times S^1$, and identify the fundamental Poisson brackets. (ii) Show that the functions $K_1 = \frac{1}{2}(x_1p_2 + x_2 p_1), K_2 = \frac{1}{2}(x_1 x_2 - p_1 p_2)$ and $K_3 = \frac{1}{4} (p_1^2 + x_1^2  + p_2^2  + x_2^2)$ represent the algebra $\mathfrak{sl}(2, \R)$. (iii) Show that $K_1, K_2$ and $K_3$ are observables and express them in terms of coordinates on the reduced state space.

  \end{exercise}

\begin{exercise}
Consider the electromagnetic field on Minkowski spacetime, interacting with an external conserved current $J^\mu=(\rho,\mathbf{J})$, with action
\begin{equation}
S[A_\mu]
=
\int d^4x \left(
-\frac14 F_{\mu\nu}F^{\mu\nu}
- J^\mu A_\mu
\right),
\qquad
F_{\mu\nu}=\partial_\mu A_\nu-\partial_\nu A_\mu ,
\end{equation}
and assume $\partial_\mu J^\mu=0$.

\begin{enumerate}
\item Perform the $3+1$ decomposition of the action, writing it in terms of the fields $A_i$ and $A_0$ and the components $J^i$ and $\rho = J^0$ of the current.   Derive the canonical Hamiltonian
\begin{equation}
H
=
\int d^3x
\left[
\frac12 \pi^i\pi_i
+\frac14 F_{ij}F^{ij}
-\!A_0(\partial_i\pi^i-\rho)
+J^iA_i
\right]
\end{equation}
Explain why $A_0$ functions as a Lagrange multiplier and identify the Gauss constraints  $g :=\partial_i\pi^i -\rho $. Are they  first-class?

\item Identify the gauge transformations generated by the smeared Gauss constraint, $g[\lambda]
=
\int d^3x\, \lambda(x)\, g(x)$.

\item Determine the reduced state space without imposing any gauge-fixing condition.  
  Decompose the fields into transverse and longitudinal parts,
\begin{equation}
A_i = A_i^{T} + \partial_i \varphi,
\qquad
\pi^i = \pi_T^i + \partial^i \psi,
\end{equation}
with $\partial_i A_T^i = 0$ and $\partial_i \pi_T^i = 0$.
Show that Gauss' law determines the longitudinal part of the momentum through Poisson's equation $\nabla^2 \psi = \rho$, 
  Conclude that the reduced state space is spanned by the transverse variables $(A_i^T,\pi_T^i)$.

\item Show that, after eliminating the constrained and pure-gauge variables, the reduced Hamiltonian is
\begin{equation}
\tilde{H }
=
\int d^3x
\left[
\frac12 \pi_T^i\pi^T_i
+\frac14 F_{ij}^T F_T^{ij}
+J^i A_i^T
\right]
+\frac12\int d^3x\, \rho\,\psi ,
\end{equation}
where $\psi$ solves $\nabla^2\psi=\rho$.  What is the meaning of the last term?
\end{enumerate}
 
\end{exercise}

\section{GR constraints and time}

\subsection{Properties of the constraints}

 First, we summarize some crucial properties of the constraints of GR.
\begin{itemize}

\item Once the lapse $N$ and shift $N^i$ are fixed, the constraints lead to a well-posed Cauchy problem for the dynamical equations \cite{HE73, FM79}.

\item Any Lorentzian metric $g$ satisfying the Einstein equations $G_{\alpha\beta}=0$ automatically satisfies the constraint equations on every spacelike hypersurface. This follows by the analysis of Sec. 4.2.

\item Conversely, if the constraint equations hold on all spacelike hypersurfaces of $\M$, then $g$ satisfies the full Einstein equations $G_{\alpha\beta}=0$.

\end{itemize}

The last result is of particular  significance. It implies that constraint fully encode the dynamical content of GR.
The proof is straightforward. The Hamiltonian constraint implies
\[
n^\alpha n^\beta G_{\alpha\beta}=0
\]
for the normal $n^\alpha$ to any hypersurface. Since the foliation can be chosen so that $n^\alpha$ coincides with any timelike vector at a point, this implies $m^\alpha m^\beta G_{\alpha\beta}=0$
for all timelike vectors $m$. Now, if $m_1$ and $m_2$ are timelike vectors, so is  $m_1+ m_2$. This implies that $m_1^\alpha m_2^\beta G_{\alpha\beta}=0$. 

We can write any spacelike vector as $v = m_1 - m_2$, in terms of two time-like vectors $m_1$ and $m_2$. Hence, for any timelike $m$,  $m^\alpha v^\beta G_{\alpha\beta}=0$. With the same procedure, we show that $v_1^\alpha v_2^\beta G_{\alpha\beta}=0$ for all spacelike vectors $v_1$ and $v_2$. By linearity, we conclude that 
\[
u^\alpha v^\beta G_{\alpha\beta}=0
\]
for all vectors $u^\alpha,v^\beta$.  Hence $G_{\alpha\beta}=0$.

This proof shows that the Hamiltonian constraint alone is sufficient; the momentum constraints need not be imposed separately.

\begin{exercise}
Show that if Gauss' law holds on any spacelike hypersurface, then Maxwell's equations follow.
\end{exercise}

\subsection{Deparameterization of the ADM Action}

A natural route to isolating the physical degrees of freedom of general relativity is to introduce {\em internal coordinates}, defined as functionals of the canonical variables. The basic idea, emphasized already by Baierlein, Sharp, and Wheeler \cite{BSW62} in the 1960s, is to construct four functionals $\X^A(x;h ,\pi]$, $A=0,1,2,3$, that can serve as spacetime coordinates associated with the point $x\in\Sigma$ and the canonical data $(h, \pi)$. These variables provide a relational notion of time and space, in which events are identified by values of the fields rather than by an external background structure.

%In a given spacetime $(\M,g)$, such functionals assign to each event $X \in \M$ coordinates
%\begin{equation}
%    X^A(X) = \X^A\big(x;h(t),p\big],
%\end{equation}
%where $(x,t)$ are defined with respect to a reference foliation. Consistency requires that the level surfaces $T=\mathrm{const}$ be spacelike and the curves %$Z^a=\mathrm{const}$ be timelike, so that the functionals $\X^A$ define an admissible coordinate system with respect to the metric $g$. In this sense, spacetime coordinates %are reconstructed from the dynamical variables themselves.

The introduction of internal coordinates allows for a canonical transformation
\begin{equation}
   \big(h_{ij},\pi^{kl}\big)\mapsto
   \big(\X^A,\PP_B;\phi^r,p_s\big), \label{cctr}
\end{equation}
where $\X^A$ are the coordinate variables, $\PP_B$ their conjugate momenta, and $(\phi^r,p_s)$, $r=1,2$, denote the genuinely dynamical degrees of freedom of the gravitational field\footnote{I use the words ``genuinely dynamical" to distinguish from the true degrees of freedom, that is, state space variables that commute with the constraints.}. 

The symplectic form on the phase space $\Gamma$ is written in the Darboux form
\bey
\omega = \int d^3 x \delta \pi^{ij}(x) \wedge \delta h_{ij}(x) = \int d^3 x \left[ \delta \PP_A(x)\wedge \delta \X^A(x) + \delta p_s(x)\wedge \delta \phi^s(x)\right],
\eey
corresponding to canonical Poisson brackets for the conjugate pairs and vanishing cross-brackets. 
 In these variables, the state space action takes the form
\begin{equation}
    S=\int dt\int_\Sigma d^3x\,
    \big(\PP_A\dot{\X}^A+\pi_r\dot\phi^r - N  \mathcal{H} - N^i \mathcal{H}_i\big), \label{sssa}
\end{equation}
where the constraints are expressed as functionals of $(\X,\PP,\phi,p)$.

The next step is to solve the constraints for the momenta $\PP_A$,
\begin{equation}
    \PP_A + h_A(\X,\phi, p)=0, \label{depar}
\end{equation}
allowing us to express the dynamics in terms of an action for  {\em parameterized field theory}
\bey
S=\int dt\int_\Sigma d^3x\ \big[\PP_A\dot{\X}^A+\pi_r\dot\phi^r - \lambda^A(\PP_A + h_A(\X,\phi, p))\big], \label{depar2}
\eey
where $\lambda^A$ are Lagrange multipliers. 

Alternatively, substituting the solution (\ref{depar}) into the  action (\ref{sssa})
  yields a reduced action depending only on the physical variables,
\begin{equation}
    S_{red}[\phi, p] =
    \int dt\int_\Sigma d^3x\,
    \big\{p_r \dot\phi^r - h_A(\X,\phi,p)\,\dot{\X}^A\big\}.
\end{equation}
where the variables $\X^A$ are no longer varied but are gauge-fixed to prescribed functions $\chi_t^A(x)$.  This step is the converse of the parameterization of the action discussed in Sec. 2.2---it constitutes the {\em deparameterization} of the theory: the coordinate variables are treated as external parameters, and the dynamics is expressed entirely in terms of the physical modes.

The obvious gauge choice is $\chi^0_t=t$, $\chi^i_t=x^i$, in which case the reduced system is governed by the Hamiltonian

\begin{equation}
    H_{true}(t) = \int_\Sigma d^3x\,
     h_0\big(x, t,\phi, p\big),
\end{equation}
and evolves according to Hamilton's equations for $(\phi^r,p_s)$.  
This construction shows that, at least formally, general relativity can be rewritten as a conventional Hamiltonian system for two physical degrees of freedom per spatial point.  

\subsection{Time functions}
Unfortunately, deparameterization does not seem to work. To understand why, we will analyze its finite dimensional analogue:  a parameterized system is described by a phase space $\Gamma$ with canonical form $\theta$, a first-class constraint $C(\xi)$ and a Hamiltonian $H(\xi) = NC$ that vanishes when the constraint is imposed; $N > 0$ is  a Lagrange multiplier.

This system is deparameterized if we can perform a canonical transformation $\xi = (y, T, P_T)$, such that 
 $\theta(y, T, P_T) = \tilde{\theta}(y) + P_T dT$, and we can solve the constraint as $P_T + h(T, y) = 0$. Then, $T$ defines a time variable for the $y$ degrees of freedom evolving under the Hamiltonian $h(T, y)$. However, not all variables $T$ are good times. A genuine {\em time function} must be strictly increasing along the solutions to the equations of motion, so we should have:
 \bey
 \{T, C\} > 0. \label{ineqtc}
 \eey
Eq. (\ref{ineqtc}) implies that a time function can never
be an observable, in the  sense of commuting with the constraints: $T$ is not a
true degree of freedom and does not live in $\Gamma_{red}$. To obtain a time coordinate at all,
one must reach outside the reduced state space. This is the precise sense in which time,
in a constrained theory, is never one of the physical variables but always an extra label
adjoined to them.
 
 We will see how this works in two examples.

\medskip

\noindent{\em The relativistic particle.}
Consider  a relativistic particle on Minkowski spacetime. In this system, $\xi = (x^{\mu}, p_{\nu})$, and $\theta = p_{\mu} d x^{\mu}$. The constraint is
\bey
C = p_\mu p^\mu + m^2 = 0.
\eey
Choosing $T = x^0$, we find
\bey
\{x^0, C\} = 2 p^0,
\eey
which is strictly positive (or negative) on each of the two disjoint branches $p^0>0$ or $p^0<0$. Restricting to one branch, $T=x^0$ is a monotonic function along all solutions. The constraint can then be solved globally as
\bey
p_0 = -\sqrt{\mathbf{p}^{\,2}+m^2},
\eey
yielding a Hamiltonian for the spatial degrees of freedom. Thus, after selecting a sector, the system admits a global deparameterization.

\medskip

\noindent{\em The harmonic oscillator.}
Consider now the parameterized harmonic oscillator with constraint
\bey
C = p_t + \frac12(p_q^2 + \omega^2 q^2) = 0.
\eey
If we choose $T = t$, the system is trivially deparameterized. However, if we attempt to use $T=q$ as a time variable, we find
\bey
\{q, C\} = p_q,
\eey
which vanishes at the turning points of the motion. Hence $q$ is not monotonic along the solutions, and cannot serve as a global time function. Correspondingly, solving the constraint for $p_q$ yields
\bey
p_q = \pm \sqrt{-2p_t - \omega^2 q^2},
\eey
which defines two branches that meet at $p_q=0$. The reduced Hamiltonian is therefore multivalued, and the deparameterization fails globally.

These two examples illustrate the key point: a deparameterization exists only if one can find a phase-space function $T$ that is globally monotonic along the constraint flow. In general, such functions need not exist, even in simple finite-dimensional systems. This obstruction becomes much more severe in general relativity, where one must find not a single time variable but a field of such functions, one at each spatial point. We will see in Section 4.4.7 that simple models where GR is restricted to a finite-dimensional submanifold of its state space cannot admit a time function.

\subsection{Deparameterization with Dust Fields}

A concrete realization of deparameterization in generally covariant systems is obtained by coupling gravity to a pressureless perfect fluid (“dust”) and using the dust variables as physical reference fields. In the Brown--Kucha\v{r} formulation \cite{BrKu}, the dust is described by four scalar fields $(T,Z^a)$, $a=1,2,3$, interpreted as the proper time and comoving spatial labels of the fluid elements. The canonical action for the coupled system takes the form
\begin{equation}
S=\int dt\int_\Sigma d^3x
\left(
p^{ij}\dot h_{ij}
+P\dot T
+P_i\dot Z^i
-N\HH
-N^i\HH_i
\right),
\end{equation}
where $(h_{ij},p^{ij})$ are the gravitational variables, $(T,P)$ and $(Z^a,P_a)$ are the dust variables, and $N$, $N^i$ are the lapse and shift. The total constraints split into gravitational (g) and dust (d) contributions: $\HH = \HH^{g}+\HH^{d}$, and $\HH_i = \HH_i^{g}+\HH_i^{d}$, where the gravitational contributions are given by Eqs. (\ref{hamc},\ref{momc})
\bey
\HH_i^{d}=P\,\partial_i T+P_a\,\partial_i Z^a \\
\HH^{d} = \sqrt{P^2 +h^{ij} \HH^d_i\HH^d_j}.
\eey
The key point here is that a canonical transformation that expresses the gravitational variables as functions of $Z^a$ enables us to write the Hamiltonian constraint in deparameterized form. This means essentially that when we select a foliation where the map ${\cal E}$ takes the dust variables as arguments. 
The crucial feature of the dust model is that the Hamiltonian constraint can be rewritten in deparameterized form. After an appropriate canonical transformation, one can solve it for the momentum conjugate to the dust time,
\begin{equation}
\HH = P + h = 0,
\end{equation}
where
\begin{equation}
h
=
- \sqrt{
\left(\HH^{g}\right)^2
-
h^{ij}\mathcal H_i^{g}\mathcal H_j^{g}
}.
\end{equation}
In this expression, $h$ depends only on the gravitational variables. The total Hamiltonian thus becomes a true Hamiltonian generating evolution with respect to the dust time $T$.

 This construction achieves a genuine deparameterization: the constraints are reduced to evolution equations, and the dynamics is generated by a non-vanishing Hamiltonian. This is made possible by the highly special properties of pressureless dust, whose flow defines a congruence of non-intersecting timelike curves along which $T$ is monotonic. 
 
 However, dust is not a fundamental field, but an idealized form of matter with very restrictive properties (absence of pressure, vorticity, and interactions), and its use introduces additional physical degrees of freedom into the theory. The resulting system is therefore not pure gravity, but gravity coupled to a very particular matter model, in which a preferred temporal structure is effectively put in by hand through the choice of dust.

\subsection{Clock relativism}

An alternative strategy, due to Rovelli, is to abandon the requirement of a
global time function altogether and define dynamics in a purely relational
manner \cite{Rovelli}. The basic idea is to treat any state-space function $T(\xi)$ as a
\emph{partial observable} that can be used as a clock, without requiring it to
be globally monotonic. Given another state-space function $f(\xi)$, one defines
the corresponding \emph{evolving constant of motion},
\bey
f_{\tau}^{(T)}(\xi) :=   f(\bar{\xi}_{\tau}), \;\; \mbox{where}\;\; \{\bar{\xi}_\tau\} = {\cal O}_{\xi}  \cap \{\xi: T(\xi) = \tau \}
\eey
where ${\cal O}_{\xi}$ is the constraint orbit that passes from $\xi$.  
The time function $T$ selects one point from the orbit  ${\cal O}_{\xi}$ for each value of $\tau$, and this point determines the value of $f_{\tau}^{(T)}$ at $\xi$.
By construction, $f_{\tau}^{(T)}$ is constant on each orbit, hence, an observable.
 In this way, dynamics is
encoded in correlations between observables, rather than in evolution with
respect to an external or preferred time parameter.

We can always choose $T$ as a coordinate on the constraint surface so that $\xi = (T, q)$ where $q$ refers to the remaining degrees of freedom. Then, the equations determining 
$f_{\tau}^{(T)}(T, q)$ are
\bey
f_{\tau}^{(T)}(T, q) = f(\tau, q), \; \;\; \{f_{\tau}^{(T)}, C\} = 0.
\eey 
It is straightforward to derive the evolution equation:
\bey
\frac{d}{d\tau} f_{\tau}^{(T)} = \frac{\{f, C\}}{\{T, C\}}. \label{timev}
\eey
\

At the classical level, this construction is well defined whenever $T$ can be
used as a local parameter along the gauge orbits. In particular, it does not
require $T$ to be a global time function in the sense discussed above. This
seems to provide a way around the obstruction to deparameterization: instead of
selecting a preferred time variable, one allows many different choices of
clock, each leading to a corresponding family of relational observables.

However, this freedom comes at a price. Different choices of clock generally
lead to inequivalent descriptions of causal orders. To see this explicitly,
consider   the parameterized free particle with phase space $\Gamma = \{(t, p_t, x, p_x)\}$ and constraint
\bey
C = p_t + \frac{p_x^2}{2}.
\eey
Let us take two admissible clock variables,
\bey
T = t, \qquad T' = t + b x,
\eey
with constant $b$. We have
\bey
\{T, C\} = 1, \qquad \{T', C\} = 1 + b p_x,
\eey
so both define valid clocks if $p_x > - b^{-1}$. The orbit ${\cal O}_{\xi}$ of $\xi = (t, p_t, x, p_x)$ is given by $(t + s, p_t, x + p_x s, p_x)$ for all $s \in \R$.
Hence, for any $\xi$, $\bar{x}_{\tau} = (\tau, p_t, x + p_x (\tau - t), p_x)$ with respect to $T$, and 
$$\bar{\xi}' = \left(t + \frac{\tau-t}{bp_x} - \frac{x}{p_x}, p_t, \frac{\tau-t}{b}, p_x\right).
$$
Consider now two events $\alpha_1$ and $\alpha_2$ in the same orbit, $\alpha_1$ defined by $\tau = \tau_1$ and $\alpha_2$ defined by $\tau = \tau_2$, with respect to the $T$ clock. The associated times for the $T'$ clock satisfy
\bey
\tau_2'-\tau_1' = (bp_x)(\tau_2 - \tau_1).
\eey
For $-1 < bp_x < 0$, the two clocks yield opposite ordering tof  events.

This example shows that Rovelli’s relational construction does not merely
reparametrize a pre-existing notion of time. Rather, different choices of clock may
define incompatible causal orders. In particular, the relation
``before'' between events is no longer invariant, but depends on the auxiliary
choice of clock variable. While each choice yields a consistent set of
gauge-invariant observables, there is in general no coherent way to combine
them into a single, clock-independent description of temporal ordering.

The conclusion is that, although the relational framework avoids the need for a
global time function, it does so by relinquishing the invariance of causal orders. In this, it contradicts classical GR, where the spacetime causal ordering is unique for a given solution to the equations of motion. If a time function fails to respect it, it is simply a bad time function. However, this is not necessarily true in quantum gravity.

\subsection{Causal relations as fundamental}
 It is important to separate the two aspects of time that the clock constructions conflate: causal
relation (the ordering of events) and measure (the duration between them). Clocks deliver
the measure, but, as the clock-relativism example above shows, they do not in general
deliver the correct causal relations.  

To preserve causal relations for the true degrees of freedom, we need to employ a formulation in which histories, rather than instantaneous states are fundamental. Histories have an intrinsic causal structure that is essentially logical, and this structure underlies all dynamical interactions.

 To this end, we use Savvidou's symplectic histories  formalism   to define the history space $\Pi$ as the space of all maps from $\R$ to $\Gamma$. 
The history space $\Pi$ is endowed with the symplectic structure (\ref{hisav}) that generates the history Poisson brackets (\ref{histpo}), lifting 
  the canonical structure is lifted from instantaneous variables to time-labelled fields. Two functionals play a central role. The first is the Liouville functional
\[
V[\xi(\cdot)] = \int dt \, \theta_{a}[\xi(t)] \dot{\xi}^a_t,
\]
which generates translations of the parameter $t$ along the history: $\xi_t^a \rightarrow \xi_{t+s}^a$. Note that the parameter $t$ is arbitrary: $V$ is invariant under a path reparameterization.

The second key functional is the history Hamiltonian constraint
\[
H_N[\xi(\cdot)] = \int dt \, N(t)  C[\xi(t)],
\]
where $N(t)$ is an arbitrary smearing function. The action functional is then given by $S_N = V - H_N$.
 
The crucial point is that $V$ and $H_N$ generate distinct transformations: the former corresponds to shifts in the \emph{ordering parameter} $t$, while the latter generates the gauge transformation of the Hamiltonian constraint.
Since $H_N$ does not depend on derivatives of $\dot{\xi}^a_t$, the symplectic reduction on $\Pi$ through the constraint $H_N = 0$, for all $N$, preserves the $t$-label. So $\Pi_{red}$ is identical with the set of paths from $\R$ to $\Gamma_{red}$. Furthermore, since $\{V, H_N\} = H_{\dot{N}}$, $\{V, H_N\}$  vanishes on the history constraint surface, so it is projected to a function $\tilde{V}$ in $\Pi_{red}$. Note that $\tilde{V}$ is defined uniquely, even if $\Gamma_{red}$ does not admit a canonical form. 

In $\Pi_{red}$, $\tilde{V}$ coincides with the action functional, so the classical equations of motion satisfy $\{\tilde{V}, \tilde{F}_t\} = 0$, for any family of observbles $\tilde{F}_t$. This means that the classical paths on $\Pi_{red}$ correspond to constants. Note, however, that the notion of time has not been lost. Any path on $\Pi_{red}$ has a well defined notion of time.
 
The conceptual outcome is that the apparent disappearance of time in the reduced phase space is an artifact of the canonical formalism. In the histories framework, temporal ordering is not something to be reconstructed from physical clocks, coordinates, or gauge fixing conditions; it is part of the kinematical structure of the theory. Dynamics then describes how physical quantities vary along this ordered structure, but does not create it. This construction generalizes to full GR \cite{Sav1, Sav2}, and we will discuss it in Chapter 6.

\subsection{Application: minisuperspace models}
 Minisuperspace models provide a symmetry-reduced version of canonical quantum gravity in which only a finite number of gravitational degrees of freedom are retained. Instead of quantising the full infinite-dimensional superspace of three-geometries, one restricts attention to highly symmetric metrics, typically homogeneous and isotropic cosmological spacetimes. The resulting theory reduces the Einstein equations to a finite-dimensional constrained dynamical system, making explicit calculations possible while preserving many conceptual features of quantum gravity, such as the Hamiltonian constraint and the Wheeler--DeWitt equation.

The most commonly used example is the Friedmann--Robertson--Walker (FRW) spacetime coupled to a homogeneous scalar field $\phi(t)$. The metric is
\begin{equation}
ds^2
=
-N^2(t)dt^2
+a^2(t)
\left[
\frac{dr^2}{1-\zeta r^2}
+r^2(d\theta^2+\sin^2\theta\, d\varphi^2)
\right],
\end{equation}
where $a(t)$ is the scale factor, $N(t)$ is the lapse function, and
$
\zeta=0,\pm1
$
determines the spatial curvature.

For a scalar field with potential $V(\varphi)$, the Einstein--Hilbert action reduces to the  Lagrangian
\begin{equation}
L
=
-\frac{3a\dot a^2}{N}
+3N \zeta a
+\frac{a^3}{2N}\dot\varphi^2
-
Na^3V(\varphi).
\label{frwlag}
\end{equation}
Note that a constant contribution to $V(\phi)$ corresponds to the cosmological constant $\Lambda$.

It is convenient to use configuration coordinates $x$ and $\phi$, where $a = \frac{3}{2}x^{2/3}$ and $\phi = \frac{3}{4}\varphi$, and to rescale the potential $V(\varphi) \rightarrow \frac{27}{4}V(\phi)$. Then,
\bey
L = -\frac{\dot{x}^2}{2N} + \frac{9N}{2}\zeta x^{2/3} + \frac{3x^2}{N}\dot \phi^2- \frac{1}{2}N x^2 V(\phi).
\eey

The conjugate momenta are $p_x = -\dot x/N$ and $p_\phi = 6 x^2 \dot \phi/N$. The Hamiltonian is $H =  N C$, where the Hamiltonian constraint is 
\bey
C = \frac{-p_x^2}{2} + \frac{p_{\phi}^2}{12 x^2} - \frac{9\zeta}{2} x^{2/3} + \frac{1}{2}x^2 V(\phi)
\eey

For generic $V(\phi)$, there is no elementary reduction of the system. However, for constant $V(\phi) = \tilde{\Lambda}$, $\{p_{\phi}, C\} = 0$, so $p_{\phi}$ is an observable.
 The constraint equation becomes
\bey
x^2p_x^2 + \upsilon(x) = \frac{1}{6}p_{\phi}^2, \label{redcon}
\eey
where $\upsilon(x) = -\tilde{\Lambda}x^4 + 9 \zeta x^{8/3}$, and $\tilde{\Lambda}$ is the cosmological constant $\Lambda$ times $27/4$.   

If $\upsilon(x) < 0$, we can parameterize the constraint surface through a parameter $u$, so that $\sqrt{|\upsilon(x)|} = \frac{|p_\phi|}{\sqrt{6}} \sinh u$ and $xp_x = \pm \frac{|p_\phi|}{\sqrt{6}} \cosh u$.  If $\upsilon(x) > 0$, we use an angular parameter $\theta$, so that $\sqrt{\upsilon(x)} = \frac{|p_\phi|}{\sqrt{6}} \sin \theta$ and $xp_x =  \frac{|p_\phi|}{\sqrt{6}} \cos \theta$. In general, the reduced state space is spanned by $p_\phi$ and its conjugate observable $f$, which is a  combination of $\phi$ and $u$ (or $\theta$).

\begin{exercise}
For $\tilde{\Lambda}> 0$ and $\zeta = 0$, show that the symplectic form  on the reduced state space $\Gamma_{red}$ is $\omega = dp_{\phi} \wedge df$, where $f = \phi \pm \frac{1}{2\sqrt{6}}\cosh u$. How many connected components does $\Gamma_{red}$ consist of?
\end{exercise}
\begin{exercise}
For $\tilde{\Lambda}= 0$ and $\zeta = 1$, show that the symplectic form on the reduced state space $\Gamma_{red}$ is $\omega = dp_{\phi} \wedge df$, where $f = \phi + \frac{1}{2\sqrt{6}}\log \tan\frac{\theta}{2}$.  
\end{exercise}

We can deparameterize the system using different variables for time $T$. For example, taking $T = x$, we solve the constraint (\ref{redcon}) to obtain
\bey
p_x \pm \sqrt{\frac{p_{\phi}^2}{6x^2} - \upsilon(x)} = 0
\eey
If $\upsilon(x) < 0$ for all $x$, then $p_x$ never vanishes and $T$ is an appropriate time function in the branch of the constraint surface with $p_x< 0$. 

York proposed that the trace of the extrinsic curvature $K = K_{ij}h^{ij}$ could place the role of time, at least in spacetimes where on can find an foliation in which $K$ is constant at each leaf \cite{York72}. In the FRW minisuperspace model, the York time is $T = \frac{p_x}{2x}$, and its conjugate is $p_T = -  x^2$. The Hamiltonian constraint in these variables becomes
\bey
C = \frac{1}{2}p_T^2\left[(2T)^2 - V(\phi)\right] -\frac{p_{\phi}^2}{12 p_T} + \frac{9\zeta}{2} p_t^{1/3}.
\eey
\begin{exercise}
Deparameterize the Hamiltonian constraint with respect to York time for $\zeta = 0$. When does York time define a good global time parameter?
\end{exercise}

\begin{exercise}
For $\tilde{\Lambda}> 0$ and $\zeta = 0$, write $x(\tau)$ as an evolving constant of a motion (i) for $T = \phi$, and (ii) for the York time.
\end{exercise}

\section{Gravity as a spin-2 interaction}
An alternative route to general relativistic dynamics starts from the dynamics of a massless spin-2 field on Minkowski spacetime. One seeks a Lorentz-invariant theory for a symmetric tensor field $\gamma_{\mu\nu}$ that propagates the two physical degrees of freedom of a massless particle. The unique linear theory with this property is the Fierz--Pauli theory \cite{FP39}, with action
\begin{equation}
S_{\mathrm{FP}}[\gamma]
=
\int d^4x \left[
-\frac12 \partial_\lambda \gamma_{\mu\nu}\partial^\lambda \gamma^{\mu\nu}
+\partial_\mu h^{\mu\nu}\partial^\lambda \gamma_{\lambda\nu}
-\partial_\mu \gamma^{\mu\nu}\partial_\nu \gamma
+\frac12 \partial_\lambda \gamma \partial^\lambda \gamma
\right], \label{FPt}
\end{equation}
where $\gamma=\eta^{\mu\nu}h_{\mu\nu}$. This action is invariant under the gauge transformation
\begin{equation}
\gamma_{\mu\nu} \mapsto \gamma_{\mu\nu} + \partial_\mu \xi_\nu + \partial_\nu \xi_\mu,
\end{equation}
which removes the unphysical degrees of freedom.

To describe gravity, the field must couple to matter. At lowest order, the only consistent coupling is
\begin{equation}
S_{\mathrm{int}} = \kappa \int d^4x \, \gamma_{\mu\nu} T^{\mu\nu},
\end{equation}
where $T^{\mu\nu}$ is a conserved stress-energy tensor. However, this linear coupling is not consistent by itself. The field $\gamma_{\mu\nu}$ carries energy and momentum, and therefore must contribute to the source. Including its stress-energy tensor modifies the equations of motion, which in turn alters the stress-energy tensor itself. This leads to an iterative process in which the field couples to its own energy-momentum.

Requiring consistency of this self-coupling uniquely fixes the nonlinear structure of the theory. As shown by Deser \cite{Deser} (following earlier insights by Feynman), the iteration resums into a generally covariant theory. Introducing
\begin{equation}
g_{\mu\nu} = \eta_{\mu\nu} + \kappa \gamma_{\mu\nu}, \label{pertg}
\end{equation}
one finds that the nonlinear completion of the equations is equivalent to the Einstein equations. 
  Equivalently, the action becomes the Einstein--Hilbert action.

The essential point is that the requirement of consistent coupling to a conserved source forces the theory to become nonlinear, and the requirement that the field couple to its own stress-energy tensor leads uniquely to general relativity. The linear gauge symmetry of the Fierz--Pauli theory is thereby promoted to full diffeomorphism invariance.

This line of argument explains why gravity, if described by a massless spin--2 field with universal coupling and Lorentz invariance, leads naturally to dynamics equivalent to those of general relativity. In this sense, it establishes a form of dynamical uniqueness for Einstein gravity. However, the construction is intrinsically perturbative: one begins with a field $\gamma_{\mu\nu}$ on a fixed Minkowski background and only subsequently reinterprets $\eta_{\mu\nu}+\kappa \gamma_{\mu\nu}$ as an effective spacetime metric.

As a result, the full conceptual content of general relativity is not present from the outset. In particular, diffeomorphism invariance appears only indirectly, and the identification of the metric with spacetime geometry is not fundamental but emergent within the perturbative scheme. Moreover, the configuration space of a symmetric tensor field is much larger than that of Lorentzian metrics. For fixed $x$, $\gamma_{\mu \nu}$ is a general $4 \times 4$ matrix, so $g_{\mu \nu}$ defined by  (\ref{pertg}) is not guaranteed to have Lorentzian signature.   Additional conditions must therefore be imposed to recover the physically relevant sector.

Thus, while the argument provides a compelling route to the dynamics of Einstein gravity, it does not by itself capture the kinematical structure of general relativity or fully account for the intimate relation between the metric and the causal and geometric properties of spacetime.

\chapter{Approaches to quantum gravity and their treatment of time}

\section{Quantisation methods}
 
 \begin{flushright}
\emph{“Quantisation is an art, not a functor.”}\\
— Anonymous 
\end{flushright}

A central issue in all approaches to quantum gravity is the choice of \emph{quantisation method}. Gravity is a constrained gauge system with diffeomorphism invariance, and the standard quantisation techniques must be adapted to handle constraints, redundancies, and the absence of a preferred time parameter.

\subsection{Quantisation rules}

Quantisation is not an a priori, uniquely
defined procedure. One cannot quantize all classical observables consistently,
so each quantisation scheme selects  a family of classical observables with a direct quantum counterpart. The rules of this selection is what
  distinguishes one quantisation programme from
another.

Heisenberg's postulate of the fundamental commutation relations raises a basic question: classical observables such as position and momentum are functions on a phase space $\Gamma$, so why should they be replaced by operators satisfying algebraic relations in quantum theory? Is this structure already present in classical mechanics?

A first systematic answer was given by Dirac \cite{Dirac25, Dirac30}, who observed a close analogy between the Poisson bracket and the operator commutator. On a classical phase space $\Gamma = \mathbb{R}^{2n}$ with coordinates $(x_a, p_a)$, the fundamental Poisson brackets are
\begin{equation}
\{x_a, x_b \} = 0, \qquad
\{p_a, p_b \} = 0, \qquad
\{x_a, p_b\} = \delta_{ab}.
\end{equation}
Dirac suggested that the similarity of the Poisson bracket to the operator commutator provides the basis for a {\em quantisation} procedure, i.e., a set of rules for associating quantum observables to classical ones. He proposed that we associate functions on the classical state space $\Gamma$ to self-adjoint  operators on the Hilbert space ${\cal H}$ according to the following rules.

\begin{enumerate}
\itemsep0em
\item The functions   $x_a$  and $p_a$ on $\Gamma$ respectively correspond to the operators $\hat{x}_a$ and $\hat{p}_a$ on ${\cal H}$.
\item Any function $f(x_a)$  on $\Gamma$ corresponds to the operator $f(\hat{x}_a)$ and any function $g(p_a)$ on   $\Gamma$ corresponds to the operator  $g(\hat{p}_a)$.
\item If the functions $F$ and $G$ on $\Gamma$ correspond to the operators $\hat{F}$ and $\hat{G}$, respectively, then the function $F +  G$ corresponds to the operator $\hat{F} + \hat{G}$.
\item The Poisson bracket $\{\cdot, \cdot\}$ on $\Gamma$ corresponds to $\frac{1}{i} [\cdot, \cdot]$, where $[\cdot, \cdot]$ is the operator commutator on ${\cal H}$. \index{Poisson bracket}
\end{enumerate}

Dirac's rules have proved very useful. For example, they allow us to specify the Hamiltonian of a particle as a Schr\"odinger operator  $\hat{H} = \frac{\hat{p}^2}{2m} +V(\hat{x})$, from the knowledge of the classical Hamiltonian $H(x, p) = \frac{p^2}{2m} + V(x)$.

However, this correspondence is not unique. For classical observables involving both $x$ and $p$, there is no canonical operator ordering. For example, the function $xp^2$ can be mapped to different operators such as
$$\frac{1}{2}(\hat{x}\hat{p}^2 + \hat{p}^2 \hat{x}), \qquad \hat{p}\hat{x}\hat{p},
$$
which differ by terms of order $\hbar$. This \emph{factor-ordering ambiguity} reflects the absence of a unique quantisation map.

More seriously, the Dirac rules are not globally consistent. As shown by Groenewold \cite{Groen45}, there exist classical identities involving Poisson brackets that cannot be preserved under the correspondence with commutators. Thus, quantisation cannot be defined as a universal algebraic homomorphism between classical and quantum observables.

\begin{exercise}
Prove that
\begin{eqnarray}
\{x^3, p^3\}+ \frac{1}{12} \{\{p^2, x^3\},\{x^2, p^3\}\} = 0. \nonumber 
\end{eqnarray}
Use the quantisation rules to express the left-hand side of the equation above in quantum mechanics.  Then, prove that
\begin{eqnarray}
\frac{1}{i}[\hat{x}^3, \hat{p}^3] + \frac{1}{12i}[\frac{1}{i}[\hat{p}^2, \hat{x}^3], \frac{1}{i}[\hat{x}^2, \hat{p}^3]] = -3, \nonumber
\end{eqnarray}
to show the inconsistency of Dirac's rules.
\end{exercise}

\medskip

A further limitation is that the canonical commutation relations rely on the assumption that the phase space $\Gamma$ is a linear space. For systems with nontrivial topology or with constraints, the Heisenberg commutation relations do not work. One can see this even in elementary systems. For a  particle moving on the half-line, 
space translations are obstructed by the boundary, and the momentum operator cannot be defined. For a particle on a circle, a self-adjoint position operator with periodic spectrum is impossible.

 In recent years, research in many fields of physics (condensed matter, high-energy physics, gravity theory) has uncovered many systems with non-trivial classical state spaces. A new set of quantisation rules is necessary. To this end, 
 two complementary approaches have been developed. The first approach aims to construct the quantum theory geometrically, i.e., through the introduction of additional geometric structures on the classical state space. The result is a formalism known as {\em geometric quantisation}. The second approach is closer to Heisenberg's logic, it tries to identify an analogue  of the Heisenberg-Weyl group (see below),  the {\em canonical group}, in each classical system. The associated quantum theory is obtained from the study of the unitary  representations of the canonical group.  
 
 To see how the latter works, consider the state space $\Gamma=\R^2$ with coordinates $(x,p)$. The functions $f_{(a,b)} = ax - bp$.
generate the canonical transformations
\begin{equation}
(x,p) \rightarrow (x - sb,\, p - sa),
\end{equation}
which corresponds to position and momentum translations. Their Poisson brackets satisfy 
\bey
\{f_{(a_1,b_1)}, f_{(a_2,b_2)}\} = a_2 b_1 - a_1 b_2.
\eey
From the corresponding quantum operators $\hat{x}$ and $\hat{p}$ in the Hilbert space ${\cal H} = L^2(\R, dx)$, we define the self-adjoint operators, $\hat{f}_{a,b} = a\hat{x} - b \hat{p}$, which satisfy
\bey
[\hat{f}_{a_1, b_1}, \hat{f}_{a_2, b_2}] = i (a_2 b_1 - a_1 b_2)\hat{I}.
\eey
They also generate position and momentum translations, since
\bey
e^{i\hat{f}_{a,b}s}\hat{x}e^{-i\hat{f}_{a,b}s} = \hat{x} - s b\hat{I},\;\;\; e^{i\hat{f}_{a,b}s}\hat{p}e^{-i\hat{f}_{a,b}s} = \hat{p} - s a\hat{I}.
\eey

The same group---the Heisenberg--Weyl group---acts on both the classical state space and the quantum phase space. Moreover, in both cases, its generators allow us to build all observables: classical observables as functions of $x$ and $p$ on $\R^2$, and quantum observables as functions of $\hat{x}$ and $\hat{p}$. This is because the classical group action on $\R^2$ is transitive, while the corresponding unitary representation on ${\cal H}$ is cyclic\footnote{Let a group $G$ act on a set (or manifold) $\Gamma$ via maps $f_g:\Gamma \to \Gamma$. The action is called \emph{transitive} if for any pair of points $\xi_1, \xi_2 \in \Gamma$, there exists an element $g \in G$ such that
$f_g(\xi_1) = \xi_2$. Equivalently, the orbit of any point coincides with the whole space,
$G \cdot \xi = \Gamma, \qquad \forall \xi \in \Gamma$.

Let now $G$ act on a Hilbert space ${\cal H}$ via a unitary representation $U:G \to {\cal U}({\cal H})$.
A vector $\psi \in {\cal H}$ is called \emph{cyclic} if the set of vectors obtained by the action of $G$ on $\psi$ spans a dense subspace of ${\cal H}$, i.e.,
\begin{equation}
\overline{\mathrm{span}\{ U(g)\psi \,:\, g \in G \}} = {\cal H}.
\end{equation}
A representation is called \emph{cyclic} if it admits at least one cyclic vector.
  
  Cyclicity provides the natural analogue of transitivity for group actions on Hilbert spaces: while unitary transformations cannot map arbitrary vectors to each other, the orbit of a cyclic vector is dense and generates the whole space.}.

This correspondence suggests that quantisation should be understood as a passage from a transitive classical group action to a cyclic unitary representation. This observation can be elevated to a general quantisation principle.
The procedure is as follows \cite{Isham83}.

\bigskip

\noindent (i) We identify the \emph{canonical group} $G$ associated to  the classical state space $\Gamma$: the smallest Lie group acting transitively on $\Gamma$ by canonical transformations, so as to capture the global structure of $\Gamma$.

\medskip

\noindent (ii) Let $\mathfrak{g}$ be the Lie algebra of $G$. Each element $T \in \mathfrak{g}$ defines a one-parameter family of canonical transformations, generated by a function $F_T$ on $\Gamma$. The map $T \mapsto F_T$ is defined up to constants, and in general realizes the Lie algebra only up to a central extension,
\begin{equation}
\{F_{T_1},F_{T_2}\} = F_{[T_1,T_2]} + z(T_1,T_2).
\end{equation}

\medskip

\noindent (iii) We construct the quantum theory by identifying the unitary irreducible and cyclic representations of $G$ (or its central extension). Each such representation defines a possible quantum kinematics.

\medskip

\noindent (iv) The generators $\hat{T}$ of the representation correspond to the classical functions $F_T$, with the correspondence
\begin{equation}
\{F_{T_1},F_{T_2}\} \;\longrightarrow\; i[\hat{T}_1,\hat{T}_2],
\end{equation}
together with linearity,
\[
F_{T_1}+F_{T_2} \rightarrow \hat{T}_1+\hat{T}_2, \qquad
\lambda F_T \rightarrow \lambda \hat{T}.
\]

\medskip

\noindent (v) Whenever many inequivalent representations exist, we select the appropriate one by demanding the incorporation of important observables, for example, the Hamiltonian, or other elements of the group of spacetime symmetries. 
\medskip

In this formulation, quantisation is not defined as a map between individual observables, but as a correspondence between classical symmetries and their quantum realizations. The role of canonical coordinates is replaced by the global action of the canonical group, allowing a consistent treatment of systems with nontrivial phase space structure.

While quantisation via the canonical group provides the most physically transparent route---since it directly relates quantum observables to generators of classical symmetries---it is not the only possible framework. In many cases, one works instead with more general algebraic structures, such as loop algebras or graded algebras. In these approaches, the fundamental objects  algebraic relations among classical observables that do not correspond to Lie algebras. Quantisation is implemented by constructing representations of these relations \cite{AshTa}. Such methods are often technically more flexible and can be applied in situations where a simple canonical group action is not available. Their drawback is that they  may obscure the correspondence with the classical state space, and thus the quantum-classical correspondence.

\subsection{Canonical quantisation of constrained systems}

The central idea in the quantisation of constrained systems is that only the true degrees of freedom carry physical significance; no probabilistic interpretation should be assigned to redundant variables.

Two distinct quantisation procedures arise, depending on whether the constraints are implemented before or after quantisation.  

\medskip

\noindent {\em 1. Solve constraints before quantisation.}  
If the constraints are solved at the classical level, one obtains the reduced state space of physical degrees of freedom. This reduced space defines a genuine Hamiltonian system, which can be quantized using standard methods. This procedure is known as \emph{reduced-state-space quantisation}.  

This approach is conceptually the most transparent, as it avoids any reference to redundant variables at the quantum level. However, it is technically difficult to implement, since the reduced state space may fail to be a smooth manifold and can exhibit singularities or complicated global structure. Although methods exist to deal with such cases, their complexity limits practical applications to relatively simple systems.

\medskip

\noindent {\em 2. Solve constraints after quantisation.}  
In this approach, one first quantizes the unconstrained system with state space $\Gamma$, obtaining a Hilbert space ${\cal H}$. The constraints are then represented by self-adjoint operators $\hat{\phi}_a$, and one requires that the Hamiltonian $\hat{H}$ preserves them,
\begin{equation}
[\hat{H}, \hat{\phi}_a] = 0.
\end{equation}

The \emph{physical Hilbert space} ${\cal H}_{phys}$ is defined as the subspace of states satisfying
\begin{equation}
\hat{\phi}_a |\psi\rangle = 0. \label{consdirac}
\end{equation}
Physical states belong to ${\cal H}_{phys}$, and probabilities are defined using its inner product. Observables are represented by self-adjoint operators $\hat{X}$ on ${\cal H}$ that commute with all constraints, so that they map ${\cal H}_{phys}$ into itself. This procedure is known as \emph{Dirac quantisation}.  

Dirac quantisation involves significant technical subtleties, particularly when the constraint operators have continuous spectrum at zero, making the definition of ${\cal H}_{phys}$ nontrivial. Nevertheless, it is the most widely used method, since the constraint equations (\ref{consdirac}) admit systematic approximation schemes (such as semiclassical or perturbative expansions) and can also be implemented in path-integral formulations.

The two routes are not interchangeable: they yield the same quantum
theory only for the simplest systems. For a theory as intricate as general relativity there is
no preferred quantisation, and no general reason for the two procedures to agree: operator-
ordering, non-uniqueness of the canonical group, and topological obstructions all enter and
can separate them. In practice the order of operations is itself part of the physical input
of a given quantum-gravity programme.

\subsection{Path-integral quantisation and the S-matrix.}

Path-integral quantisation shares the broad aim of the canonical approach---a Hilbert
space with observables---differing only in that the basic tool is a path integral rather than
the commutation relations.  

The challenge in path integral quantisation is the consistent treatment of gauge variables. A 
 naive path-integration over all configurations leads to overcounting. The Faddeev--Popov method provides a systematic way to define a consistent measure by restricting the integral to physically distinct configurations \cite{FaPo, FaSl}.

\medskip

Consider a system with phase space variables $(q^i, p_i)$, first-class constraints  $\phi_a(q,p)$, and gauge fixing conditions   $\chi_a(q,p) = 0$ that satisfy the condition (\ref{condgauge}).
The transition amplitude can then be written in the form
\begin{equation}
\langle \text{out}|S|\text{in}\rangle
=
\int D\mu [q(\cdot), p(\cdot)] \exp\left\{ \frac{i}{\hbar} \int_{-\infty}^{\infty}
\left( \sum_i p_i \dot{q}^i - H \right) dt \right\}
\end{equation}
where the measure is defined as
\begin{equation}
D\mu [q(\cdot), p(\cdot)] = \Delta(\chi_a) \Delta(\phi^a) \, P[\det \left| \{\chi_a,\phi_b\} \right|] \, Dq(\cdot) Dp(\cdot).
\end{equation}
Here, we defined the path delta function $\Delta(f) = \prod_t \delta[f(p(t), q(t)]$, and the path product P[f] = $\prod_t f[(p(t), q(t)]$.
The delta functions impose the conditions $\varphi^a=0$ and $\chi_a=0$, while the determinant factor ensures that each physically distinct configuration is counted once. 

This determinant is the \emph{Faddeev--Popov determinant}.
The role of this determinant can be understood by rewriting the measure in adapted coordinates $(\bar{p}, \bar{q}, p^*, q^*)$, where  $\bar{p}_a = \chi_a$ and  $\{\chi_a, \phi_b\} = -\partial \phi_b/\partial \bar{q}_a$. It follows that
\bey
\Delta(\chi_a) \Delta(\phi^a) P[\det \left| \{\chi_a,\phi_b\}\right|] = \Delta(\bar{p}_a) \Delta[\bar{q}_a - \bar{q}_a(p^*, q^*)]. \nonumber
\eey
Hence, after integrating out the redundant variables, the path integral involves only  degrees of freedom in $\Gamma_{red}$
\begin{equation}
\int \exp\left\{ \frac{i}{\hbar} \int
\left( \sum_i p_i^* \dot{q}^{*i} - H^* \right) dt \right\}Dq^*(\cdot) Dp^*(\cdot).
\end{equation}

\noindent {\em Ghost fields and local formulation.}
The Faddeev--Popov determinant can be represented as a functional integral over auxiliary anticommuting (Grassmann) fields $c^a$ and $\bar{c}_a$, known as \emph{ghost fields},
\begin{equation}
\det \left| \{\chi_a,\varphi^b\} \right|
=
\int \mathcal{D}\bar{c}\,\mathcal{D}c\;
\exp\left\{ i \int \bar{c}_a \{\chi_a,\varphi^b\} c_b \, dt \right\}.
\end{equation}
This representation replaces the nonlocal determinant by a local contribution to the action. As a result, the full path integral can be written with an effective action involving the original variables together with the ghost fields.

This formulation is particularly useful in perturbation theory, since it leads to a local Lagrangian and hence to standard Feynman rules. The ghost fields contribute additional propagators and interaction vertices, ensuring that the perturbative expansion correctly accounts for the structure of the measure.

 \medskip
 
The Faddeev--Popov construction can be systematically generalized by the Batalin--Vilkovisky (BV) formalism \cite{BV1, BV2}, which extends the field space to include ghosts, antifields, and higher-order ghost structures, and encodes the algebraic structure of the theory in an extended action $S_{\text{BV}}$.
Gauge fixing is implemented through a gauge fermion, yielding a consistent path-integral formulation even when the algebra of constraints is   field-dependent---as is the case in GR. The BV framework provides a unifying and conceptually complete extension of the Faddeev--Popov method and underlies modern perturbative treatments.

However, both the Faddeev--Popov and BV constructions rely on the assumption that the gauge fixing $\chi_a$ intersects each orbit uniquely. In general, this assumption fails, because of the  Gribov ambiguity. The functional integral may therefore overcount configurations. This limitation reflects a global obstruction in the definition of the measure and indicates that the quantisation procedure is not completely under control beyond perturbation theory.
 
 By contrast, Dirac quantisation does not rely on such auxiliary conditions and therefore is not affected by this type of ambiguity. Its difficulties are of a different nature, arising instead from the definition of the physical Hilbert space and the implementation of the constraint equations. Reduced state space quantisation also does not suffer from the Gribov problem, as long as the reduced state space is identified globally and not via gauge fixing. 
 
\section{Approaches to quantum gravity}
There exists a wide range of approaches to quantum gravity. Some have generated tens of thousands of papers, while others are represented by only a handful of works. Remarkably, after more than fifty years of sustained effort, none of these programs has acquired clear epistemic priority, largely due to the absence of decisive experimental input or explanatory success. This situation is likely to appear striking to future historians of science.

There is a well-known Indian parable in which several blind men attempt to describe an elephant after each has touched only one part of it. One concludes that the elephant is like a snake, another like a pillar, another like a fan. The story suggests that apparently conflicting descriptions may ultimately correspond to different aspects of a single underlying reality. Quantum gravity research   {\em is almost certainly not}  of this kind. There is no compelling reason to expect that the existing approaches are merely partial views of a deeper unified framework, and more than fifty years of work have produced primarily divergence rather than convergence. A better analogy is with natural philosophy in Newton’s time, when Aristotelian element theory, Paracelsian and other alchemical theories, corpuscular accounts of matter, and competing ether hypotheses coexisted without an agreed conceptual framework.  These approaches were not complementary descriptions of a deeper structure, but largely incompatible attempts to understand the constitution of matter.

We classify approaches to quantum gravity into four broad types, distinguished by their fundamental assumptions. Such a classification is necessarily approximate and admits borderline cases, but it serves to organize the conceptual landscape of the subject. The four types are:
\begin{description} 
\item[Type I:] Spin-2 approaches,
\item[Type II:] Geometry quantisation,
\item[Type III:] Non-quantum gravity,
\item[Type IV:] Structural reformulations.
\end{description}

\subsection{Type I approaches: spin-2, QFT, and the S-matrix}

Type I strategies take as their starting point the identification of gravity with a \emph{massless spin-2 field} propagating on a fixed background spacetime, as suggested by the Pauli–Fierz construction. At the linearized level, the Pauli–Fierz field exhibits a gauge symmetry corresponding to linearized diffeomorphisms. The requirement that this symmetry be preserved in the interacting theory plays a central role in determining its structure. The guiding idea—going back to consistency arguments of Steven Weinberg in the 1960s \cite{Wein1, Wein2}—is that a Lorentz-invariant interacting massless spin-2 field must couple universally to a conserved stress tensor, and that consistency uniquely fixes its nonlinear self-interactions. This universality of coupling ensures that the field interacts with all forms of energy and momentum, and leads to general relativity as the resulting theory. In this sense, GR is not taken as fundamental, but as the \emph{infrared completion} of a spin-2 quantum field theory.

One expands the metric around a background (typically Minkowski),
\begin{equation}
g_{\mu\nu}=\eta_{\mu\nu}+\kappa\, \gamma_{\mu\nu}, \nonumber 
\end{equation}
and quantizes $\gamma_{\mu\nu}$ as a field on this fixed spacetime. The full classical dynamics is encoded in the Einstein–Hilbert action, while higher-order terms in the field define the interaction Lagrangian $\mathcal{L}_{\text{I}}[\gamma_{\mu\nu}]$, a local function of the field that generates, upon expansion, an infinite tower of interaction vertices for the graviton. This construction is inherently {\em background-dependent}, as it assumes a fixed spacetime geometry around which the field is expanded.

The interaction Lagrangian enables us to construct the $S$-matrix. Physically, the $S$-matrix formulation bypasses the need for a detailed description of local time evolution by focusing instead on \emph{asymptotic observables}, defined in terms of free particle states at past and future infinity. This makes it particularly well suited to perturbative quantum field theory. At the same time, it reveals an intrinsic limitation: the construction presupposes a well-defined notion of asymptotic time, and is therefore difficult to extend to cosmological settings or strongly curved spacetimes where such a notion is absent.

Power counting shows that Newton’s constant has negative mass dimension in $d=4$,
$$
[G] = -2,
$$
with the consequence that loop corrections generate an infinite tower of higher-curvature counterterms. The resulting effective action takes the form
\begin{equation}
\Gamma_{\text{eff}}[h_{\mu\nu}] = \int d^4x\,\sqrt{-g}\left[
\frac{1}{16\pi G}R
+ c_1 R^2 + c_2 R_{\mu\nu}R^{\mu\nu}
+ c_3 R_{\mu\nu\rho\sigma}R^{\mu\nu\rho\sigma}
+ \cdots
\right],
\end{equation}
illustrating that the theory is non-renormalizable in the traditional sense, but admits a well-defined low-energy expansion as an effective field theory.

Several directions attempt to complete or improve the spin-2 framework. These approaches retain the basic structure of a spin-2 description of gravity, while modifying its ultraviolet behavior through additional symmetry, extended structure, or nonperturbative dynamics:

\begin{itemize}

\item \textbf{Supergravity:} 
Supersymmetry relates bosonic and fermionic degrees of freedom and produces cancellations among their loop contributions, thereby improving the ultraviolet behavior of the theory. In extended supergravity, these cancellations can postpone divergences to higher loop orders and may be stronger than those implied by standard symmetry arguments alone \cite{FrPr,vanNi}. Supergravity theories are nevertheless non-renormalizable by conventional power counting. Explicit divergences are known in several cases, but the ultraviolet behavior of highly extended supergravity theories—including whether and at which loop order divergences first occur in four dimensions—remains an open question\footnote{Extended supergravity refers to supergravity theories with more than one supersymmetry generator ($\mathcal{N}>1$). Increasing $\mathcal{N}$ enlarges the field content and imposes strong symmetry constraints on the interactions, leading to improved ultraviolet behavior through cancellations between bosonic and fermionic degrees of freedom.}.

\item \textbf{Superstrings:} 
In string theory \cite{GSW,Pol}, gravity arises naturally at low energies from a massless spin-2 excitation of the closed string. The extended nature of strings softens short-distance interactions, and perturbative superstring amplitudes are generally understood to be ultraviolet finite at each order in the genus expansion. This has been established explicitly at low orders and is supported by general worldsheet arguments, although a completely general all-genus proof remains lacking
\footnote{The genus expansion organizes string perturbation theory according to the topology of the worldsheet. A connected closed oriented worldsheet of genus $g$, namely one with $g$ handles, carries the topological weight $g_s^{2g-2}$, where $g_s$ is the string coupling. The leading contribution, $g=0$, comes from the sphere and corresponds to tree level, while higher-genus surfaces describe loop corrections.   Thus, the perturbative expansion is organized by worldsheet topology rather than by individual Feynman diagrams in spacetime.} At energies well below the string scale, the interactions of the massless spin-2 mode are governed, to leading order, by the Einstein–Hilbert action, coupled in general to additional massless fields and supplemented by higher-derivative corrections.  Perturbative amplitudes may nevertheless exhibit infrared divergences, and a complete nonperturbative formulation applicable to general background spacetimes is still lacking.

\item \textbf{Asymptotic safety:} 
The asymptotic safety scenario proposes that gravity may remain well defined at arbitrarily high energies if its effective strength approaches a stable value. In four spacetime dimensions, Newton’s constant has mass dimension $[G]=-2$, so one introduces a dimensionless coupling
\begin{equation}
g(\mu) = G(\mu)\,\mu^2,
\end{equation}
where $\mu$ is an energy scale characterizing the resolution at which the theory is probed. The key idea is that, as $\mu$ increases, $g$ may approach a constant value rather than growing without bound \cite{Wein79, Percacci}. If this occurs, the theory can remain predictive at all scales, despite being non-renormalizable in the usual perturbative sense. Evidence for such behavior has been found in approximate calculations, although its validity is still under investigation\footnote{More precisely, the renormalization group flow of   $g(\mu)$---or any other dimensionless coupling---is governed by a beta function $\beta(g)=\mu\,\frac{dg}{d\mu}$. Asymptotic safety corresponds to the existence of a nontrivial ultraviolet fixed point $g_*$ satisfying $\beta(g_*)=0$. Predictivity requires that only a finite number of directions in coupling space are relevant at the fixed point, so that the high-energy behavior is controlled by a finite set of parameters.}.

\end{itemize}

It is worth keeping in mind  that every Type I construction is connected to observed gravity solely through the
spin-2 field. Supergravity and superstrings introduce large numbers of
new, as-yet-unobserved fields and particles, and asymptotic safety reorganizes the high-
energy behaviour; but their sole point of contact with the gravitational force we measure
is the massless spin-2 excitation. Without the interpretation that this excitation generates
gravity, these frameworks would say nothing about gravity at all.

\subsection{Type II approaches: quantisation of the geometry}

Type II approaches take a conceptually different starting point from Type I: rather than deriving gravity from a spin-2 field on a fixed background, they treat spacetime geometry itself—or, in canonical formulations, spatial geometry—as the fundamental variable to be quantized. The guiding principle is that general relativity, as a dynamical theory of geometry, should be quantized analogously to other classical field theories, while preserving (as far as possible) diffeomorphism invariance.

A central consequence is that time is no longer an external parameter but part of the dynamical structure. As a result, both observables and the notion of evolution become nontrivial.

Two broad directions can be distinguished, corresponding to different treatments of time and observables.

\subsubsection*{Direction I: Canonical quantisation on a hypersurface.}

Canonical approaches begin with a $3+1$ decomposition of spacetime into hypersurfaces $\Sigma_t$. The basic variables are fields on $\Sigma$, and dynamics is encoded in constraints rather than standard Hamiltonian evolution.

Two main lines arise, depending on the choice of variables.

\medskip

\noindent
\textbf{Geometrodynamics.}
In the metric formulation \cite{DeWitt, Kuchar, MTW}, the basic variables are the spatial metric $h_{ab}(x)$ and its conjugate momentum $\pi^{ab}(x)$,
\begin{equation}
[h_{ab}(x), \pi^{cd}(y)] = i \delta_{(a}^{\,c} \delta_{b)}^{\,d} \, \delta^{(3)}(x-y).
\end{equation}
However, these commutation relations are not fully natural, since they do not preserve the positivity of $h_{ab}$. This motivates alternative structures, such as affine commutation relations, where $\pi^{ab}$ is replaced by $\pi^a_{\ b} = \pi^{ac} h_{cb}$, yielding transformations compatible with the geometric nature of the configuration space.

\medskip

\noindent
\textbf{Connection dynamics (loop quantum gravity).}
An alternative formulation uses connection variables \cite{Ashtekar}. The canonical pair is an $\mathrm{SU}(2)$ connection $A_a^i$ and densitized triad $E^a_i$,
\begin{equation}
[A_a^i(x), E^b_j(y)] = i \delta_a^{\,b} \delta^i_{\ j} \, \delta^{(3)}(x-y).
\end{equation}
Loop quantum gravity \cite{Rovelli2, Thiemann} instead employs holonomies along curves and fluxes across surfaces. The holonomy is
\begin{equation}
h_\gamma[A] = \mathcal{P} \exp\left( \int_\gamma A \right),
\end{equation}
and the flux is
\begin{equation}
E(S,f) = \int_S E^a_i f^i n_a \, d^2\sigma.
\end{equation}
These generate the holonomy–flux algebra, replacing canonical commutation relations. States are cylindrical functions of holonomies, yielding a Hilbert space with discrete spectra for geometric operators such as area and volume.

\paragraph{Implementation of the constraints.}

Dynamics is encoded in the diffeomorphism and Hamiltonian constraints. Reduced state space quantisation is rarely feasible, as it requires full deparameterization; thus Dirac quantisation is typically employed.

In the metric formulation, one works on superspace $\text{Riem}(\Sigma)/\DS$. Wave functionals $\Psi[h_{ab}]$ automatically satisfy the momentum constraints, while the Hamiltonian constraint yields the Wheeler–DeWitt equation
\begin{equation}
\hat{\mathcal{H}}(x)\,\Psi[h_{ab}] = 0.
\end{equation}
Explicitly,
\begin{equation}
\left[
-2\kappa \, G_{abcd}(h)\,\frac{\delta^2}{\delta h_{ab}\delta h_{cd}}
+ \frac{\sqrt{h}}{2\kappa}\,R
\right]\Psi[h_{ab}] = 0, \label{wdw}
\end{equation}
where $G_{abcd}$ is the DeWitt supermetric. Its indefinite signature reflects the pseudo-Riemannian structure of superspace. The Wheeler–DeWitt equation is only formal: operator ambiguities, regularization issues, and the absence of a well-defined inner product remain unresolved.

\medskip

In loop quantum gravity, the constraints are expressed in holonomy–flux variables. The diffeomorphism and Gauss constraints can be implemented rigorously, leading to states labeled by spin networks, which form a discrete basis of quantum geometry. The Hamiltonian constraint is more subtle: a well-defined operator exists (e.g., Thiemann \cite{QSD}), but questions remain about its interpretation, algebra, and semiclassical limit.

\subsubsection*{Direction II: Path-integral and sum-over-histories approaches.}

An alternative is to define a sum over geometries,
\begin{equation}
G(h_f, \Sigma_f \,|\, h_i, \Sigma_i) = \int \mathcal{D}g_{\mu\nu}\, e^{i S_{EH}[g]},
\end{equation}
assigning amplitudes to spacetime histories interpolating between boundary data. This formulation is manifestly covariant and does not rely on a preferred time slicing; topology change may also be included.

Its interpretation depends on boundary conditions: with two boundaries it defines a transition amplitude, with one a wave functional, and without boundaries a partition function,
\begin{equation}
Z = \int \mathcal{D}g\, e^{i S_{EH}[g]}.
\end{equation}

Several realizations exist:

\begin{itemize}

\item \textbf{Euclidean quantum gravity:}
One considers the path integral in Euclidean time \cite{GiHa, Hawking79}.
\begin{equation}
Z = \int \mathcal{D}g\, e^{- S_E[g]},
\end{equation}
 While this improves convergence, the Euclidean action is unbounded below due to the conformal factor, rendering the integral ill-defined. Moreover, Wick rotation is not generally well defined, and the connection to Lorentzian physics remains unclear. The approach is mainly useful in semiclassical approximations.

\medskip

\item \textbf{Dynamical triangulations:}
The path integral is replaced by a sum over simplicial geometries,
\begin{equation}
Z = \sum_T e^{-S[T]},
\end{equation}
with curvature encoded via Regge calculus \cite{Regge}. In causal dynamical triangulations (CDT) \cite{AmbjornLoll, CDT}, only causally well-behaved triangulations are included, leading to improved behavior and evidence for emergent four-dimensional spacetime and a viable continuum limit.

\medskip

\item \textbf{Spin foams:}
Histories are combinatorial structures labeled by group data \cite{Baez, Perez},
\begin{equation}
Z = \sum_{\mathcal{F}} \prod_f A_f \prod_e A_e \prod_v A_v.
\end{equation}
These provide a covariant formulation of loop quantum gravity, with amplitudes implementing the constraints. In contrast to triangulations, spin foams sum over algebraic data encoding quantum geometry \cite{EPRL, FK}.

\end{itemize}

In all these approaches, the fundamental object is a \emph{sum over histories} rather than time evolution. Nevertheless, the framework remains that of standard quantum theory: the path integral defines transition amplitudes between boundary states and presupposes an underlying Hilbert space structure.

\subsection{Type III approaches: non-quantum gravity}
Type III approaches depart from the assumption that gravity must be quantized. Instead, they treat gravity as a fundamentally \emph{classical} field, coupled to quantum matter. In this framework, the gravitational field retains its classical character, while matter fields are described by quantum theory. This viewpoint is motivated by the possibility that gravity is either an emergent phenomenon or intrinsically non-quantum, and therefore does not admit a fundamental quantisation  as the  other interactions.

Two broad directions can be distinguished.
\subsubsection{Emergent gravity}
\begin{itemize}

\item \textbf{Induced gravity:}
Originally proposed by Sakharov, induced gravity treats the metric as a classical, non-dynamical background field in the matter action \cite{Sakharov}. One considers quantum matter fields $\phi$ on a prescribed geometry $g_{\mu\nu}$ and integrates them out,
\begin{equation}
e^{i \Gamma[g]} =
\int \mathcal{D}\phi \,
e^{i S_{ matter}[\phi,g]} .
\end{equation}
The resulting effective action contains metric terms, including an induced Einstein--Hilbert term,
\begin{equation}
\Gamma[g] =
\int d^4x\,\sqrt{-g}\left[
\Lambda_{*}
+\frac{1}{16\pi G_{*}}R
+\cdots
\right]
\end{equation}
where $\Lambda_*$ and $G_*$ are the induced cosmological and gravitational constant, respectively.
Gravitational dynamics is then obtained by extremizing $\Gamma[g]$, so that gravity emerges from quantum fluctuations of matter rather than being fundamental.

\item \textbf{Thermodynamic gravity:}
In thermodynamic approaches, due to Jacobson and Padmanabhan, the Einstein equations are interpreted as equations of state \cite{Jacobson, Padmanabhan}. The key inputs are horizon entropy,
\begin{equation}
S = \frac{A}{4G},
\end{equation}
and the Unruh relation between temperature and acceleration,
\begin{equation}
T = \frac{a}{2\pi}.
\end{equation}
Jacobson showed that applying the Clausius relation $\delta Q = T\,\delta S$ to local Rindler horizons yields the Einstein equations when entropy is proportional to area. Padmanabhan develops a related perspective based on horizon thermodynamics.

A related proposal is Verlinde’s entropic gravity \cite{Verlinde}, where gravity appears as an entropic force arising from information changes on holographic screens. In all these approaches, spacetime geometry and gravity are emergent, thermodynamic phenomena rather than fundamental fields.

\end{itemize}

\subsubsection*{Fundamentally classical gravity.}

A more radical possibility is that gravity is fundamentally classical, even at the microscopic level, and that its coupling to quantum matter requires a modification of standard quantum theory. The simplest realization is the M{\o}ller--Rosenfeld semiclassical theory \cite{Moller,Rosenfeld}, in which the metric remains classical and is sourced by the expectation value of the stress tensor,
\begin{equation}
G_{\mu\nu}[g] = 8\pi G\, \langle \hat T_{\mu\nu} \rangle .
\end{equation}
In the Newtonian limit this leads to Schr\"odinger--Newton equations, where the gravitational potential is sourced by the mass density $|\psi|^2$. These equations are nonlinear and raise difficulties with statistical interpretation, measurement, and signalling, and are therefore unlikely to define a consistent fundamental theory. Nonetheless, they remain paradigmatic as the simplest realization of a classical gravitational field coupled to quantum matter.

More general hybrid theories seek to couple classical and quantum degrees of freedom within a single framework \cite{Aleksandrov1981,BlanchardJadczyk1993,BlanchardJadczyk1995}. A central lesson is that purely deterministic couplings are difficult to maintain: consistency conditions such as positivity and absence of signalling typically require stochastic dynamics. This motivates models in which gravity induces decoherence in quantum matter. Di\'osi’s proposal \cite{Diosi1984,DiosiTilloy} is a key example, where the gravitational field effectively monitors mass density and suppresses spatial superpositions.

A recent realization is Oppenheim’s post-quantum theory of classical gravity \cite{Oppenheim}, which can be viewed as a modern hybrid framework adapted to geometrodynamical variables. Here spacetime remains classical, but its coupling to quantum matter is described by a stochastic, completely positive dynamics. While this avoids the inconsistencies of semiclassical models, extending such hybrid dynamics to fully relativistic settings remains challenging, as one must reconcile stochastic classical evolution with locality, covariance, conservation laws, and the probabilistic structure of quantum theory.

\subsection{Type IV approaches: modification of geometry and quantum structure}
Type IV approaches encompass a heterogeneous set of frameworks that go beyond both quantisation of gravity and classical–quantum hybrids, and instead propose a more radical modification of the underlying structures of physics. In these approaches, the standard notions of spacetime, geometry, or even quantum theory itself are replaced or generalized. Prominent examples include:

\begin{itemize}

\item \textbf{Noncommutative geometry:}
In noncommutative geometry, due to Connes, spacetime is described not by a manifold but by an algebra of noncommuting operators \cite{Connes}. The fundamental structure is a \emph{spectral triple} $(\mathcal{A},\mathcal{H},D)$, consisting of an algebra $\mathcal{A}$, a Hilbert space $\mathcal{H}$, and a Dirac operator $D$, from which geometric information (such as distance and curvature) can be reconstructed. The gravitational action emerges from spectral data via the spectral action principle. However, there is no clear notion of quantisation in this framework: geometry is reformulated rather than quantized.\footnote{Noncommutative geometry replaces the algebra of functions on a space by a noncommutative algebra, motivated by the idea that a space can be reconstructed from its algebra of functions. In the commutative case, this recovers ordinary geometry; in the noncommutative case, it leads to generalized ``quantum'' spaces without underlying points.}

\item \textbf{Twistor theory:}
In twistor theory, introduced by Penrose, the basic objects are elements of a complex vector space (twistor space), from which spacetime geometry is reconstructed \cite{PenroseTwistor}. This framework is particularly well adapted to the description of massless fields and scattering amplitudes, where it provides remarkable simplifications. However, the relation to quantum gravity remains indirect. 

\item \textbf{Histories-based approaches:}
In histories formulations, the primary objects are histories rather than instantaneous states. The consistent histories approach of Gell-Mann and Hartle defines a generalized quantum mechanics based on decoherence functionals \cite{Hartle}. Isham developed the History Projection Operator (HPO) formalism, giving a logically structured framework for histories in quantum theory \cite{Ish94}, and even generalizations that do not employ a Hilbert space structure. Savvidou further extended this program by introducing a two-time structure and a canonical history quantisation scheme, in which histories carry both kinematical and dynamical time transformations \cite{Sav10}. These developments build progressively toward a formulation of quantum theory compatible with generally covariant systems.

\item \textbf{Causal set theory:}
In causal set theory, pioneered primarily by Sorkin \cite{BLMS87, Sorkin05}, spacetime is fundamentally discrete and is modeled as a partially ordered set, where the order relation encodes causal structure. The continuum spacetime geometry is expected to emerge as an approximation to this underlying structure. The emphasis on causal order places causal set theory close to histories-based approaches: the fundamental object is a spacetime history endowed with a discrete causal structure rather than a spatial configuration evolving in time. Quantisation is supposed to proceed by the definition of a decoherence functional on causal set configurations.

\end{itemize}
 
 \section{The treatment of time in approaches to quantum gravity}

An alternative way to classify approaches to quantum gravity is by how they treat time. Three main attitudes can be distinguished:

\begin{itemize}

\item \textbf{Attitude A:}
Time is \emph{assumed} to behave as in ordinary physics, providing a background parameter with respect to which evolution and causality are defined. Conceptual difficulties are set aside at the outset.

\item \textbf{Attitude B:}
Time becomes a \emph{burden}: since it is part of the dynamical geometry, its absence as an external parameter leads to difficulties in defining observables, causality, and evolution.

\item \textbf{Attitude C:}
Time is taken as \emph{central}: it is built into the fundamental variables—through histories, causal order, or related structures—rather than emerging from dynamics.

\end{itemize}

These attitudes reflect not only different views about time, but also different choices about what is most important in constructing a theory of quantum gravity.

The four types discussed above can be broadly aligned with them. Type I and Type III approaches largely fall under Attitude A, as they assume a standard notion of time. Type II approaches correspond primarily to Attitude B, where the absence of time gives rise to the {\em problem of time}. Type IV approaches---excepting non-commutative geometry---are closest to Attitude C, placing time or causal structure at the center of the framework.

It is widely expected that a theory of quantum gravity will require a revision of our usual concept of time. This implies that even approaches falling under Attitude A must eventually confront this issue. The gap between their current formulations and a satisfactory account of time is therefore significant.

In what follows, we examine how this tension manifests itself in each of the approaches discussed above. While the underlying goal is shared, the specific choices made in each research program lead to distinct conceptual and technical difficulties, many of which can be traced to how time is treated.

\subsection{Problems of Type I approaches with respect to time}

Despite their successes, Type I approaches face a basic difficulty:
the temporal and causal structures employed in their quantisation are
those of the background spacetime, not those of the geometry that is
supposed to emerge from the quantum spin-2 field. This difficulty takes
several related forms.

\smallskip

 \noindent \textbf{Signature and causal structure.}
In the spin-2 framework, the fundamental quantum variable is a symmetric
tensor field $\hat{\gamma}_{\mu\nu}$ propagating on a fixed Lorentzian
background. It is not a metric constrained to take values in the space
of nondegenerate Lorentzian metrics. Consequently, its kinematical
quantisation does not impose the nonlinear, pointwise requirement that
\[
g_{\mu\nu}=\eta_{\mu\nu}+\kappa\gamma_{\mu\nu}
\]
remain nondegenerate and Lorentzian.

Indeed, if $\hat{\gamma}_{\mu\nu}$ satisfies the canonical commutation
relations of a free spin-2 field, every nonzero, noncentral physical
linear observable
\[
\hat{\gamma}(f):=\int d^4x\,
\hat{\gamma}_{\mu\nu}(x)f^{\mu\nu}(x)
\]
has spectrum equal to the full real line.%
\footnote{On Minkowski spacetime, the condition
$\partial_\mu f^{\mu\nu}=0$ makes $\hat{\gamma}(f)$ invariant under the
linearized gauge transformation
$\gamma_{\mu\nu}\mapsto\gamma_{\mu\nu}
+\partial_\mu\xi_\nu+\partial_\nu\xi_\mu$, assuming suitable boundary
conditions. The statement applies after quotienting out smearings that
vanish by the field equations or lie in the kernel of the physical
symplectic form.}
The quantum field algebra therefore contains no intrinsic restriction
corresponding to fixed Lorentzian signature. Outside the perturbative
regime, there is no general reason for an effective tensor such as
\[
g^{\mathrm{eff}}_{\mu\nu}
=\eta_{\mu\nu}
+\kappa\langle\psi|
\hat{\gamma}_{\mu\nu}|\psi\rangle
\]
to remain nondegenerate and Lorentzian. Since the expectation value is
distributional, this statement must in practice be formulated in terms
of suitable averaged or renormalized quantities.

This causes no immediate difficulty in perturbation theory, where the
field is assumed to remain a small fluctuation around the background
metric. Beyond that regime, however, the signature and light-cone
structure of the putative physical metric are not secured by the
kinematics. They must emerge dynamically from the quantum spin-2 field,
yet no existing spin-2 field formulation provides a concrete mechanism
that guarantees this emergence or ensures that the resulting geometry
remains everywhere nondegenerate and Lorentzian.

\smallskip

 \noindent \textbf{Microcausality and background dependence.}
In quantum field theory, microcausality is defined with respect to the
causal structure of a fixed background metric. For a spin-2 field on
Minkowski spacetime, one writes
\begin{equation}
[\hat{\gamma}_{\mu\nu}(x),
 \hat{\gamma}_{\rho\sigma}(y)]=0
\qquad\text{if}\qquad
\eta_{\alpha\beta}(x-y)^\alpha(x-y)^\beta>0,
\end{equation}
subject to the usual qualifications concerning gauge-dependent fields.
More precisely, causal commutativity applies to gauge-invariant local
observables or suitably smeared physical fields. It follows for the free
spin-2 field and is preserved order by order in perturbative
constructions. The decisive point, however, is that spacelike separation
is defined by the background metric.

The formalism supplies no analogous intrinsic condition for the
putative full metric operator $\hat{g}_{\mu\nu}
=\eta_{\mu\nu}\hat{I}
+\kappa\hat{\gamma}_{\mu\nu}$. Microcausality cannot
simply be reformulated using $\hat{g}_{\mu\nu}$, because the condition
determining whether a commutator should vanish would itself be
operator-valued.

  Standard spin-2 field theory
provides no background-independent replacement for microcausality that
is derived from, and guaranteed to agree with, the emergent geometry.
This is a central tension in interpreting a quantum spin-2 field as a
quantum theory of spacetime geometry.

\smallskip

 \noindent \textbf{Asymptotic time and the $S$-matrix.}
An asymptotic formulation bypasses some of these difficulties without
resolving them. In an asymptotically flat spacetime, the $S$-matrix
relates states prepared at past infinity to states registered at future
infinity. Its notions of time, particles, and causal propagation are
defined relative to the asymptotic geometry. It can therefore describe
scattering without supplying an intrinsic notion of time or causal
localization in the intervening dynamical geometry.

This is sufficient for asymptotic transition probabilities, but it does
not by itself assign probabilities to events occurring at finite times
and finite locations. Nor does it apply to spacetimes without the
required asymptotic regions. The $S$-matrix thus removes finite-time
causal structure from the description rather than explaining how it
emerges.

The situation is different in asymptotically AdS spacetimes, where an
ordinary $S$-matrix is generally unavailable. The timelike conformal
boundary carries a fixed causal structure and provides a boundary time
with respect to which boundary correlation functions and transition
amplitudes may be defined. In holographic formulations, this boundary
time is identified with the time parameter of the boundary quantum
theory. Nevertheless, it remains an asymptotic notion: the reconstruction
of spacetime localization and causal relations from boundary data is
indirect, may be state-dependent, and is generally model-dependent.
The existence of boundary time therefore does not by itself supply an
intrinsic notion of physical time or causality in a fluctuating spacetime
geometry.

\subsection{Problems of Type II approaches with respect to time}

In Type II approaches, time is not supplied by an external background
but must be reconstructed from the theory itself. In canonical quantum
gravity, this gives rise to the well-known \emph{problem of time}. A
complete deparametrization would resolve its central dynamical aspect:
one would identify a globally valid internal clock and express the
remaining degrees of freedom as evolving with respect to it; see
Sec.~6.1. However, such constructions are possible only in restricted models.
No global
deparametrization of this kind is known for general relativity, and there are strong structural reasons to doubt that one exists without introducing special matter or restricting the solution space \cite{Kuchar,Torre}.

Canonical approaches must therefore recover time relationally,
semiclassically, or within a limited region of phase space. A particular
degree of freedom may be selected as a clock, or an approximate time
parameter may emerge along a semiclassical gravitational trajectory.
Neither construction is unique. Different choices of clock can lead,
after quantisation, to inequivalent notions of evolution and even to
inequivalent quantum theories. This is the \emph{multiple-choice
problem}: the formalism does not in general determine which internal
variable, if any, represents physical time.

The problem is not merely the absence of a preferred clock. In the canonical theory, time evolution is apparently lost after the imposition of the constraints. 
Consequently, one must explain not only how an effective clock emerges,
but also how the spacetime distinction between past and future and the
causal ordering of events are recovered. An approximate clock does not,
by itself, provide this reconstruction.

These questions arise independently of the technical difficulty of
defining the Hamiltonian constraint and constructing the physical Hilbert
space. Even if both problems were solved, one would still have to identify
physical temporal observables, establish their relations under different
choices of clock, and recover the causal structure of general relativity
in an appropriate semiclassical limit.

In path-integral or sum-over-histories approaches, the problem of time is
less explicit but remains present. These approaches define amplitudes
between boundary data by summing over possible geometries. Each geometry
has its own causal structure, but different geometries need not agree on
the temporal ordering of intermediate events. There is therefore no
single causal order for the full quantum process. Moreover, amplitudes
between boundaries do not by themselves give probabilities for
intermediate events. It remains necessary to explain how temporal order,
physical time, and probabilities for spacetime events emerge from the
sum over geometries.

\subsection{Problems of Type III approaches with respect to time}

Type III approaches attempt to retain a classical notion of spacetime while coupling it to quantum matter or deriving it as an emergent phenomenon. However, this strategy inherits significant difficulties concerning the role of time.

In induced gravity, the situation is closely analogous to Type I approaches. The formulation presupposes a background spacetime in order to define quantum fields, in particular through conditions such as microcausality and the existence of a well-defined temporal structure. These notions depend on a background metric. However, the same metric is then expected to emerge dynamically from the effective action obtained by integrating out matter fields. This raises a conceptual tension: the metric appears both as an input required to define the theory and as an output to be derived from it. A sharper   formulation is required to avoid   circularity.

More generally, emergent gravity scenarios are expected to lead to hybrid classical–quantum systems, in which a classical spacetime geometry interacts with quantum matter. Even if the metric is not quantized, it will typically exhibit fluctuations induced by the quantum state of matter. 
The problem of time remains, because, as we explained in Section 4.1, it originates from the breakdown of determinism, not necessarily from quantum theory.

There are also problems specific to hybrid and emergent theories. In
general relativity, spacetime dynamics is encoded in the constraints.
Their algebra represents deformations of spatial hypersurfaces and is
essential for ensuring that evolution defines a consistent spacetime
rather than depending on an arbitrary choice of foliation. The inclusion
of matter changes the form of the constraints but not this structural
role: the total gravitational and matter constraints must satisfy the
same hypersurface-deformation algebra. A hybrid theory must recover this
structure, or provide a consistent replacement for it, despite treating
matter and gravity as fundamentally different types of system.\footnote{The case of postquantum classical
gravity is illustrative. Oppenheim and Weller-Davies constructed hybrid generalizations
of the Hamiltonian and momentum constraints, but found that, for a broad
class of realizations, their algebra does not close without additional
constraints \cite{OWD22}. Subsequent covariant
path-integral formulations impose diffeomorphism invariance directly
\cite{OWD23}; for full general relativity, however,
there remains a tension between complete positivity and the
implementation of the complete gravitational dynamics.}.

\subsection{Problems of Type IV approaches with respect to time}

Type IV approaches place time at the center of the theoretical framework, often by building it directly into the fundamental structures. While this avoids some of the difficulties encountered in other approaches, it introduces challenges of a different kind.

In noncommutative geometry, time is not fundamental in the usual sense, but must emerge from the underlying algebraic structure. In this respect, the situation is similar to Type II approaches: time is not given a priori and must be recovered from more basic ingredients. An additional problem is that  the formulation of Lorentzian noncommutative
geometry is considerably less developed than its Riemannian counterpart.

Other programs incorporate particular aspects of temporal or causal
structure more directly. Histories approaches take temporally ordered
histories as fundamental; causal set theory takes causal
order as a primitive element; and twistor theory encodes the structure of the light-cone in its
basic geometry. These approaches avoid both the treatment of time as an internal clock variable and the reliance on a background spacetime structure. They must still derive the more familiar aspects of time:  durations, clocks, localized events,  and explain the emergence of 
the effective Lorentzian geometry of general relativity. The form taken
by this problem differs substantially between the programs. I will discuss examples in Chapter~6.

\bigskip

Taken together, these considerations show that the problem of time is not confined to a particular class of approaches, but permeates all attempts at quantum gravity. The differences between the four types lie not in whether time poses a difficulty, but in how and when this difficulty is confronted. Approaches that assume time gain technical control but defer the problem; those that treat it as a problem confront deep conceptual obstacles; and those that place it at the center must reconstruct familiar notions of dynamics and spacetime from more primitive structures. In all cases, the challenge is to recover a notion of time that is both conceptually coherent and empirically adequate.

\chapter{Notable ideas and results about time in quantum gravity}

\section{Deparametrization and its limitations}

We have already discussed deparametrization in general relativity and the difficulties associated with it. The basic idea is to identify a set of internal variables that can serve as time and spatial coordinates, so that the dynamics can be rewritten in terms of evolution with respect to these variables. If such a construction were available in full generality, it would effectively eliminate the problem of time.

Historically, this idea is closely tied to the canonical formulation of general relativity. Beginning in the late 1950s and early 1960s, with the work of Dirac \cite{Dirac64} and of Arnowitt, Deser, and Misner \cite{ADM1}, the canonical structure of general relativity was often viewed as that of a \emph{parameterized field theory}. In such theories, one enlarges the phase space by introducing embedding variables—spacelike hypersurfaces in spacetime—and their conjugate momenta, thereby rendering the theory generally covariant. Kucha\v{r} later developed this perspective in detail \cite{Kuchar2}, suggesting that these embedding variables are implicitly present in geometrodynamics, disguised among the canonical variables. From this viewpoint, the Hamiltonian and momentum constraints acquire a natural interpretation: they generate deformations of the hypersurfaces and encode the energy and momentum densities of the true dynamical degrees of freedom.

This paradigm was particularly attractive because it provides a clean conceptual separation between gauge and physical degrees of freedom. Classically, it simplifies the Cauchy problem by identifying freely specifiable data, while at the quantum level it appears to offer a resolution of the problem of time. If the embedding variables can be identified, they supply an internal notion of “rods and clocks,” allowing one to formulate the theory in terms of a functional Schr\"odinger equation.

If deparametrization were possible, the constraints of general relativity would be recast as functional Schr\"odinger equations involving a \emph{many-fingered time} variable,
\begin{equation}
i \frac{\delta}{\delta T(x)} \, \Psi[\phi, T] = \hat{h}_0(x)\, \Psi[\phi, T].
\end{equation}
Here $T(x)$ assigns an independent time parameter to each spatial point, reflecting the fact that evolution in general relativity corresponds to local deformations of hypersurfaces rather than evolution with respect to a single global time.

In this formulation, $\Psi[\phi, T]$ represents the probability amplitude for the physical degrees of freedom $\phi$ on a hypersurface specified by the embedding variables $T(x)$, while $\hat{h}_0(x)$---the quantum analogue of $h_0$ in Eq. (\ref{depar}) generates local deformations of that hypersurface.

A reduction to a single global time parameter,
\begin{equation}
i \frac{\partial}{\partial T} \Psi[\phi, T] = \hat{H}\, \Psi[\phi, T],
\end{equation}
would require the existence of a preferred foliation and constitutes a much stronger condition. Such a reduction is not implied by deparametrization itself and is only available in highly special cases.

However, this picture relies crucially on the possibility of deparameterizing the theory, and this possibility is highly nontrivial. First, even if deparametrization were possible, there is no guarantee of uniqueness: different choices of internal time generally lead to inequivalent descriptions. This ambiguity already appears in simple finite-dimensional systems and becomes much more severe in field theory.

More fundamentally, there are strong indications that general relativity does not admit a global deparametrization. A key argument, due to Torre \cite{Torre}, is that the constraint surface of general relativity does not have the structure required for a parameterized field theory. In particular, the presence of symmetric solutions (spacetime geometries possessing Killing vectors) leads to degeneracies that prevent the constraint surface from being a smooth manifold. By contrast, in parameterized field theories the constraint surface is a regular manifold, allowing a clean separation between embedding variables and true degrees of freedom. This structural difference suggests that general relativity cannot, in general, be cast into the form of a parameterized theory.

This argument is not entirely definitive, since the problematic configurations form a set of measure zero. Nevertheless, it highlights an important structural difference: the constraint surface of general relativity differs from that of parametrized field theories. While this suggests that deparameterization is not a generic feature of GR,  it is not  a proof that deparametrization is impossible. Indeed, deparametrization can be achieved in special cases—for example, when gravity is coupled to dust—and one cannot exclude the possibility that particular matter couplings or modest extensions of general relativity may render it viable more generally.

\section{The Wheeler--DeWitt equation}
 
 Next, we will analyze the Wheeler-DeWitt (WDW) equation  (\ref{wdw}),
 \bey
 \hat{\HH}(x) \Psi(h) =  
\left[
-2\kappa \, G_{abcd}(h)\,\frac{\delta^2}{\delta h_{ab}\delta h_{cd}}
+ \frac{\sqrt{h}}{2\kappa}\,R
\right]\Psi[h] = 0, \nonumber
\eey
obtained from Dirac quantisation in geometrodynamics. 
 Its definition involves a simple, if arbitrary, choice of ordering in which all momentum variables are placed to the right of the metric variables. Furthermore, a rigorous definition would involve proper regularization, as it involves  products of functional derivatives evaluated at the same spatial point, which generate  divergences.

A major difficulty with the WDW is the lack of an appropriate inner product, leading to a lack of a direct probability interpretation. One might attempt to consider a naive inner product 
\begin{equation}
\langle \Psi|\Phi\rangle 
=
\int {\cal D}h\,\Psi^*[h]\Phi[h],
\end{equation}
but this is not satisfactory, because it is defined on arbitrary functionals of the three-metric, not specifically on solutions of the constraints.  
With this inner produce, one would interpret $|\Psi[h]|^2$ as a probability density on superspace. This requires that the metric $h_{ij}$ defines an observable acting multiplicatively on $\Psi[h]$. However, the metric does not commute with the constraint: so if $\Psi[h]$ is a solution to the WDW equation, $h_{ij}(x)\Psi[h]$ is not; the metric does not define a physical observable to which a probability $|\Psi[h]|^2$ can be assigned.

 Writing $h_{ab}(x)$ as $q^A$ and the De Witt supermetric   $G_{abcd}(h)$ as $G^{AB}(q)$, we bring  the WDW equation into the form  
 \bey
\left[ -\frac{1}{2\mu} G^{AB}\frac{\partial^2}{\partial q^A \partial q^B} + \mu V(q)\right]\Psi(q),\label{WDE}
 \eey
where we set $\mu = (4\kappa)^{-1}$ and $V = 2\sqrt{h}R$. The DeWitt metric has indefinite signature, so Eq. \eqref{WDE} resembles a functional Klein--Gordon equation on superspace rather than on spacetime.

\subsection{Inner product for solutions to Klein-Gordon equation}
Consider the Klein-Gordon (KG) equation on a spacetime $(\M, g)$, 
\bey
g^{\mu \nu}\nabla_{\mu}\nabla{\nu}\Psi(x) - V(x)\Psi(x) = 0.
\eey
The KG equation can be viewed as the Dirac quantisation condition   for a  particle on $(\M, g)$, described by the symplectic form $\omega = dp_{\mu}\wedge dx^{\mu}$ and 
 subject to the constraint $\HH = g^{\mu \nu} p_{\mu}p_{\nu} + V(x) = 0$.

Let $\Phi$ and $\Psi$ be a pair of solutions. For any Cauchy surface $\Sigma$ with normal $n_{\mu}$, the quantity 
\bey
(\Phi, \Psi) = \frac{1}{2i}\int_{\Sigma}\sqrt{h}d^3x  n^{\mu} (\Psi^*\nabla_\mu\Phi - \Phi\nabla_\mu \Psi^*), \label{indsi}
\eey
 is   independent of $\Sigma$. However, $(\Phi, \Psi)$ does not   define an inner product: $(\Psi, \Psi)$ vanishes for all real $\Psi$, and it may even take negative values.

The problem can be resolved, if the spacetime admits an everywhere timelike Killing vector field $t^{\mu}$---i.e., a timelike vector field that satisfies $\nabla_{\mu}t_{\nu}+\nabla_{\nu}t_{\mu} = 0$. This allows us to define an energy function $E = - t^{\mu}p_{\mu}$ classical. It is straightforward to show that
\bey
 \{g^{\mu \nu} p_{\mu}p_{\nu}, t^{\rho}p_{\rho}\} = 0. \label{enee}
\eey
If $t^{\mu}$ also leaves the potential invariant, $t^{\mu}\partial_{\mu}V = 0$, then $E$ is an observable: $\{E, \HH \} = 0$. 

 \begin{exercise}
Prove that the product $(\Phi, \Psi)$ is independent of the choice of $\Sigma$. Prove Eq. (\ref{enee}). 
 \end{exercise}

The quantum version of the energy function is an operator $\hat{E} =  i t^{\mu}\partial_{\mu}$ that commutes with the constraint $[\hat{E}, \hat{\HH}] = 0$. Then, we can find simultaneous eigenfunctions for $\hat{E}$ and $\hat{\HH}$.

Let $t$ be the time parameter   associated to the Killing vector $t^{\mu}$, so that $t^{\mu} \partial_{\mu}t = 1$. Assuming, for simplicity, that the spacetime is static, 
we can choose the surface $\Sigma$ in  Eq. (\ref{indsi}) to be defined by the requirement that $t$ is constant.  Then, $t^{\mu} = N n^{\mu}$ in terms of the lapse function $N > 0$.

For any eigenvector $\Psi$ of $\hat{E}$ with eigenvalue $E$, we find that 
\bey
(\Phi, \Phi) = \int d^3x \frac{\sqrt{h}}{N}E |\Phi|^2,
\eey
hence the inner product is well defined for $E > 0$. 
  
Using the $t$ coordinate, we write $\hat{E} = i \partial_t$ and the KG equation as $(\partial_t^2 - \hat{A})\Phi = 0$,
where 
\bey 
\hat{A} = -\frac{N}{\sqrt h}\partial_i
\left(
N\sqrt h\,h^{ij}\partial_j
\right)
+
N^2V(x).
 \eey
The positive-energy condition requires that $\hat{A}$ is positive, so that $\hat{A}^{1/2}$ exists. Then, we can write   Schr\"odinger equation $i  \partial_t \Psi = \hat{A}^{1/2} \Psi$ with positive Hamiltonian $\hat{A}^{1/2}$.  
Since the Laplacian term defines a positive operator, a sufficient condition for this is that $V(x)$ is positive. In contrast,
if $V(x)$ becomes sufficiently negative, $\hat{A}$   can acquire negative spectrum.
In that case there is no positive-frequency decomposition for the full solution space, and the KG product cannot be made positive definite in the above way.

\bigskip

The crucial question is whether we can apply the same reasoning as in the KG equation, to define an inner product for its ``positive energy" solutions. This would require finding  a Killing vector field on superspace, or equivalently, a functional $X_{ab}[h,x)$ such that the classical function
\bey
E = -i \int d^3x X_{ab}[h,x) \pi^{ab}(x),
\eey
has vanishing Poisson bracket with the Hamiltonian constraint. Then, one may try to define the quantum version of $E$ as a quantum energy operator.  
However, Kucha\v{r} proved that no such Killing functional $X_{ab}[h,x)$  exists \cite{Kuchar81}.  
  
Even if such a vector existed, the potential term $V(h)$ can take negative values; it is, in fact, not bounded from below. So the positivity condition cannot be satisfied. In any case, the restriction to positive energy solutions makes sense for particles propagating in spacetime, but it has no meaning in superspace. The corresponding ``energy" would have nothing to physical energy, so a restriction to  positive values would be completely {\em ad hoc}.

Hence, there is no natural definition of an inner product for physical states. A probability assignment to general solutions of the WDW equation requires the introduction of additional structure.

Note that the absence of an inner product is a difficulty of geometrodynamics, not an absolute obstruction to canonical quantisation. Loop quantum gravity, built on connection and
holonomy–flux variables rather than the metric, formulates the dynamics differently and evades this particular version of the problem.

 \subsection{From no-time to semiclassical time}

The WDW contains no reference to time, causal ordering, or anything temporal. For Wheeler, this was a good thing:

\begin{quote}
{\small ..the “time ordering of events” is a notion devoid of meaning
[...] the concept of spacetime and time itself are not primary
but secondary ideas in the structure of physical theory. These
concepts are valid in the classical approximation. However, they
have neither meaning nor application under circumstances when
quantum-geometrodynamic effects become important. Then one
has to forgo that view of nature in which every event, past, present,
or future, occupies its preordained position in a grand catalog called
“spacetime”. There is no spacetime, there is no time, there is no
before, there is no after. The question what happens “next” is
without meaning.} \cite{Wheeler}
\end{quote}

Many researchers think otherwise. The WDW equation is practically useless unless one explains how to connect its solutions to our world, where time evolution and causal order exists. 

Perhaps the most popular approach to time in the canonical framework, is to recover time at a semiclassical limit. To this end, we extend the WDW equation (\ref{WDE}) to include matter fields $\phi$ as

\begin{equation}
\left[
-\frac{1}{2\mu}G_{AB}(q)
\frac{\delta^2}{\delta q_A\delta q_B}
+\mu V(q)
+\hat H_m(q,\phi)
\right]\Psi[q,\phi]=0,
\end{equation}
where $\hat H_m$ is the matter Hamiltonian density. Then, we make the WKB-type ansatz
\begin{equation}
\Psi[q,\phi]
=
C_0(q) \exp\left(i \mu S_0[q]\right)
\psi[q,\phi]. \label{ansatz}
\end{equation}
The key idea is to expand the WDW equation in powers of $\mu$. At first order in $\mu$, we obtain
\begin{equation}
\frac12 G_{AB}
\frac{\delta S_0}{\delta q_A}
\frac{\delta S_0}{\delta q_B}
+V(q)=0.
\end{equation}
This is the 
  Hamilton--Jacobi equation for classical gravity in vacuum---$\frac{\delta S_0}{\delta q_A}$ corresponds to the conjugate momentum $p^A$ for the solution. Thus $S_0[q]$ defines a family of classical spacetime backgrounds. 
  
At zero-th order in $\mu$, we obtain
\bey
\frac{i}{2C_0^2} G^{AB}\frac{\partial}{\partial h^A}\left(C_0^2\frac{\partial}{\partial h^B}\right)\psi_0 + iG^{AB}\frac{\partial S_0}{\partial h^A}\frac{\partial}{\partial h^B} \psi - H_m\psi = 0. \label{WKB2}
\eey
Following standard WKB practice, we separate the contribution $C_0$, by imposing the continuity equation  $G^{AB}\frac{\partial}{\partial h^A}\left(C_0^2\frac{\partial}{\partial h^B}\right) = 0$. 
Then, we define the   WKB time derivative   along the corresponding classical gravitational trajectory by
\begin{equation}
\frac{\delta}{\delta T(x)}
:= G^{AB}\frac{\partial S_0}{\partial h^A}\frac{\partial}{\partial h^B} =   G_{ijkl}[h, x)\pi^{ij}(x) \frac{\delta}{\delta h_{kl}(x)}.
\end{equation}
With this definition, Eq. (\ref{WKB2}) becomes  an approximate Schr\"odinger equation for matter fields propagating on the WKB background,
\begin{equation}
i\hbar \frac{\delta \psi}{\delta T(x)}
=
\hat H_m \psi .
\end{equation}
Thus time reappears only after a semiclassical gravitational background has been selected. It is not present at the level of the Wheeler--DeWitt equation itself, but is extracted from the WKB phase of the gravitational part of the wave functional. In this regime, one recovers an effective Schr\"odinger equation for matter, and any inner product with respect to which $\hat{H}_m$ is self-adjoint can be used to define probabilities for $\psi$.

This construction is powerful: it explains how quantum field theory in curved spacetime emerges as an approximation to quantum geometrodynamics. However, it relies on highly restrictive assumptions.

The key limitation is that it presupposes a state of the form \eqref{ansatz}, i.e. a single WKB branch with a well-defined phase $S_0[q]$. A generic solution of the Wheeler--DeWitt equation is instead a superposition,
\begin{equation}
\Psi = \sum_\alpha C_\alpha \, e^{\frac{i}{\hbar} S_\alpha[q]} \,\psi_{\alpha}[q,\phi],
\label{eq:superposition}
\end{equation}
and in this case no unique phase function exists. Each branch $\alpha$ defines its own time parameter $t_\alpha$, and there is no natural way to select one. The interference terms between different phases obstruct the construction of a single Hamilton--Jacobi function, and hence of a global time variable.

One may appeal to decoherence to suppress these interference terms and recover an approximate branch-dependent notion of time. For a closed universe, however, there is no external environment: decoherence must arise from tracing over internal degrees of freedom or from an appropriate coarse-graining of histories. The required system--environment split and coarse-graining are not uniquely determined by the WDW equation, while the existence of sufficiently decoherent WKB branches imposes strong conditions on the state. Decoherence may therefore explain the stability of a semiclassical notion of time within a suitable sector, but it does not by itself explain why the physical state of the Universe belongs to that sector.

Hence, the semiclassical construction does not derive time from a timeless framework; it relocates the problem. Temporality appears only for states with a well-defined WKB phase, and the formalism offers no explanation for why the physical state of the universe should belong to this class. In this sense, semiclassical time does not resolve the problem of time, but assumes, in disguised form, the very structure it seeks to explain.

\subsection{Proposals for cosmological initial conditions}

A major line of research in quantum cosmology attempts to specify boundary or initial conditions for the wave function of the universe. In minisuperspace models, the Wheeler--DeWitt equation,
\begin{equation}
\hat{\mathcal H}\Psi=0,
\end{equation}
admits many solutions, and additional conditions are required in order to select physically relevant states. The most influential proposals are those of Hartle and Hawking \cite{HartleHawking} and of Vilenkin \cite{Vilenkin84,Vilenkin86}.

The Hartle--Hawking no-boundary proposal defines the wave function through a Euclidean path integral over compact regular geometries,
\begin{equation}
\Psi[h_{ij},\phi]
=
\int \mathcal D g \, \mathcal D \phi \;
e^{-S_E[g,\phi]/\hbar},
\end{equation}
where the integral is taken over Euclidean four-geometries that smoothly interpolate between a compact geometry and the final three-geometry $(h_{ij},\phi)$. In minisuperspace, this selects the solution that is regular at $a=0$. For a closed de Sitter universe, the semiclassical wave function takes the form
\begin{equation}
\Psi_{HH}(a)
\sim
\exp\left[
\frac{1}{3H^2}
(1-H^2a^2)^{3/2}
\right]
\end{equation}
in the classically forbidden region, and becomes oscillatory for large $a$. The resulting state is a real superposition of expanding and contracting semiclassical universes.

Vilenkin's tunneling proposal instead imposes the condition that only outgoing WKB modes are present at large scale factor. In the semiclassical regime,
\begin{equation}
\Psi_V(a)
\sim
e^{-iS(a)},
\end{equation}
corresponding to an expanding universe. The proposal interprets the universe as tunneling from a classically forbidden region near $a=0$ into an expanding Lorentzian spacetime. The distinction between incoming and outgoing modes is defined through the Klein--Gordon-type probability current on minisuperspace.

Other proposals also exist. DeWitt suggested imposing the boundary condition
\[
\Psi(a=0)=0,
\]
in order to suppress singular geometries. Linde proposed a variant of the Hartle--Hawking wave function with opposite Euclidean weighting, favoring inflationary universes \cite{Linde}. 

Despite their differences, these proposals presuppose that the Wheeler--DeWitt wave function can ultimately be given a meaningful probabilistic interpretation, even though canonical quantum gravity possesses no preferred positive-definite inner product. In practice, probabilities are generally assigned only within a semiclassical regime, using WKB branches, Klein--Gordon-type currents, or decoherent-histories arguments.   Such proposals may select particular solutions of the Wheeler--DeWitt equation, but they do not by themselves provide those solutions with a complete physical interpretation. They do not resolve  the questions of probability, observables, and the emergence of time.

\section{Relational time}
 
\subsection{The Page--Wootters Construction}
  
The Page--Wootters (PW) construction provides a   realization of the idea that time can emerge from correlations in a globally stationary quantum state \cite{PW83, Wootters84}. 

The key idea in the PW approach is that the Universe is in an eigenstate $|\Psi\rangle$ of the world Hamiltonian $\hat{\HH}$, conveniently taken as a zero eigenstate: $\hat \HH|\Psi\rangle = 0$.   Hence, $\HH$ is  analogous to the Hamiltonian constraint operator in Dirac quantisation. 

The fundamental idea is to express all predictions in terms of the fundamental conditional probabilities (\ref{condprob})
\bey
\mbox{Prob}(b|a) = \frac{\langle \Psi|\hat{P}_a \hat{Q}_b \hat{P}_a|\Psi\rangle}{\langle \Psi| \hat{P}_a |\Psi\rangle}, \label{condprop2}
\eey
where $\hat{P}_a$, $\hat{Q}_b$ are spectral projectors corresponding to values  $a$ and $b$ for observables $\hat{A}$ and $\hat{B}$, respectively. 

PW take the observable $\hat{A}$ to correspond to a clock and $a = t$ is the reading of the clock. Hence, ``time evolution" is nothing but the conditioning of the probabilities   $\mbox{Prob}(b|t)$ on the reading of the clock. 

In the simplest model, the Hilbert space of the system---prior to the imposition of the constraint $\HH|\Psi\rangle = 0$---factorizes as ${\cal K}_T \otimes{\cal K}_S$, where ${\cal K}_T$ is the Hilbert space of the clock and ${\cal K}_S$ is the Hilbert space of the system of interest. The clock Hilbert space  ${\cal K}_T$ supports a time observable $\hat{T}$ with generalized eigenvectors $|t\rangle$, and a Hamiltonian $\hat{\Omega}$ such that $[\hat{T}, \hat{\Omega}] = i \hat{I}$. Note that $\hat{\Omega}$ acts on time wavefunctions $\phi(t) = \langle t|\phi\rangle$ as $\hat{\Omega}\phi(t) = - i \frac{d \phi(t)}{dt}$.

Then, we take the total Hamiltonian to be $\hat{\HH} = \hat{\Omega} \otimes \hat{I} + \hat{I} \otimes \hat{H}_S$.  Let $|q\rangle$ be any basis on ${\cal K}_S$. The condition $\langle t, q|\hat{\HH}|\Psi\rangle = 0 $ yields 
\bey
i\frac{\partial}{\partial t}\Psi(t, q) = \hat{H}_S \Psi(t, q),
\eey
essentially Schr\"odinger's equation. The difference is that $t$ is now not an external parameter,  but a random variable, so the natural normalization is $\int dt \int dq |\Psi(t, q)|^2 = 1$.

However, $|\Psi(t, q)|^2$  is not the direct analogue of the standard quantum probabilities.
 The PW proposal is that the latter are obtained by  conditioning $|\Psi(t, q)|^2$
for fixed values of the time reading,
\bey
\mbox{Prob}(q|t) = \frac{|\Psi(t, q)|^2}{\int dq'|\Psi(t, q')|^2}.
\eey

Equivalently, one may introduce the normalized conditional $|\psi(t)\rangle_S$ by
$$
\langle q|\psi(t)\rangle_S
 =
 \frac{\langle t, q|\Psi\rangle}
 {\sqrt{\langle\Psi|
 |t\rangle\langle t|\otimes\hat I 
 |\Psi\rangle}},
$$
so that $\mbox{Prob}(q|t) =  |\langle q|\psi(t)\rangle_S|^2$. 
  
%The PW formalism provides an interesting demonstration how a system that is static according to an external observer can still incorporate changes when viewed from the %internal perspective \cite{Moreva}.  

A traditional objection to the PW construction is that the clock operator
$\hat T$ does not commute with the constraint $\hat{\HH}$ and is therefore not
a Dirac observable. Similarly, the kinematical projector
$|t\rangle\langle t|\otimes\hat Q_b$ does not preserve the physical Hilbert
space. If it is interpreted as an ordinary projective measurement followed by
state collapse, it maps a physical state into a state that generally fails to
satisfy the constraint.  

Recent work by H\"ohn, Smith, and Lock has shown that this objection does not
invalidate the PW construction \cite{HSL1, HSL2}.
The clock reading $t$ should be understood as a relational variable rather
than as a Dirac observable by itself. A gauge-invariant relational observable
can be constructed by averaging the clock-conditioned operator along the gauge
orbit. Schematically,
\bey
\hat F_{Q|T}(t)
 =
 \int_{\mathbb R} ds\,
 e^{-is\hat{\HH}}
 \bigl(|t\rangle\langle t|\otimes\hat Q_b\bigr)
 e^{is\hat{\HH}},
\eey
and this operator satisfies $[\hat F_{Q|T}(t),\hat{\HH}] = 0$.
 
When evaluated with the physical inner product, the probabilities associated
with these relational Dirac observables coincide with the PW conditional
probabilities. The PW description can therefore be interpreted as a
gauge-fixed representation of an underlying gauge-invariant relational
theory, rather than as a projective measurement that takes the state outside
the physical Hilbert space.  

%A separate question is whether the clock is actually measured by a physical
%apparatus. The original conditional-probability formula does not describe such
%a measurement interaction and should not, by itself, be interpreted as a
%collapse produced by reading the clock. To represent an operational clock
%measurement, the measuring apparatus and its interaction with the clock must
%be included in the constrained system. Recent models have shown that such
%clock measurements can be incorporated while retaining a stationary global
%state and, under suitable conditions, unitary relational evolution
%\%cite{HausmannEtAl25,KuypersRijavec25}. For finite-resource clocks, however,
%different implementations of measurement need not be equivalent and may lead
%to non-unitary or temporally non-local effective dynamics.
 The original PW conditional-probability rule does not, by itself, provide a
prescription for sequential measurements. Suppose that $\hat Q_{a_1}$ and
$\hat P_{a_2}$ represent properties of the system at clock readings $t_1$ and
$t_2$, respectively. A naive attempt to construct a two-time probability by
successively inserting the clock-conditioned projectors
\[
|t_1\rangle\langle t_1|\otimes\hat Q_{a_1},
\qquad
|t_2\rangle\langle t_2|\otimes\hat P_{a_2},
\]
fails because ideal clock states at distinct times are orthogonal: $\langle t_2|t_1\rangle=\delta(t_2-t_1)$.

There are two principal responses to this problem. The first uses
gauge-invariant relational observables. H\"ohn, Smith, and Lock showed that
appropriately constructed relational Dirac observables reproduce the correct
transition probabilities and permit an arbitrary number of relational
conditionings \cite{HSL1}. Thus, the failure of the naive product of
clock projectors does not imply that relational dynamics itself is
inconsistent. It shows instead that kinematical clock projectors cannot simply
be treated as ordinary projectors describing successive physical
measurements.

The second response, proposed by Giovannetti, Lloyd, and Maccone (GLM), models
the measurements dynamically and introduces auxiliary systems that store their
outcomes \cite{GLM15}. Suppose that measurements take place when the clock
reads $t_1<t_2$. Memory systems $M_1$ and $M_2$ are coupled to the measured
system $Q$ at the corresponding clock readings. In the conditional
description, this is represented by
\[
\hat H_{Q M_1M_2}(t)
 =
 \hat H_Q
 +\delta(t-t_1)\hat h_{QM_1}
 +\delta(t-t_2)\hat h_{QM_2},
\]
where $\hat h_{QM_i}$ correlates the measured quantity with the corresponding
memory. Equivalently, these clock-controlled interactions may be incorporated
into the stationary Hamiltonian constraint of the enlarged system.

After both interactions have occurred, the joint probability for the recorded
outcomes is obtained from a single conditional probability,
$$
\mbox{Prob}(a_1,a_2|t)
 =
 \frac{
 \langle\Psi|
 \bigl(
 |t\rangle\langle t|
 \otimes\hat I_Q
 \otimes\hat P^{M_1}_{a_1}
 \otimes\hat P^{M_2}_{a_2}
 \bigr)
 |\Psi\rangle}
 {\langle\Psi|
 \bigl(
 |t\rangle\langle t|
 \otimes\hat I_{QM_1M_2}
 \bigr)
 |\Psi\rangle},
\qquad t>t_2.
$$
For an ideal clock and ideal measurement interactions, this construction
reproduces the standard probabilities for sequential measurements. The
multi-time question is thereby represented as a single-time question about
the records retained by the memories.

Note that the order of the measurement events is encoded in the clock-controlled
interactions: one coupling is supported at $t_1$, another at $t_2$, and their
ordering is fixed by the already specified order of the clock readings,
$t_1<t_2$. One therefore recovers the statistics of a sequence
whose temporal order has been supplied as part of the model. The ordering is not recovered from the stationarity of the global state.

Thus,  properly constructed relational observables or explicit
memory systems recover the standard transition probabilities. They do not,
however, show that temporal succession itself emerges from clock--system
correlations. The PW mechanism explains dynamics relative to an ordered family
of clock readings, while the order of those readings remains part of the
structure presupposed by the construction.

\subsection{Evolving constants of the motion}

Rovelli's proposal \cite{Rovelli90, Rovelli} follows from the classical analysis of clock relativism in Sec. 4.4. The idea is to construct the analogue of the evolving constant of the motion in Dirac quantisation. Let ${\cal H}$ be the Hilbert space prior to the solution of the constraints, and let $\hat{C}$ be the Hamiltonian constraint operator. 

We will use the result that for any self-adjoint operator $\hat{A}$ on ${\cal H}$, the operator 
\bey
\hat{A}_{ph} = \int ds e^{i\hat{C}s}\hat{A}e^{-i\hat{C}s},
\eey
commutes with the constraint operator $\hat{C}$. The proof is elementary since by construction $e^{i\hat{C}\tau}\hat{A}_{ph}e^{-i\hat{C}\tau} = \hat{A}_{ph}$\footnote{The proof is only formal, because often the integral may not exist, but the overall logic is sound and can be used by suitable regularization procedures.}. 

Let $\hat{T}$ be the self-adjoint operator on ${\cal H}$ that defines the clock. We denote its spectral projectors  $\hat{E}_t = \delta(\hat{T} - t)$. Then, for any self-adjoint operator $\hat{F}$ on ${\cal H}$, we define the corresponding evolving constants of the motion as
\bey
\hat{F}_{\tau} = \int ds e^{i\hat{C}s}\hat{E}_{\tau}\hat{F} \hat{E}_{\tau} e^{-i\hat{C}s}. \label{qevc}
\eey
This expression may require regularization of the projectors $\hat{E}_t$   if $\hat{T}$ has continuous spectrum, but otherwise it is a precise implementation of Rovelli's idea at the quantum level\footnote{Eq. (\ref{qevc}) is not found in the original papers, where the account is more descriptive than mathematically formal, but it is natural in the context of later developments in the canonical quantisation of constrained systems.}.

The relation between this construction and the Page--Wootters formalism has
been clarified by H\"ohn, Smith, and Lock \cite{HSL1,HSL2}. They showed that Rovelli-type 
  Dirac observables, PW conditional states, and quantum deparametrization provide three equivalent representations of relational quantum dynamics. 

\subsubsection*{Implementation: the two-oscillator model.}

Consider two oscillators with number operators $\hat N_1,\hat N_2$ and constraint
\begin{equation}
\hat C=\omega_1 \hat N_1-\omega_2 \hat N_2 .
\end{equation}
For simplicity, take $\omega_1/\omega_2=k$ a positive integer, so that the physical Hilbert space is spanned by states $|n\rangle_{ph} = |n\rangle \otimes |k n\rangle$.
 
We use the phase of oscillator 2 as the clock. We introduce phase states
\begin{equation}
|\theta\rangle=\frac{1}{\sqrt{2\pi}}\sum_m e^{-im\theta}|m\rangle ,
\end{equation}
so that the spectral projectors are $\hat{E}_{\theta} = \hat{I}\otimes|\theta\rangle\langle \theta|$. 
 
Let $\hat Q$ be an operator acting on oscillator 1.  We have $\hat{E}_{\theta}\hat{Q}\otimes \hat{I} \hat{E}_{\theta} = \hat{Q} \otimes |\theta\rangle\langle \theta|$, where we regularize by dropping an infinite term $\langle \theta|\theta\rangle$\footnote{A more precise treatment would employ 
  a covariant phase POVM, but this refinement does not affect the conceptual
point of the example.}. Hence, Eq. (\ref{qevc}) yields
\bey
\hat{Q}_{\theta} = \int_{0}^{2 \pi} ds e^{i\omega_1 \hat{N}_1s} \hat{Q} e^{-i\omega_1 \hat{N}_1s}\otimes |\theta + \omega_2 s\rangle \langle \theta + \omega_2 s|,
\eey
where the integration is restricted to $[0, 2 \pi)$ due to the periodicity of the system.

We evaluate the matrix elements of $\hat{Q}_{\theta}$ on physical states, 
\bey
{}_{ph}\langle n| \hat{Q}_{\theta}|m\rangle_{ph} = \langle n|\hat{Q}|m\rangle e^{i\omega_1/\omega_2\theta (n - m)},
\eey
 which is equivalent to a Heisenberg-time evolution  rule $\hat{Q}_\tau = e^{i\hat{N}_1\omega_1(\tau/\omega_2)}\hat{Q}e^{-i\hat{N}_1\omega_1(\tau/\omega_2)}$.

\bigskip

The problem of evolving constants persists in the quantum theory. As shown in Sec.~4.4, different clocks can assign incompatible temporal orders to the same events, and each family of operators $\hat F_\tau$ inherits the order defined by its chosen clock. Rovelli's natural response is that temporal order, like evolution, is relational: different clocks define different correlations, with no requirement that they combine into a single ordering. This is unproblematic for parameter or coordinate order, but causal order is different. In classical general relativity, the precedence of timelike-related events is fixed by the light-cone structure, independently of any clock. In quantum gravity, where the metric itself fluctuates, a relational theory must either identify clocks that recover a common causal order or explain how the clock-independent order of the classical limit emerges.

This exposes a foundational divide: is causal order a fundamental or an emergent structure? If it is emergent, incompatible clock-relative orderings may be regarded as different perspectives rather than contradictions, provided that they do not yield inconsistent predictions for the same physical situation. If causal order is fundamental, a theory that fails to recover a consistent ordering is at best incomplete. Relational programmes take the first path; the histories formulations considered next provide a natural setting in which to explore the second.

\section{Histories-based theories}

We first encountered the decoherent histories approach in Section 3.4, where we gave the basic definitions, and explained that probabilities for histories are defined only if specific decoherence conditions are satisfied. The latter are expressed through the decoherence functional, a complex-valued function of pairs of histories.

 Here, we will elaborate on this framework as a candidate for addressing the problem of time in quantum gravity. It is necessary 
  to provide a few useful definitions. 
 
\begin{itemize}
\item  A {\em homogeneous history\/}
is any time-ordered sequence $\oph{\alpha}$ of
projection operators.

\item  A homogeneous history  $\beta:=\oph{\beta}$ is {\em coarser\/} than
another history $\alpha:=\oph{\alpha}$ if, for every
$t_i$, $\op\alpha_{t_i}\leq\op\beta_{t_i}$ where $\leq$
denotes the usual ordering operation on the space
of projection operators---$\op \alpha \leq \op\beta$ means that
the range of $\op \alpha$ is a subspace of the range of
$\op\beta$. This relation
on the set of homogeneous histories is a partial
ordering.

\item Two homogeneous histories  $\alpha:=\oph{\alpha}$ and $\beta:=\oph{\beta}$ are
{\em disjoint\/} if, for at least one time point
$t_i$, $\op\beta_{t_i}$ is disjoint from
$\op\alpha_{t_i}$.

\item In calculating a decoherence functional it may be necessary to
go outside the class of homogeneous histories to
include {\em inhomogeneous\/} histories. A history
of this type arises as a logical OR (denoted
$\vee$) operation on a pair of disjoint
homogeneous histories $\alpha:=\oph{\alpha}$ and
$\beta:=\oph{\beta}$. Such a history $\alpha \vee\beta$ is
generally   not homogeneous, 
but, when computing the decoherence functional, it
is represented by the operator
$\op{C}_{\alpha \vee\beta}:=\op{C}_\alpha+\op{C}_\beta$. The
coarse-graining relations $\alpha \leq \alpha \vee\beta$ and
$\beta \leq\alpha \vee\beta$ are deemed to apply to this
disjoint OR operation. The NOT operation
$\neg$ also usually turns a homogeneous history
into an inhomogeneous history, with
$\op{C}_{\neg\alpha}:=\op{I}-\op{C}_\alpha$.  
\end{itemize}

\subsection{ The Gell-Mann-Hartle-Isham axioms}

  Gell-Mann and Hartle  
postulated a new approach to quantum theory in
which the notion of history has a fundamental
role \cite{hartlelo}:   a ``history"
is an irreducible entity in its own right, not
necessarily derived from time-ordered strings of
single-time propositions.  In the context of quantum gravity, a histories theory shifts the central question from ``Which variable should be used as time?'' to ``What structures define possible temporal sequences of events, and how are probabilities assigned to them?''

Gell-Mann and Hartle  formulated histories theory in terms of a set of axioms. These axioms were generalized and strengthened by Isham \cite{Ish94}, who developed the History Projection Operator (HPO) approach in which individual histories are represented by projection operators  on an appropriate Hilbert space. 

The basic structure of the Gell-Mann-Hartle-Isham (GHI) axiomatization of histories   follows.

\subsubsection{The space of histories}

The set $\mathcal{U}$ consists of all possible histories (called “history filters” by Isham).  A history is a generalization of the notion of homogeneous histories in the standard case. Abstracting from the properties of homogeneous histories, we assume that  $\mathcal{U}$ is equipped with the following structure.

\begin{enumerate}

\item {\em Partial ordering.} $\mathcal{U}$ is equipped with a partial order $\leq$:    $\alpha \leq \beta$ implies that $\beta$ is a \emph{coarser} description of the same physical situation as $\alpha$.  

\item There are two special histories:
\begin{itemize}
\item the \emph{unit history} $I$, corresponding to “something happens” (no restriction),
\item the \emph{null history} $0$, corresponding to an impossible history.
\end{itemize}
Every history lies between these: $0 \leq \alpha \leq I$. A history if {\em fine-grained} if the only histories $\beta$ such that $\beta \leq \alpha$ are either $0$ or $\alpha$.

\item {\em Logical AND}.  
There is an operation $\wedge$ that represents logical AND. The history $\alpha \wedge \beta$ represents the situation in which both $\alpha$ and $\beta$ hold.

\item {\em Temporal composition.}  
Histories can also be composed in time. If it makes sense to say that $\beta$ occurs after $\alpha$, we can form a composite history $\alpha \circ \beta$. We then say that $\beta$ \emph{follows} $\alpha$, or $\alpha$ \emph{precedes} $\beta$.  

For homogeneous histories, this corresponds to concatenating two time-ordered sequences. When this composition is defined, it agrees with the logical AND operation.
In temporal logic, $\circ$ corresponds to the logical connective AND THEN.

\end{enumerate}

\subsubsection{The space of temporal supports}
 
In addition to the logical structure of histories, one must specify their temporal structure. This is done by introducing a second space, denoted by $\mathcal{S}$, whose elements represent the \emph{temporal supports} of histories. For a homogeneous history, $\alpha:=\oph{\alpha}$, the temporal support is the set of time points $\{t_1, t_2, \ldots, t_n\}$. In general, a temporal support may be an open set of the real line, or even be defined without reference to a moment of time. 

Typically $\mathcal{S}$ is identified with a collection of subsets of the
space ${\cal T}$ of time points. In the standard case ${\cal T}=\R$, but this
is not necessary. The only requirement is that ${\cal T}$ carries a partial
order $\po$, which distinguishes temporal precedence. The formalism can therefore be used to describe systems with exotic temporal structures.

The relation between histories and their temporal supports is given by a map
$
\sigma: \mathcal{U} \rightarrow \mathcal{S},
$
  that  
 assigns to each history its temporal support. 

The partial order on ${\cal T}$ is a primitive kinematical element of the
histories theory. It is not derived from the dynamics, from correlations
between physical clocks, or from the decoherence functional. Rather, it
determines which propositions can be combined into a temporally ordered
history in the first place. In a relativistic theory, this order is naturally
identified with causal order: timelike- or null-related events can be arranged
in succession, whereas spacelike-separated events are generally incomparable.
The causal structure therefore precedes the assignment of probabilities. It
defines the admissible histories on which the dynamical and probabilistic
structures of the theory are subsequently constructed.

%\[
%\alpha = \alpha^1 \circ \alpha^2 \circ \cdots \circ \alpha^N. \label{dechis}
%\]
%A history that admits no such decomposition is called {\em nuclear}. Similarly, a support $s$ is nuclear if it cannot be decomposed into a sequence of two supports.
%In physical terms, nuclear histories generalize single-time propositions and nuclear supports  of individual moments of time.

%Finally, we note that a decomposition (\ref{dechis}) 
%is said to be \emph{irreducible} if each component $\alpha^i$ is nuclear. For homogeneous histories, this corresponds to expressing a history as a sequence of single-time %propositions ordered in time.

\subsubsection*{The space of history propositions.}

The space $\mathcal{U}$ of histories introduced above is not sufficient to describe all meaningful statements about a system. In particular, one often needs to consider logical combinations of histories, such as alternatives (“either $\alpha$ OR $\beta$”) or negations (NOT $\alpha$). For this reason, $\mathcal{U}$ is embedded in a larger space $\mathcal{UP}$ of \emph{history propositions}.

The space $\mathcal{UP}$ is equipped with the structure of a logical lattice, in which histories can be combined using the operations: $\wedge$ (AND), $\vee$ (OR) and $\neg$ (NOT).
These operations extend the corresponding structure already present in $\mathcal{U}$, and allow one to form more general propositions about histories.

In particular, any element of $\mathcal{UP}$ can be constructed from histories in $\mathcal{U}$ by applying a finite (or countable) number of these logical operations. Thus, $\mathcal{UP}$ represents the set of all propositions “about” histories.

Two history propositions $\alpha$ and $\beta$ are said to be \emph{disjoint} (denoted $\alpha \perp \beta$) if $\alpha \leq \neg \beta$, meaning that they cannot both occur. A set of propositions is \emph{exclusive} if its elements are pairwise disjoint, and \emph{exhaustive} if their disjunction yields the unit proposition:
\[
\alpha^1 \vee \alpha^2 \vee \cdots \vee \alpha^N = I.
\]
Such sets play the role of alternative histories to which probabilities may be assigned.

A particularly important realisation of this structure is obtained by representing history propositions as projection operators on a Hilbert space. In the case of homogeneous histories with fixed times $\{t_1,\ldots,t_n\}$, each history
\[
\alpha = (\alpha_{t_1}, \alpha_{t_2}, \ldots, \alpha_{t_n})
\]
can be represented by the tensor product operator
\begin{equation}
\tilde{\alpha} := \alpha_{t_1} \otimes \alpha_{t_2} \otimes \cdots \otimes \alpha_{t_n},
\label{Def:th_clean}
\end{equation}
acting on the tensor-product Hilbert space
\[
\mathcal{H}_{t_1} \otimes \mathcal{H}_{t_2} \otimes \cdots \otimes \mathcal{H}_{t_n}.
\]
This is the   \emph{history projection operator} (HPO) representation.
In this representation, logical operations correspond to operations on projection operators, and the lattice structure of $\mathcal{UP}$ is realised concretely within standard quantum theory.

The extension from $\mathcal{U}$ to $\mathcal{UP}$ is essential for formulating a probabilistic theory of histories. While $\mathcal{U}$ encodes individual possible histories, $\mathcal{UP}$ allows one to describe alternatives, more general coarse-grainings, and logical combinations of histories.

At the same time, this construction introduces a non-trivial enlargement of the Hilbert space, since histories are represented on tensor products of single-time Hilbert spaces. This highlights an important feature of the formalism: temporal structure is encoded not only in the ordering of events, but also in the algebraic structure of the space of history propositions.

\subsubsection*{The space of decoherence functionals.}

The final ingredient in the histories framework is the notion of a
\emph{decoherence functional}. A decoherence functional assigns a
complex number $d(\alpha,\beta)$ to each pair of history propositions,
and measures the degree of interference between the histories
$\alpha$ and $\beta$. The collection of all decoherence functionals
is denoted by $\mathcal{D}$.

The decoherence functional contains both the dynamical
information of the theory and the probabilistic structure associated
with histories.

A decoherence functional must satisfy a number of general conditions.

\begin{itemize}

\item {\em Hermiticity:} $d(\alpha,\beta)=d(\beta,\alpha)^*$.

\item {\em Positivity:} $d(\alpha,\alpha)\geq 0$.

\item {\em Null triviality:} $d(0,\alpha) = 0$.

\item {\em Additivity:} If $\alpha \perp \beta$, then $d(\alpha\vee\beta,\gamma)
=
d(\alpha,\gamma)+d(\beta,\gamma)$.

\item{\em  Normalisation:} 
$
d(I,I)=1 .
$

\end{itemize}

The current understanding of the theory is that probabilities can be defined only for  an exhaustive and exclusive set of histories which satisfy the decoherence condition
\[
d(\alpha,\beta)=0
\qquad
\text{for}
\qquad
\alpha\neq\beta .
\]
In this case, the diagonal elements
\[
p(\alpha)=d(\alpha,\alpha)
\]
satisfy the Kolmogorov probability rules and can be interpreted as
ordinary probabilities.

The histories approach is usually thought of as a realist interpretation of quantum mechanics---meaning that it attempts to describe quantum systems without reference to measurement. However, it also has an operational interpretation, the history structure enabling the construction of  observables with complex temporal structure. In this interpretation, the off-diagonal elements of the decoherence functional can be reconstructed through phase measurements \cite{Anhil}.

\subsubsection{Beyond discrete-time histories}

The GHI axioms enable the formulation of histories theories beyond the basic case of sequences of projection operators at discrete times.
Examples include the following.

\begin{enumerate}
\item The first major step in this direction was taken by Hartle in his spacetime formulation of generalised quantum mechanics \cite{hartlelo}.  Hartle 
analyzed configuration space histories. For a system characterized by configuration space $Q$, a fine-grained history corresponds to a path $q: [0, T]\rightarrow Q$. 
If $\Pi_Q$ is the space of such paths, a general coarse-grained history $\alpha$ is a subset of $\Pi_Q$. The logical connectives between histories are defined through set-theoretic operations on $\Pi_Q$.

In this approach, the decoherence functional is written directly in path-integral form,
\begin{equation}
d(\alpha',\alpha)
=
\int_{\alpha'} \mathcal{D}q'(\cdot)
\int_{\alpha} \mathcal{D}q(\cdot) \,
e^{\frac{i}{\hbar}(S[q'(\cdot)]-S[q(\cdot)])}
\rho(q'_0,q_0),
\end{equation}
where path integration is over all paths such that $q(0) = q_0, q'(0) = q_0'$ and $q(T) = q'(T) = q_f$.

This formulation has the important advantage that it no longer requires histories to be defined with respect to a preferred foliation of equal-time surfaces. History propositions about time-extended properties, time-of-arrival, and crossing of spacetime regions   can be formulated in this framework. 

\item Isham and Linden constructed a history Hilbert space ${\cal V}$ in which the histories---represented by projection operators---have temporal support on open subsets of the real line $\R$ \cite{IL95}. The idea is that ${\cal V}$ carries a representation of the {\em history group}, the history analogue of the canonical group. For a particle on a line, the defining equations for the history group are
    \bey
   [\hat{x}_t, \hat{x}_{t'}] = 0, \;\;  [\hat{p}_t, \hat{p}_{t'}] = 0,\;\; [\hat{x}_t, \hat{p}_{t'}] = i\delta(t, t') \hat{I},
    \eey
The history group has an infinity of possible representations. The  appropriate one  is selected by the requirement that an appropriately time-averaged Hamiltonian operator $\int dt \lambda(t) \hat{h}_t$ exists. This result has been generalized  to quantum fields, thus describing  histories with support on open subsets of Minkowski spacetime \cite{Sav03}.

\item Anastopoulos analyzed continuous-time phase space histories subject to the GHI axioms with no prior reference to a Hilbert space structure \cite{An03}. He showed that the standard Hilbert space description of quantum theory can be reconstructed, if we assume that the system is described by decoherence functionals  that carry no memory and are time-reversible.
\end{enumerate}

\subsection{Histories canonical gravity and spacetime symmetries}
 
Savvidou showed that the intrinsic temporal structure  of  histories is unaffected by the implementation of constraints, so a fundamental notion of causal ordering may persist even at the level of quantum gravity \cite{Sav1, Sav2}.  

  The starting point is the classical space of histories $\Pi = T^*\mbox{Lor}(\M)$ of Lorentzian metrics on a manifold $M$ of topology $\R\times \Sigma$. This space is equipped with the natural symplectic structure
    \bey
    \Omega
=
\int d^4X \,
\delta \pi^{\mu\nu}(X)\wedge \delta g_{\mu\nu}(X), \label{sympgr}
    \eey
where $g_{\mu\nu}(X)$ is the spacetime metric and $\pi^{\mu\nu}(X)$ its conjugate momentum density. The space $\Pi$ carries a symplectic action of the $\DM$ group with generators $V_W = \int d^4X \pi^{\mu \nu}{\cal L}_W g_{\mu\nu}$, for any vector field $W^{\mu}$ on $\M$.

  One introduces a foliation functional, ${\cal E}(g)$, that is a functional that assigns a spacelike foliation $\mathcal{E}[g]: \mathbb{R}\times \Sigma \rightarrow \M$ to each metric $g$. One then implements a symplectic transformation from the covariant variables $g_{\mu \nu}(X), \pi^{\rho \sigma}(X)$, to the canonical history variables $h_{ij}(t,x), N(t, x)$,  $N_i(t, x)$, and their conjugate momenta. 
    The key result is that the constraints of GR, defined in terms of the canonical variables are $\DM$-invariant, if the foliation functional satisfies a consistency condition. Hence, the transition from $\Pi$ to the reduced history space $\Pi_{red}$---described in Sec. 4.4.6---is implemented in a $\DM$-invariant procedure. The reduced history space contains histories ordered by a time parameter $t$.
 
  The same procedure can work in quantum theory. This requires the introduction of a history Hilbert space ${\cal V}$ through a representation of the history group that corresponds to (\ref{sympgr}); ${\cal V}$ also carries a representation of the $\DM$ group. The constraint operators are $\DM$ invariant, and they preserve the time-ordering parameter $t$ of histories. This parameter does not correspond to physical evolution generated by the Hamiltonian constraint, but to the intrinsic ordering structure of histories.
   
  Events as described by histories have a well-defined causal order, but they are expressed in terms of non-local variables that commute with the constraints. One may then reconstruct an effective spacetime notion of time by adjoining either gauge variables or physical material clocks. The crucial point is that causal ordering is taken to be more primitive than metric time itself. The histories parameter provides an intrinsic ordering relation between events, independently of any specific choice of clock variable or spacetime foliation. Spacetime time then emerges as a secondary structure reconstructed from these ordered events and their correlations.

  The main difficulties  of this program come from the  technical aspects of the quantisation procedure. Constructing the appropriate  Hilbert space for histories and the associated operators is more difficult than doing so for the canonical case. It is already a difficult task even for toy models with a finite number of degrees of freedom.

\section{The causal set approach to quantum gravity}

The causal set programme is based on the idea that the fundamental structure underlying spacetime is causal rather than geometric \cite{BLMS87,Sorkin05}. Its central postulate is that spacetime is fundamentally discrete, with causal ordering more primitive than metric or topological structure. A spacetime is therefore replaced by a \emph{causal set}: a locally finite partially ordered set whose elements represent elementary spacetime events.

A causal set is a pair
$(C,\prec),
$
where $C$ is a set, and $\prec$ is a causal precedence relation satisfying transitivity and irreflexivity,
\[
x\prec y,\quad y\prec z \Rightarrow x\prec z,
\qquad
x\nprec x,
\]
together with local finiteness: for any pair $x,z\in C$, the set
\[
\{y\in C\,|\,x\prec y\prec z\}
\]
contains finitely many elements. Local finiteness implements spacetime discreteness by ensuring that finite spacetime regions contain only finitely many elementary events.

The underlying motivation comes from the fact that the causal structure of a globally hyperbolic spacetime determines its metric up to a conformal factor. Supplementing causal order with volume information allows, in principle, the reconstruction of geometry.
The causal order determines the conformal structure, while the number of elements corresponds to spacetime volume.

A continuum spacetime is approximated through a random Poisson \emph{sprinkling}, in which points are distributed with density proportional to the spacetime volume element and inherit the causal order of the continuum metric. The randomness is essential: regular lattices violate Lorentz invariance by selecting preferred directions, whereas Poisson sprinklings are statistically Lorentz invariant.

One of the main open problems in the causal set programme is the formulation of dynamics. In the sequential growth dynamics of Rideout and Sorkin \cite{Rideout99}, causal sets grow by the stochastic addition of new elements together with their causal relations. The resulting structure resembles a discrete causal history rather than a spatial geometry evolving in time. The growth parameter itself is treated as gauge, so that the physical content resides only in the causal ordering and the number of elements.

Macroscopic spacetime geometry is expected to emerge statistically from sufficiently large causal sets. Geometric quantities can be reconstructed approximately from the order structure: the longest chain between two elements approximates proper time, the abundance of relations determines dimension, and discrete analogues of differential operators can be defined directly from the causal relations.

The causal set programme admits a natural interpretation in terms of histories. A causal set can be viewed as a \emph{fine-grained history}, and the space of all causal sets defines the space of possible histories. Coarse-grained histories correspond to classes of causal sets sharing specified structural properties.

The central problem then becomes the construction of a decoherence functional
$
d(\alpha,\beta)
$
on pairs of coarse-grained causal histories. Such a decoherence functional would encode interference between alternative causal structures and provide the fundamental dynamical object of the theory. In this perspective, one does not quantise a single causal set, but formulates quantum theory directly on the space of causal histories.

The causal set programme is another implementation of the idea that causal order is more primitive than spacetime geometry. Temporal succession is built directly into the fine-grained histories, while spacetime geometry emerges only after coarse-graining and decoherence.

The main problem of this approach is that the recovery 
of 
the Einstein equations is not automatic: a causal-set dynamics can be engineered so
that they emerge in a suitable limit, but this is not a generic feature of the construction. Second, and more severe, is the difficulty common to all fundamentally discrete
approaches—reconstructing a smooth four-dimensional continuum from the underlying
discrete structure remains an open mathematical problem.  
\section{The complex geometry of the light-cone}
Penrose's introduction of twistors was motivated by the idea that the fundamental structure of spacetime is not metric geometry, but the causal and conformal structure encoded in light cones \cite{Penrose67, Penrose72, PenroseRindler}.   Penrose proposed that these null structures, rather than spacetime points themselves, should be regarded as primary. Twistor theory implements this idea by replacing spacetime events with geometric objects associated to null rays. In this framework, the basic entities are spinorial variables naturally adapted to the propagation of massless particles, and spacetime geometry emerges indirectly from relations between them.  

In this section, we explain the basic definitions of twistors and how they relate to spacetime points. 
\subsection{Spinors and null vectors}
First, we analyze the group $SL(2,\C)$, which forms the double cover of the Lorentz group. Its defining representation is as a matrix on $\C^2$. We refer to the vectors of this representation as \emph{spinors}. 

We denote a spinor by $\lambda\in\mathbb{C}^2$, with components $\lambda^A$, $A=1,2$. Under $\alpha\in SL(2,\mathbb{C})$, $\lambda \mapsto \lambda'=\alpha \lambda$.
The complex conjugate spinor $\bar{\lambda}$ transforms with the conjugate matrix,
\begin{equation}
\bar{\lambda} \mapsto \bar{\lambda}'=\alpha^* \bar{\lambda}.
\end{equation}
We will write the coordinates of the conjugate spinor as  $\bar{\lambda}^{A'}$.

There exists a natural map between matrices on $\C^2$ and four-vectors on Minkowski spacetime. We can map each four-vector  $\xi^{\mu}$ to a self-adjoint $2 \times 2$ matrix
\begin{eqnarray}
\tilde{\xi} := \xi^{\mu} \sigma_{\mu} = \left( \begin{array}{cc} \xi^0 + \xi^3& \xi^1-i \xi^2\\ \xi^1+i\xi^2& \xi^0-\xi^3 \end{array}\right), \label{fundiso}
\end{eqnarray}
where    $\sigma_{\mu} := (I, \sigma_1, \sigma_2, \sigma_3)$. We will refer to $\tilde{\xi}$ as the {\em spin matrix} associated to the four-vector $\xi$. It transforms as $\tilde{\xi}\rightarrow \alpha \tilde{\xi}\alpha^{-1}$ under the $SL(2,\C)$ group.

  It  is straightforward to   invert the map (\ref{fundiso}), and to express $\xi^{\mu}$ in terms of its spin matrix $\tilde{\xi}$,
\begin{eqnarray}
 \xi_{\mu} = \frac{1}{2} Tr(\tilde{\xi} \bar{\sigma}_{\mu}), \label{invertxim}
\end{eqnarray}
where $\bar{\sigma}_{\mu} := (I,  -\sigma_1, -\sigma_2, -\sigma_3)$.
%Θα χρησιμοποιήσουμε το συμβολισμό $\tilde{x}_{AA'}$, όπου η σημασία των κεφαλαίων λατινικών γραμμάτων και των τονούμενων γραμμάτων  θα φανεί στη συνέχεια.
 
Eq.  (\ref{fundiso}) implies the crucial identity
\begin{eqnarray}
\det \tilde{\xi} = (\xi^0)^2 - (\xi^1)^2 - (\xi^2)^2 -(\xi^3)^2 = \xi^{\mu} \xi_{\mu}, \label{detiso}
\end{eqnarray}
The eigenvalues of the spin matrix  $\tilde{\xi}$ are $\xi^0 \pm |\mathbf{\xi}|$. If $\xi^0 \geq 0$, then  $\tilde{\xi}$ is a positive matrix, for all timelike vectors $\xi^{\mu}$. The converse also holds.

For any non-zero spinor $\lambda$, the matrix $\tilde{p}:=\lambda\lambda^\dagger$ is Hermitian, positive, and has vanishing determinant.
 Hence it corresponds to a future-directed null vector $p^\mu$,
\begin{equation}
p_\mu=\frac{1}{2} \lambda^\dagger \bar{\sigma}_\mu \lambda
\end{equation}
Conversely, any future-directed null vector $p^\mu$ can be written in this form, with $\lambda$ unique up to a phase: $\lambda \sim e^{i\theta}\lambda$.

The antisymmetric matrix
\begin{equation}
\epsilon=i\sigma_2
=
\begin{pmatrix}
0&1\\
-1&0
\end{pmatrix}
\end{equation}
is used to form invariant contractions of spinors. If $\lambda$ and $\mu$ are spinors, then $\langle \lambda,\mu\rangle:=\lambda^T\epsilon\mu = \epsilon_{AB}\lambda^a \mu^B$
is invariant under $SL(2,\C)$, because
$\alpha^T\epsilon \alpha=\epsilon$. We use $\epsilon_{AB}$ to raise and lower spinor indices: $\lambda_{A} = \epsilon_{AB} \lambda^B$.

\subsection{Massless particles and twistors}

The state space of a relativistic particle in Minkowski spacetime $\M$ is $\Gamma = \{ (x^{\mu}, p_{\nu})\} = T^*\M$, with symplectic form $\omega = dp_{\mu}\wedge dx^{\mu}$ subject to the constraint
\begin{equation}
C = \frac{1}{2}p^\mu p_\mu=0, 
\end{equation}
and the condition $p_0 > 0$. The constraint generates the orbits $(p_{\mu}, x^{\mu}) \rightarrow (p_{\mu}, x^{\mu} + p^{\mu}s)$, with $s \in \R$.

We write the symplectic form on the constraint surface in terms of the spinor $\lambda_A$ associated to $p_{\mu}$,
\bey
\theta = p_{\mu} d x^{\mu} = \bar{\lambda} \sigma_{\mu} \bar{\lambda}^{\dagger} d x^{\mu} = \bar{\lambda} (d \tilde{x}) \lambda.
\eey
Introducing the spinor $\mu_{A'}:= i\tilde{x}_{A'A}\lambda^A$, we obtain
\bey
\theta = i (\bar{\mu}_{A} d\lambda^A - \bar{\lambda}^{A'} d \mu_{A'}),
\eey
with symplectic form
\bey
\omega = i (d\bar{\mu}_{A} \wedge  d\lambda^A + d\bar{\mu}^{A'} \wedge d \lambda_{A'}). \label{symtw}
\eey
We  note that $\mu$ is an observable, since it is invariant under the action of the constraint. We also note that the phase transformation $\lambda \rightarrow e^{i\theta} \lambda$ induces   $\mu \rightarrow e^{i\theta} \mu$.

The hermiticity of $\tilde{x}$ implies that 
\bey
\bar{\lambda}^{A'}\mu_{A'} + \bar{\mu}_A \lambda^A = 0. \label{reality}
\eey

We define a {\em twistor} as a pair of spinors
\begin{equation}
Z = (\lambda^A, \mu_{A'})
\in \mathbb{C}^4.
\end{equation}
  In coordinate notation, we express the twistor as $Z^{\alpha}$. Its conjugate  $\bar{Z}_{\alpha} = (\mu_{A'}, \lambda_A)$. 

There is a natural inner product  on the  twistor space $\T$, given by $\bar{W}\cdot Z = \bar{W}_{\alpha}Z^{\alpha}$. In terms of this inner product, Eq. (\ref{reality}) becomes $\bar{Z}\cdot Z = 0$, and the symplectic form (\ref{symtw}) becomes simply $i(d\bar{Z}_{\alpha}\wedge dZ^{\alpha})$. 

We conclude that the reduced state space for a relativistic particle is the submanifold $\T_0 = \{Z\in \T, \bar{Z}\cdot Z = 0\}$ of twistor space, modulo the phase transformation $Z \rightarrow e^{i\theta}Z$: $\Gamma_{red} = \T_0/\sim$.

In this way, twistors arise directly from the spinorial description of massless particles, combining 
  position and momentum into a single geometric object. For Penrose, spinors are not merely a convenient representation of Lorentz transformations; they reveal the underlying twistor structure of null phase space.

A central aspect of the proposal is that classical spacetime should emerge only as a secondary or approximate notion. In twistor theory, spacetime points correspond not to fundamental objects, but to certain submanifolds in twistor space. Quantum amplitudes are then expected to be formulated directly in terms of holomorphic structures on twistor space rather than path integrals over metrics. This viewpoint naturally emphasises causal structure and massless propagation, and it suggests that locality itself may be emergent rather than fundamental. 

Although we are far from a twistor-based theory of quantum gravity, twistor methods have had major influence in mathematical physics. More generally, Penrose's programme remains one of the clearest attempts to formulate quantum gravity starting from causal and conformal structure rather than from quantised spacetime geometry.

\chapter{The weak gravity limit}
In this chapter, we will analyze linearized gravity and the quantisation of the Newtonian gravitational field. This serves, first, as a non-trivial testing ground for the methods presented earlier, since linearized gravity combines gauge structure, constraints, and causal propagation in a particularly subtle manner. Second, it is motivated by the growing interest in genuinely quantum effects mediated by gravity, especially proposals in which gravitational interactions generate entanglement between massive systems \cite{Bose17, Vedral17}. Such scenarios require a careful identification of the quantum degrees of freedom of the gravitational field and of the observables accessible to localized measurements.

\section{Naive linearized gravity}

In the Hamiltonian formulation,
the phase space $\Gamma$ for matter and gravity consists of points $(h_{ij}, \pi^{ij}, q_a, p^a)$, where $h_{ij}$ is a three metric on a manifold $\Sigma$, $\pi^{ij}$ is the conjugate momentum, $q_a$ are matter fields and $p^a$ their conjugate momenta.

The symplectic form on the phase space $\Gamma$ is
\begin{eqnarray}
\omega = \int d^3 x \delta \pi^{ij}(x) \wedge \delta h_{ij}(x) + \delta p^a(x) \wedge \delta q_a(x),
\end{eqnarray}
and the fundamental Poisson brackets are
\bey
\{ h_{ij}({\pmb x}), \pi^{kl}({\pmb x}')\} = (\delta_i^k \delta_j^l + \delta_i^l\delta_j^k) \delta^3({\pmb x}, {\pmb x}'), \hspace{0.5cm} \{q_a({\pmb x}), p^b({\pmb x}')\} = \delta_a^b \delta^3({\pmb x}, {\pmb x}').
\eey

Assuming that the coupling of matter to gravity involves no derivatives of the metric,
the Hamiltonian is
\begin{eqnarray}
H = \int d^3 x [ N(x) {\cal H}(x) + N^i(x) {\cal H}_i(x)], \label{Ham31}
\end{eqnarray}
where ${\cal H}$ is the Hamiltonian constraint and ${\cal H}_i$ is the momentum constraint,  
\begin{eqnarray}
{\cal H} &=& \kappa \frac{\pi^{ij}\pi_{ij} - \frac{1}{2} \pi^2}{\sqrt{h}} - \frac{\sqrt{h}}{\kappa} \; {}^3R + {\cal V} (h, q, p), \label{superH}\\
{\cal H}_i &=& -2 \nabla_j \pi^j{}_i + {\cal V}_i(h, q, p). \label{superM}
\end{eqnarray}
${\cal V}$ and ${\cal V}_i$ are the matter contribution to the constraints; they are  local functionals of the  matter field variables and   of  the three-metric. The constraint algebra is not affected by the addition of matter.

Next, we proceed to an analysis of the Hamiltonian around Minkowski spacetime, with no matter ($h_{ij} = \delta_{ij}, N = 1, N^i = 0, \pi^{ij} = 0, q_a = 0, p^a=0$). We introduce the formal expansion parameter $\lambda$, and we expand
\begin{eqnarray}
 h_{ij} = \delta_{ij}+ \lambda  \gamma_{ij},\hspace{0.3cm}
  N = 1 + \lambda n, \hspace{0.3cm}
  N_i = \lambda n_i, \hspace{0.3cm}
  \pi^{ij} \rightarrow \lambda \pi^{ij}. \label{linerr}
\end{eqnarray}
This expansion  leads to a description for the linearised gravitational field interacting with matter that is similar to the  electromagnetic theory. In fact, this gives the Hamiltonian descriptio of the zero-th order Fierz-Pauli Lagrangian (\ref{FPt}). The expansion (\ref{linerr})  breaks gauge invariance, because it incorporates {\em partial gauge-fixing} in the very first step, when it specifies the lowest order form of $N$ and $N_i$. The resulting system is not  a genuine parameterized system, so the symmetry of the full theory is lost.

For the matter degrees of freedom, we substitute ${\cal V}$ and ${\cal V}_i$ with $\lambda {\cal V}$ and $\lambda {\cal V}_i$. We keep terms up to $\lambda^2$ in the Hamiltonian
\begin{eqnarray}
H = \lambda \int d^3x {\cal E} + \lambda^2 \int d^3x \left[\kappa(\pi^{ij}\pi_{ij} - \frac{1}{2} \pi^2) + \kappa^{-1} V(\gamma)  \right. \nonumber \\
\left. +\kappa^{-1} n [\partial^2\gamma - \partial_i \partial_j
\gamma^{ij} + \kappa {\cal E} ] +n_i(-2
\partial_j \pi^{ji} + {\cal P}^i) + \gamma_{ij} I^{ij}
\right] \label{hamillin}
\end{eqnarray}
where $\gamma = \gamma_{ij}\delta^{ij}$,
\begin{eqnarray}
\EE(q, p) := {\cal V}(\delta_{ij}, q, p), \hspace{0.3cm}
{\cal P}_i(q, p) := {\cal V}_i(\delta_{ij}, q, p), \hspace{0.3cm}
I^{ij}(q, p) := \frac{\partial {\cal V}}{\partial h_{ij}}(\delta_{ij}, q, p),
\end{eqnarray}
and
\begin{eqnarray}
V := - \frac{1}{2} \partial_k\gamma_{ij} \partial^i \gamma^{kj} -   \frac{1}{4} \partial_k \gamma \partial^k \gamma + \frac{1}{2} \partial_i \gamma \partial_k \gamma^{ik} + \frac{1}{4} \partial_k \gamma_{ij} \partial^k \gamma^{ij}.
\end{eqnarray}
 Now the system is characterised by the constraints
 \begin{eqnarray}
 {\cal C} &=& \kappa^{-1}(\partial^2\gamma - \partial_i \partial_j
\gamma^{ij}) +  \EE = 0 \label{const1a}\\
{\cal C}_i &=& -2
\partial_j \pi^{ji} + {\cal P}^i = 0, \label{const2a}
 \end{eqnarray}
which coincide with the  superhamiltonian and supermomentum constraints to first order in $\lambda$.

We complete linearisation by setting $\lambda = 1$. Hence, we obtained
 a new constrained system with symplectic form
\begin{eqnarray}
\omega = \int d^3 x [ \delta \pi^{ij}(x) \wedge \delta \gamma_{ij}(x) + \delta p^a(x) \wedge \delta q_a(x)],
\end{eqnarray}
and Hamiltonian given by
\begin{eqnarray}
H =  \int d^3x \left[ \EE  + \kappa(\pi^{ij}\pi_{ij} - \frac{1}{2} \pi^2) + \kappa^{-1} V(\gamma)  + \gamma_{ij} I^{ij} +  n {\cal C}
+n_i  {\cal C}^i \right]. \label{hamillin2}
\end{eqnarray}
The Hamiltonian does not vanish. The gauge choice implicit in the linearisation has selected the background coordinate $t$ as the parameter of Hamiltonian time evolution.

\subsection{The reduced state space}

Next, we split the metric perturbation into components, in a way that simplifies the constraint equations. We express the Fourier transform $\tilde{\gamma}_{ij}({\pmb k})$ as
\begin{eqnarray}
\tilde{\gamma}_{ij} = -i\kappa (k_i \tilde{\chi}_j + k_j \tilde{\chi}_i) + \sqrt{2 \kappa} \tilde{w}_{ij} + \frac{1}{2} \tilde{\phi} \Pi_{ij}, \label{split}
\end{eqnarray}
where $\Pi_{ij} := \delta_{ij} - \frac{k_ik_k}{|{\pmb k}|^2} $ is the projector into the transverse subspace of $\tilde{\gamma}_{ij}$, $\tilde{w}_{ij}$ is the transverse-traceless (TT) component of $\tilde{\gamma}_{ij}$ ($\Pi^{ik} \Pi^{jl}w_{kl} = w^{ij}, \Pi^{ij}w_{ij} = 0$), and $\phi := \Pi^{ij}\gamma_{ij}$. Note that the uneven dependence of the components of $\gamma_{ij}$ on $\kappa$ does not affect physics, because $\kappa$ is not an expansion parameter. We have chosen this expression in order to simplify the correspondence of our results with the Newtonian description.

We substitute Eq. (\ref{split}) into the gravitational component of the symplectic potential $\theta_g = \int d^3 x \pi^{ij} \delta \gamma_{ij} = \int \frac{d^3k}{(2\pi)^3} \tilde{\pi}^{ij} \delta \tilde{\gamma}_{ij}$, to obtain
\begin{eqnarray}
\theta_g = \int \frac{d^3k}{(2\pi)^3} \left[(-2i\kappa k_i\tilde{\pi}^{ij})\delta \tilde{\chi}_j + \sqrt{2 \kappa} \tilde{\pi}^{kl} (\Pi_k^i\Pi_k^j - \Pi^{ij}\Pi_{kl})\delta \tilde{w}_{ij} + \frac{1}{2}\tilde{\pi}^{ij} \Pi_{ij}\delta \tilde{\phi}\right]
\end{eqnarray}
Hence, we identify the Fourier conjugate variables $\tilde{\pi}_\chi^{i}:= -2i\kappa k_i\tilde{\pi}^{ij}, \tilde{\pi}_{w}^{ij} := \sqrt{2 \kappa}\tilde{\pi}^{kl} (\Pi_k^i\Pi_l^j - \Pi^{ij}\Pi_{kl})$, and $\tilde{\pi}_{\phi} :=  \frac{1}{2}\tilde{\pi}^{ij} \Pi_{ij}$ to $\tilde{\chi}_i, \tilde{w}_{ij}$ and $\tilde{\phi}$, respectively. We Fourier-transform back to obtain
 \begin{eqnarray}
\theta_g = \int d^3x [\pi_{\chi}^{j}(x)\delta \chi_j(x) + \pi^{ij}_{w}(x) \delta w_{ij}(x) + \pi_{\phi}(x) \delta \phi(x)].
\end{eqnarray}
The constraints simplify when expressed in terms of the new variables
 \begin{eqnarray}
{\cal C} = \kappa^{-1}\nabla^2 \phi + \EE = 0 \label{constraint1}\\
{\cal C}^i = \kappa^{-1} \pi_{\chi}^i +{\cal P}^i = 0.
\end{eqnarray}
We also define the smeared constraints ${\cal C}(L) = \int d^3x L(x) {\cal C}(x)$ and ${\cal C}(\overrightarrow{L}) = \int d^3x L_i(x) {\cal C}^i(x)$

Next, we evaluate the various terms of the Hamiltonian on the constraint surface
\bey
\pi^{ij} \pi_{ij} - \frac{1}{2} \pi^2 = \frac{1}{2\kappa} \pi_w^{ij} \pi_{w ij} + \frac{1}{2\kappa^2} \pi_\chi^i \Delta_{ij} \pi_\chi^j + \frac{1}{\kappa} \pi_\phi \nabla^{-2} \partial_i \pi_\chi^i + \partial_i R^i, \\
 V(\gamma) = \frac{\kappa}{2}  \partial_kw_{ij} \partial^k w^{ij} -\frac{1}{8}  (\nabla \phi)^2   + \kappa^2 \partial_iK^i,
\eey
where we wrote 
\bey
\Delta_{ij} = - \nabla^{-2} \left( \delta_{ij} - \frac{3}{4} \partial_i \partial_j \nabla^{-2} \right) \\
\eey
and
$$
\nabla^{-2} f(x) = -  \int d^3x' \frac{f({\pmb x}')}{4\pi|{\pmb x} - {\pmb x}'|} .
$$
The expressions of the vector fields $R^i$ and $K^i$ are rather complex and will not be given here. They can be found in Ref. \cite{ALS}.

The constraint surface $C$ is spanned by the variables $w_{ij}, p_w^{ij}, \chi_i, \pi_{\phi}, q_a, p^a$. The TT components $w_{ij}, p_w^{ij}$ commute with the constraints, hence, they are true degrees of freedom. The vector field $\chi_i(x)$ satisfies
\bey
\{\chi_i(x) ,  {\cal C}(L) + {\cal C}(\overrightarrow{L})\} =  \kappa^{-1} L_i(x), \label{chill}
\eey
i.e., it is essentially a parameter of the gauge orbit generated by the supermomentum constraint, i.e., spatial diffeomorphisms.

Similarly, we find
\begin{eqnarray}
\{ \pi_\phi (x), \mathcal{C}(L) + \mathcal{C} (L) \} = - \kappa^{-1} \nabla^2 L(x)
\end{eqnarray}
This implies that the function
\begin{eqnarray}
\tau (x) = - \kappa \nabla^{-2} \pi_\phi (x)
\end{eqnarray}
satisfies
\begin{eqnarray}
\{\tau(x),          {\cal C}(L) + {\cal C}(\overrightarrow{L})  \} =  L(x), \label{tll}
\end{eqnarray}
i.e., $\tau(x)$ is a coordinate of the gauge orbit generated by the superhamiltonian constraint.

To understand the physical meaning of the quantities $\tau$ and $\chi_i$, we note that for the matter variables
\bey
\{q_a(x),  {\cal C}(L) + {\cal C}(\overrightarrow{L}) \} = \{q_a(x), H(L, \overrightarrow{L}) \}\\
\{p^a(x),  {\cal C}(L) + {\cal C}(\overrightarrow{L}) \} = \{p^a(x), H(L, \overrightarrow{L}) \}
\eey
where $H(L, \overrightarrow{L}) = \int d^3x [L(x)\EE(x) + L_i{\cal P}^i(x)]$ is the Hamiltonian for a parameterised relativistic scalar field in flat spacetime. It corresponds to evolution of the scalar field  along arbitrary foliations of the flat spacetime, each choice of foliation being in correspondence with a choice of $L$ and $L_i$.

Let $Q_a(x, t)$ be a solution to the field equations along a Lorentzian foliation; this solution is a functional of $q_a, p^a$, which define the initial conditions at $t = 0$. Consider a foliation with time-coordinate defined by the surface $t:=\tau(x)$ and space coordinates $y_i := \chi^i(x)$, and express the same solution as a functional $Q_a(t, y; \tau, \chi, q, p)$.
 By Eqs. (\ref{chill}, \ref{tll}), the action of the constraints on $q_a, p^a$ (evolution along the foliation) compensates on the action on $\tau, \chi$, so that the  functional $Q_a(t, y; \tau, \chi, q, p)$ commutes with the constraints. Hence, the true degrees of freedom for matter correspond to solutions of the equations of motion for matter, modulo all possible parameterisations of spacetime.

\subsection{Gauge fixing}

The meaning of the quantities $\pi_{\phi}$ and ${\cal \chi}_i$ as parameters specifying the choice of space-time coordinates
was discovered by Arnowitt, Deser, and Misner (ADM) \cite{ADM2}. In
an effort to
identify the proper dynamical degrees of freedom of
the gravitational field, they proposed a gauge-fixing condition
\bey
t = \tau(x), \hspace{0.5cm}  \chi_i(x) = 0 \label{ADMgauge}
\eey
that defines a coordinate system (a spacelike foliation) that is  as close as possible
to a Cartesian system of coordinates in the flat
space-time. Hamilton's equation of motion for $\pi^{ij}$ and $\gamma_{ij}$ with this condition imply that $n = n_i = 0$.

Eq. $t = \tau(x)$ defines the spacelike surfaces $\Sigma_t$ of constant $t$, in a spacetime described by coordinates $(t, x^i)$. On the other hand $\chi^i(x)$ defines to  leading order a deformation of the (flat) spatial coordinates by $x^i \rightarrow x^i + \chi^i(x)$, due to the presence of the gravitational field. Hence, the  ADM gauge implies the use of the background spatial coordinate system.

In the ADM gauge, the Hamiltonian becomes a sum of three terms
\bey
H_{ADM} = \int d^3x  \left[ \EE(x) +   {\cal H}_{gsi}(q, p)+{\cal H}_{GW}(w, \pi_w, q, p) \right], \label{hadm}
\eey
where
\bey
{\cal H}_{gsi} = \frac{\kappa}{2} \left[ -\frac{1}{4} \partial_i \nabla^{-2} \mathcal{E} \partial^i \nabla^{-2} \mathcal{E} + \mathcal{P}^i \Delta_{ij} \mathcal{P}^j - \Pi_{ij} \nabla^{-2} \mathcal{E} I^{ij} \right],
\eey
is the Hamiltonian density of gravitational self-interaction, and
\bey
{\cal H}_{GW} =   \frac{1}{2}  \pi_w^{ij} \pi_{wij} +  \frac{1}{2}  \partial_k w_{ij}\partial^k w^{ij} + \sqrt{2 \kappa} w_{ij}I^{ij}
\eey
 is the Hamiltonian density for gravitational waves, including a term for the interaction of gravitational waves with matter.

In the ADM gauge, the lapse vector $N = 1$ and the shift vector $N_i = 0$. However, the expression for the three metric is rather complex. Even for $w_{ij} = 0$,
so that the spacetime metric is a non-local functional of the Newtonian potential  $\phi(x) = -\kappa \nabla^{-2}\EE$,
\bey
ds^2 = - dt^2 + dx^i dx^j  \left[\delta_{ij} + \frac{1}{2} \Pi_{ij}\phi \right]. \label{metricADM}
\eey
The three metric is simpler in the {\em isotropic gauge}, determined by
\bey
\chi_i - \frac{1}{4} \nabla^{-4} \partial_i \EE = 0,  \hspace{0.3cm}  \tau + \frac{\kappa}{4} \nabla^{-4}\nabla_i {\cal P}^i = t.
\eey
%The Hamiltonian in the isotopic gauge is
%\bey
%H_{iso} = H_{ADM} - \frac{\kappa}{16}\int d^3x (\nabla \cdot {\cal P}) \nabla^{-4} (\nabla \cdot {\cal P}) - \frac{\kappa}{2} \int d^3 x \nabla^{-4}\EE \partial_i \partial_j I^{ij}. \label{hiso}
%\eey
%In Eq. (\ref{hiso}), $H_{ADM}$ stands for the formal expression (\ref{hadm}). However, the fields entering $H_{ADM}$ is (\ref{hiso}) are different as they are expressed with respect to the ADM coordinate systems.
In the isotropic gauge, the three-metric for $w_{ij} = 0$ is
\bey
h_{ij} = \delta_{ij} \left(1 + \frac{1}{2} \phi \right).
\eey

The coordinate transformation between the gauges is
\bey
t_{iso} = t_{ADM} +\frac{\kappa}{4} \nabla^{-4}\nabla_i {\cal P}^i,
\hspace{1cm}
x^i_{iso} = x^i_{ADM}- \frac{1}{4} \nabla^{-4} \partial^i \mathcal{H},
\label{isoadm}
\eey
where $\mathcal H$ and $\mathcal P_i$ are the energy and momentum densities. Interestingly, the transformation depends on the matter degrees of freedom themselves.

In classical geometrodynamics, this poses no problem. Different gauge choices correspond to equivalent Hamiltonian descriptions because the spacetime metric provides a common geometrical structure relating the different canonical evolutions \cite{ADM1}. In quantum theory, however, the situation is much less clear. There is no gauge-independent spacetime geometry that can serve as a common standard of comparison between different Heisenberg evolutions. Equation (\ref{isoadm}) becomes particularly problematic: the quantities $\mathcal H$ and $\mathcal P_i$ are now operators, whereas spacetime coordinates are, by definition, $c$-numbers. There is no mathematically consistent way of defining a transformation between the two coordinate systems.

\section{Weak gravity as a parameterised field theory}

The problem with the linearisation expansion of Sec. 7.1 is that it involves a partial gauge fixing, as it fixes  $N$ and $N_i$ to lowest order. The time-reparameterisation symmetry of the initial Hamiltonian is thereby lost.   This procedure breaks general covariance and misrepresents the causal structure of the system.
An alternative approach to  linearised gravity that preserves time-reparameterisation symmetry originates from Kucha\v{r} \cite{Kuchar70}---see also \cite{ALS}.

In this expansion scheme is that we expand the constraints (\ref{superH}, \ref{superM}) with respect to the perturbations of the metric and the momentum around the phase space point $(\delta_{ij}, 0)$. We do not expand the lapse and a shift variables around a background value. Thus, we avoid partial gauge fixing, and the Hamiltonian still vanishes due to the first-class constraints. The difference is that we have to keep terms to second order in the $\lambda$ expansion   to obtain meaningful weak-field dynamics.

We write $h_{ij} = \delta_{ij} + \lambda \gamma_{ij}$, $\pi^{ij} \rightarrow \lambda \pi^{ij}$, where $\lambda$ is an expansion parameter. We keep terms up to the first two orders in $\lambda$, to obtain
 \bey
{\cal H} &=& \lambda  \left[ \kappa^{-1}(\partial^2\gamma - \partial_i \partial_j
\gamma^{ij})
+  \EE\right]
\nonumber \\
&+& \lambda^2 \left[ \kappa(\pi^{ij}\pi_{ij} - \frac{1}{2} \pi^2) + \kappa^{-1} V(\gamma)  + \gamma_{ij} I^{ij} + \kappa^{-1} \partial_i J^i(\gamma) \right], \label{constr1}
\\
{\cal H}^i &=& \lambda \left( -2
\partial_j \pi^{ji} + {\cal P}^i \right)+ \lambda^2 \left[ L^i + M^{ijk} \gamma_{jk}\right], \label{constr2}
\end{eqnarray}
where
\bey
L^i = -\pi^{ij} \partial_j\gamma -2\partial^k \left( \gamma^{ij} \pi_{jk} \right) + \pi^{jk} \partial^i \gamma_{jk},
\eey
 $M^{ijk} = \partial {\cal V}^i/\partial h_{ij}$, and
\bey
J^i = \frac{1}{2} \gamma \partial^i\gamma + \frac{1}{2} \gamma \partial_j \gamma^{ij}  -\gamma^{jk}\partial^i\gamma_{jk}   + \partial_j(\gamma^{jk}\gamma^{i}{}_k -\gamma \gamma^{ij}).
\eey

We make a canonical transformation from $\phi, \pi_{\phi}$ to $  \tau = - \kappa \nabla^{-2} \pi_\phi$ and $ \pi_\tau = \kappa^{-1} \nabla^2 \phi$. By Eqs. (\ref{constr1}---\ref{constr2}), $ \pi_\tau + \mathcal{E} = O(\lambda) $, and $\pi_{\chi}^i + \kappa {\cal P}^i = O(\lambda)$, so we can remove $\pi_{\tau}$ and $\pi_{\chi}^i$ from the $\lambda^2$ terms in Eqs. (\ref{constr1}, \ref{constr2}) up to terms of order $\lambda^3$.

Hence, up to terms of order $\lambda^3$, the two constraints can be solved for $\pi_{\tau}$ and $\pi_{\chi}^i$, as
\begin{eqnarray}
\pi_{\tau} + {\cal G}^0(\chi, \tau, w, \pi_w, q, p) = 0\\
\pi_{\chi}^i + {\cal G}^i(\chi, \tau, w, \pi_w, q, p) = 0
\end{eqnarray}
in terms of the functions
\begin{eqnarray}
 {\cal G}^0(\chi, \tau, w, \pi_w, q, p) = \EE+ {\cal H}_{gsi}+{\cal H}_{GW}
 - 2 \kappa \partial_i \chi_j I^{ij} + \nabla^2 \tau \nabla^{-2}\partial_i{\cal P}^i + \kappa^{-1} \partial_i \bar{J}^i \nonumber 
 \\
{\cal G}^i(\chi, \tau, w, \pi_w, q, p) = \kappa {\cal P}^i + \kappa L^i + \sqrt{2} \kappa^{3/2} M^{ijk} w_{jk} - 2 \kappa^2 M^{ijk} \partial_j \chi_k - \kappa^2 M^{ijk} \Pi_{jk} \nabla^{-2}\EE \nonumber
\end{eqnarray}
where
\bey
\bar{J}^i = J^i +\kappa R^i +\kappa^2 K^i.
\eey

The symplectic potential on the constraint surface is
\begin{eqnarray}
\theta = \int d^3x \left[ p^a(x)\delta q_a(x) + \pi_w^{ij}(x)\delta w_{ij}(x) - {\cal G}^0 \delta \tau -{\cal G}_i \delta \chi^i\right], \label{symplecticpot}
\end{eqnarray}
and it describes a  parameterised field theory.  This means that the deparameterization program that---most probably---fails in full GR, is applicable to the weak gravity limit.
 
To deparameterize, 
we set $\tau$ and $\chi^i$ equal to specific functions of $x$ and $t$, say, $T_t(x)$ and $X^i_t(x)$. The surfaces of constant $t$ are defined by the condition $\tau = T_t(x)$, while the spatial coordinates at each $\Sigma_t$ are given by $x^i + X^i_t(x)$.
Then, the action for the reduced system
is
\bey
S[q,p, w, \pi_w] = \int dt d^3 x \left[\pi_w^{ij}\dot{w}_{ij} + p^a\dot{q}_a - \dot{T}_t(x)  {\cal G}^0 -\dot{X}^i_t {\cal G}_i\right]. \label{action}
\eey

The associated Hamiltonian is
\bey
H = \int d^3 x \left[\dot{T}_t(x)  {\cal G}^0  + \dot{X}^i_t {\cal G}_i\right]. \label{hamtsont}
\eey
Physical predictions should be the same irrespective of gauge fixing. Each deparameterisation leads to a different spacetime picture for the solutions to the equations of motion. However, physics is not affected. In classical GR, solutions  that correspond to different deparameterisations must be related by  a spacetime diffeomorphism.
Note that if we choose $X_t(x)$ to be $t$-independent, then only ${\cal G}^0$ contributes in the action. If also $T_t(x) = t$, then the Hamiltonian (\ref{hamtsont}) coincides with the ADM Hamiltonian.

The action (\ref{action}) contains complete information about the interaction of weak gravity with matter.  The transverse-traceless part that corresponds to gravitational waves is insignificant  if no gravitational waves exist at the initial moment of time and if the density and acceleration of matter is sufficiently low.

The Newtonian regime  is obtained by keeping only the terms of energy density for matter, and dropping all terms that involve ${\cal P}^i, I^{ij}$ and $M^{ijk}$. Then,

\begin{eqnarray}
 {\cal G}^0(\chi, \tau, w, \pi_w, q, p) &=& \EE -\frac{\kappa}{8} \partial_i \nabla^{-2}\EE\partial^i \nabla^{-2}\EE  + \kappa^{-1} \partial_i \bar{J}^i\label{G0N}\\
{\cal G}^i(\chi, \tau, w, \pi_w, q, p) &=& \kappa L^i.
\end{eqnarray}

In  the ADM   gauge,
\bey
H = \int d^3x \EE(x) - \frac{G}{2} \int d^3x d^3x' \frac{\EE(x)\EE(x')}{|{\pmb x} - {\pmb x}'|}, \label{intham}
\eey
as expected in Newtonian gravity.

\section{Effective QFT description}

\subsection{Quantisation in the ADM gauge}

Weak gravity interacting with matter is one of few cases where deparameterizationcan be meaningfully implemented. In the ADM gauge, quantisation leads straightforwardly to an effective QFT. The fundamental variables are local fields $q_a(x), p^a(x), w_{ij}(x), \pi_w^{ij}(x)$, that enter the Hamiltonian (\ref{hadm}). This system can be treated by standard perturbative QFT quantisation.

The unperturbed system consists of the matter fields $\hat{q}_a(x), \hat{p}^a(x)$ evolving under the Hamiltonian $\hat{H} = \int d^3x \EE([\hat{q}(x), \hat{p}(x)]$ in the Hilbert space ${\cal H}_{mat}$ and the gravitons under the Hamiltonian
\bey
\hat{H}_{gr} := \int d^3x \left[\frac{1}{2}  \hat{\pi}_w^{ij} \hat{\pi}_{wij} +  \frac{1}{2}  \partial_k \hat{w}_{ij}\partial^k \hat{w}^{ij} \right],
\eey
in the Hilbert space ${\cal H}_{grav}$.
The unperturbed system is then described by the Hilbert space ${\cal H}_{mat} \otimes {\cal H}_{grav}$.

There are two interaction terms. The term $\hat{w}_{ij} \hat{I}^{ij}$ generates the coupling of gravitons to matter and it is consistent up to  tree level. The second term is a non-local self-interaction term $\hat{ H}_{gsi}$ that corresponds to the gravitational self-interaction of matter. It involves components of the stress-energy tensor that are not well defined as operators in standard QFT, thus requiring appropriate regularisation.

In the Newtonian regime, the graviton field decouples from matter and matter can be treated using non-relativistic QFT. For concreteness, assume the matter fields to be fermions. In the non-relativistic limit, they are described by field operators $\hat{\psi}_a(x)$ and $\hat{\psi}_a^{\dagger}(y)$, such that
\bey
[\hat{\psi}_a({\pmb x}), \hat{\psi}^{\dagger}_b({\pmb x}') ]_+ = \delta_{ab}  \delta^3({\pmb x}, {\pmb x}').
\eey
The non-relativistic fields can be expressed in terms of fermionic creation and annihilation operators,
\bey
\hat{\psi}_a({\pmb x}) = \int \frac{d^3p}{(2\pi)^3} e^{i{\pmb p} \cdot{\pmb x}} \hat{c}_a({\pmb p}), \hspace{0.5cm}\hat{\psi}^{\dagger}_a({\pmb x}) = \int \frac{d^3p}{(2\pi)^3} e^{-i{\pmb p} \cdot{\pmb x}} \hat{c}_a^{\dagger}({\pmb p})
\eey
that satisfy the canonical anti-commutation relations.

Given the quantum fields above, we define the particle density functions
\bey
\hat{n}_a({\pmb x})  = \hat{\psi}^{\dagger}_a({\pmb x}) \hat{\psi}_a({\pmb x}).
\eey
The matter Hamiltonian is
\bey
\hat{H}_{mat} = - \sum_a \int d^3x \frac{1}{2m_a}  \hat{\psi}^{\dagger}_a({\pmb x})  \nabla^2 \hat{\psi}_a({\pmb x}) + \frac{1}{2}\sum_{a \neq b}\int d^3x d^3x' V_{ab}({\pmb x} - {\pmb x'}) \hat{n}_a({\pmb x}) \hat{n}_b({\pmb x}),
\eey
where $m_a$ is the mass of a particle of type $a$ and $V_{ab}({\pmb r})$ is a short-range interaction potential between the different matter particles.

The gravitational self-interaction Hamiltonian $\hat{H}_{gsi}$ is expressed in terms of the mass density
\bey
\hat{\mu} ({\pmb x}) = \sum_a m_a \hat{n}_a({\pmb x}), \label{massdens}
\eey
because in the non-relativistic regime the mass density is the dominant term in the energy density. Then,
\bey
\hat{H}_{gsi} = - \frac{G}{2} \int d^3x d^3x' \frac{\hat{\mu} ({\pmb x})\hat{\mu} ({\pmb x}')}{|{\pmb x} - {\pmb x}'|}. \label{hgsiq}
\eey
The Hamiltonian must be regularised through the introduction of a cut-off $\ell$ in the denominator so that $|{\pmb x} - {\pmb x}'|$ is substituted by $\sqrt{\ell^2 +|{\pmb x} - {\pmb x}'|^2}$. Then, there is a self-energy contribution $-\frac{Gm_a^2}{2\epsilon}$ to each particle that is absorbed into a mass renormalisation of the Hamiltonian.

This expression from the Hamiltonian can also be derived in non relativistic physics, by interpreting the gravitational interaction as an inter-particle potential.

\subsection{Bipartite systems}

The presence of an external constraints may allows us to express a subset of the matter Hilbert space into  a tensor product ${\cal H}_1 \otimes {\cal H}_2$, where ${\cal H}_1$ describes degrees of freedom localised in a region $C_1$ and ${\cal H}_2$ degrees of freedom in a region  $C_2$. This can be achieved by adding to the matter Hamiltonian an external potential $U(x)$ that vanishes only in $C_1$ and $C_2$ and takes very large values everywhere else. Hence, the subspace of sufficiently small eigenvalues of the Hamiltonian can be approximated by the Fock space for fields on $C_1 \times C_2$, hence, to be of the form ${\cal H}_1 \otimes {\cal H}_2$---see, Ref. \cite{AnHu20} for an example.
Then, we can write the number densities of each component as $\hat{\nu}_a^{(1)}$ and $\hat{\nu}_a^{(2)}$, and the associated mass densities  $\hat{\mu}^{(1)}$ and $\hat{\mu}^{(2)}$.

Assuming that the separation between the subsystems is much larger than the range of the interaction potential, the bipartite system is described by the Hamiltonian,
\bey
\hat{H} = \hat{H}_1 \otimes \hat{I} + \hat{I} \otimes \hat{H}_2 + \hat{H}_{int}, \label{Hamilbi}
\eey
where
\bey
\hat{H}_i &=&  - \sum_a \int_{C_i} d^3x \frac{1}{2m_a}  \hat{\psi}^{\dagger}_a({\pmb x})  \nabla^2 \hat{\psi}_a({\pmb x}) + \frac{1}{2}\sum_{a \neq b}\int d^3x d^3x' V_{ab}({\pmb x} - {\pmb x'}) \hat{n}_a^{(i)} ({\pmb x}) \hat{n}^{(i)} _b({\pmb x})\nonumber \\
 &-& \frac{G}{2}  \int_{C_i} d^3x \int_{C_i} d^3x' \frac{\hat{\mu}^{(i)} ({\pmb x})\hat{\mu}^{(i)} ({\pmb x}')}{|{\pmb x} - {\pmb x}'|} \\
\hat{H}_{int} =   &-&G \int_{C_1} d^3x  \int_{C_2} d^3x' \frac{\hat{\mu}^{(1)} ({\pmb x})\hat{\mu}^{(2)} ({\pmb x}')}{|{\pmb x} - {\pmb x}'|}. \label{hint}
\eey
The Hamiltonian (\ref{Hamilbi}) provides the foundation for any model about the gravitational interaction between separated quantum systems.  

\section{Spacetime properties}
\subsection{Proper time operators}
In the Newtonian regime, gravity possesses no independent dynamical degrees of
freedom. The metric perturbations are determined by the gravitational
constraints as functionals of the matter variables. After quantisation of the
matter system, these relations may be interpreted as Heisenberg-type operator
equations. Thus, the metric perturbations become operators, but they act
entirely on the Hilbert space of matter: no independent gravitational Hilbert
space or graviton degrees of freedom are introduced.

Consider the general linearized metric
\begin{equation}
d\hat{s}^{\,2}
=
-\left(1+2\hat n\right)dt^2
+2\hat N_i\,dt\,dx^i
+\left(\delta_{ij}+\hat\gamma_{ij}\right)dx^i dx^j .
\label{generalmetric}
\end{equation}
Here, $\hat n$, $\hat N_i$, and $\hat\gamma_{ij}$ are operator-valued
functionals of the matter mass density, momentum density, and stress tensor.
Their explicit form depends on the choice of gauge.  

In the ADM gauge considered here,
\begin{equation}
\hat n=0,\qquad \hat N_i=0,\qquad
\hat\gamma_{ij}
=
8\pi G\,\Pi_{ij}\nabla^{-2}\hat\mu,
\label{metricADMop}
\end{equation}

The metric operator   generates observables associated with spacetime measurements. Consider a
test particle whose trajectory in a chosen coordinate system is
$\pmb x(t)$, with endpoints $\pmb x_0$ at $t=0$ and $\pmb x_f$ at $t=T$.
For the general metric (\ref{generalmetric}), the proper time along this
trajectory is, to first order in the metric perturbations,
\bey
\hat\tau
={}&
\tau_0\hat I
+
\int_0^Tdt\,
\frac{1}{\sqrt{1-\dot{\pmb x}^{\,2}}}
\left[
\hat n
-\hat N_i\dot x^i
-\frac12\hat\gamma_{ij}\dot x^i\dot x^j
\right]_{\pmb x=\pmb x(t)},
\label{ptgen}
\\
\tau_0
={}&
\int_0^Tdt\,\sqrt{1-\dot{\pmb x}^{\,2}} .
\eey
Thus, the proper-time correction is a time-smeared matter observable,
$$
\delta\hat\tau:=\hat\tau-\tau_0\hat I .
$$
Operationally, it could be recorded by an internal degree of freedom carried
along the trajectory, for example, through the phase accumulated by a quantum
system acting as a clock.

 \begin{exercise}
Evaluate $\delta\hat\tau$ for the ADM gauge.
\end{exercise}

Although classical proper time is a coordinate-invariant quantity, it does
not follow immediately that the operator (\ref{ptgen}) is gauge independent.
Classical invariance requires the metric, the trajectory, and its endpoints
to be transformed together. By contrast, Eq.~(\ref{ptgen}) is constructed
by evaluating a gauge-fixed metric operator along an externally prescribed
coordinate trajectory $\pmb x(t)$. Unless this trajectory and its endpoints
are specified relationally in terms of physical matter degrees of freedom,
they do not themselves define gauge-invariant observables.

This can be seen directly. Write the metric as
$g_{\mu\nu}=\eta_{\mu\nu}+h_{\mu\nu}$ and let $z^\mu(\tau_0)$ denote the
background trajectory. Its linearized proper-time correction is
\begin{equation}
\delta\tau[h;z]
=
-\frac12\int d\tau_0\,
h_{\mu\nu}(z)u^\mu u^\nu ,
\qquad
u^\mu=\frac{dz^\mu}{d\tau_0}.
\end{equation}
Under an infinitesimal gauge transformation,
\begin{equation}
h_{\mu\nu}
\longrightarrow
h_{\mu\nu}
-\partial_\mu\xi_\nu-\partial_\nu\xi_\mu,
\end{equation}
while keeping the coordinate trajectory fixed, its variation is
\begin{equation}
\delta_\xi(\delta\tau)
=
\left[u^\mu\xi_\mu\right]_{0}^{T}
-
\int d\tau_0\,a^\mu\xi_\mu,
\qquad
a^\mu=\frac{du^\mu}{d\tau_0}.
\label{gaugetau}
\end{equation}
This does not vanish for a general trajectory and a general gauge
transformation. It vanishes under restricted conditions, for example when
the background trajectory is inertial and the gauge transformation vanishes
at the endpoints.

At the quantum level, the situation is more delicate because the
transformation between two gauge-fixed representations of the metric may
itself depend on the matter operators. The lapse, shift, and spatial metric
are redistributed under such a transformation, while the prescribed
coordinate trajectory need not possess a corresponding operator
transformation. There is therefore no general reason to expect
Eq.~(\ref{ptgen}), viewed as an operator on the matter Hilbert space, to be
gauge independent.

\subsection{Possible resolutions}

The preceding analysis establishes a \emph{prima facie} case that temporal quantum observables become gauge dependent in the presence of macroscopic quantum sources. There are three possible responses.

\subsubsection*{Preferred coordinate systems}

One may postulate a preferred class of coordinates, such as the ADM coordinates, and quantise matter relative to the corresponding approximately Minkowskian foliation. Heisenberg operators defined in this gauge would then be interpreted directly as physical observables. This preserves the ordinary framework of QFT and avoids functional evolution between arbitrary foliations.

The price is an additional spacetime structure that is difficult to reconcile with GR. The ADM coordinates are defined by the dynamical metric and therefore depend on the matter distribution; they cannot simply be identified with inertial coordinates of the background spacetime. Moreover, complete gauge fixing leaves no residual symmetry from which the gauge transformations of the linearised theory can be reconstructed. The physical status of the preferred foliation therefore remains unclear.

\subsubsection*{Gauge-invariant observables}

A second possibility is to admit only operators that commute with the
constraints. Proper times assigned to coordinate trajectories would not
themselves be observables; experiments would instead have to be described
through relational Dirac observables.

This is the viewpoint of Dirac quantisation and the evolving-constants
programme. Deparameterisation may introduce internal clock variables, but
physical predictions must ultimately be expressed through quantities that are
invariant under the gauge transformations generated by the constraints. A
local field may serve as a clock, but its value at a coordinate point is not
itself a Dirac observable; the corresponding relational observable is
generally a nonlocal functional of the canonical data. Time evolution then
becomes relational, and spacetime coordinates lose their direct operational
significance. Recovering the approximately local, time-resolved description
used in laboratory QFT is consequently difficult and depends non-uniquely on
the choice of clocks and reference systems.

\subsubsection*{An enlarged quantum framework}

A third possibility is to employ parametrised quantum field theory, treating
the embedding variables ${\cal X}^{\alpha}(\pmb x)$ as canonical variables
and quantising them together with the matter and gravitational degrees of
freedom. They satisfy
\begin{equation}
[\hat{\cal X}^{\alpha}(\pmb x),
 \hat P_{\beta}(\pmb x')]
=
i\delta^\alpha_{\beta}\delta^3(\pmb x-\pmb x'),
\end{equation}
with analogous commutation relations for the remaining canonical pairs. In
the Schr\"odinger representation, the constraints formally take the
many-fingered-time form
\begin{equation}
i\frac{\delta\Psi}{\delta{\cal X}^{\alpha}(\pmb x)}
=
{\cal G}_{\alpha}
\left(
{\cal X},q,
-i\frac{\delta}{\delta q}
\right)\Psi ,
\end{equation}
describing changes of the quantum state under deformations of the embedding.

This formulation makes transformations between arbitrary spacelike
foliations part of the quantum theory and could provide a framework for
comparing different representations of temporal observables. Quantising the
embedding variables does not, however, by itself make spacetime observables gauge
invariant: the constraints must still be imposed and physical quantities
must commute with them.

There is also a serious representational obstruction. Even for a free scalar
field in Minkowski spacetime, evolution between generic spacelike embeddings
is not unitarily implementable in the standard Fock representation in more
than two spacetime dimensions
\cite{Kuchar88,Kuchar89,ToVa1,ToVa2,Var06}. This does not exclude every
possible quantisation of the parametrised theory, but it implies that a
consistent implementation of arbitrary-foliation evolution would have to
depart substantially from the conventional Fock-space framework of
perturbative QFT.

%Several possibilities have been explored. Polymer quantisation methods related to loop quantum gravity provide one candidate framework \cite{LaVa, TT}, though these %representations are non-continuous and do not admit ordinary Hamiltonian generators. My preference is for a histories theory formulation, in which one constructs the Hilbert %space for histories \cite{Sav10}, by considering the spacetime version of the commutation relations (\ref{DDD1}---\ref{DDD3}).

\section{The weak-field challenge}

A predictive and spacetime-covariant description of quantum experiments in
linearized gravity constitutes an important test for any programme in quantum
gravity. A viable theory must recover not only the dynamics of quantum matter,
but also a coherent operational notion of spacetime and concrete predictions
for spacetime observables in regimes far removed from the Planck scale.

Different approaches may lead to inequivalent predictions for macroscopic
quantum phenomena, particularly for spacetime observables and the role of
reference frames. Although such effects may lie beyond present experimental
capabilities, increasingly precise control of macroscopic quantum systems
appears more accessible than direct probes of Planck-scale physics. Weak-gravity
quantum experiments may therefore provide a realistic window into the
conceptual structure of quantum gravity.

More fundamentally, the preceding analysis shows that the problems of time and
observables are not confined to the Planck regime, black holes, or cosmology.
They arise already in nearly flat spacetime whenever quantum matter is used to
define clocks, reference frames, and spacetime measurements. Any proposed
resolution must therefore account for the weak-gravity regime in which ordinary
quantum experiments can, at least in principle, be performed. Progress in such
experiments may ultimately help distinguish between competing approaches and
constrain the space of viable solutions.

\chapter{Outlook}

The problem of time in quantum gravity is often reduced to a single contrast. In ordinary quantum theory, time is an external parameter with respect to which states evolve; in general relativity, it is part of the dynamical geometry. This contrast is fundamental, but it is only one manifestation of a broader conflict. Quantum theory and general relativity differ across the full range of structures associated with time: the definition of events, the relation between dynamics and causal order, the status of clocks, and the applicability of operational concepts.

The distinction developed in these notes between causal order, temporal measure, and the present reveals the extent of this conflict. Causal order determines which events can influence which others. Temporal measure assigns durations by means of physical clocks. The present concerns the occurrence of events and, plausibly, the distinction  between possible outcomes and actual records. Established theories combine these three elements in different ways. General relativity incorporates causal order and temporal measure into a dynamical Lorentzian geometry, but contains no distinguished present. Quantum theory presupposes an external temporal order for its dynamics, while measurement produces definite outcomes at particular stages of an experiment. Quantum field theory adds relativistic causality, but ordinarily inherits its causal structure from a fixed background.

Much of the literature addresses the most visible manifestations of this conflict: the vanishing Hamiltonian, the Wheeler--DeWitt equation, the choice of an internal clock, and the recovery of a Schr\"odinger equation in a semiclassical approximation. These are important problems, but they are not necessarily the deepest ones. Any satisfactory treatment of the problem of time must ultimately incorporate an account of events: it must explain how definite occurrences are represented, how they are temporally ordered, and how probabilities are assigned to them. This is not to suggest that a theory of events, by itself, solves the problem of time. It does imply, however, that recovering some form of time evolution is not sufficient. The problem of time is inseparable from questions in quantum foundations, particularly the emergence of definite outcomes.

The four types of approach examined in these notes respond differently to this conflict.

Type I approaches retain the temporal and causal structure of ordinary quantum field theory. A spin-2 field is quantized on a background spacetime, and its dynamics, commutators, time ordering, and asymptotic states are defined relative to that background. This framework is highly successful perturbatively, but it presupposes the temporal structure that a quantum theory of geometry should explain. The background light cone organizes the quantum theory, while the light cone of the putative physical metric is expected to emerge from it. No established mechanism guarantees this emergence beyond the perturbative regime. The $S$-matrix bypasses questions about finite-time events and dynamical causal structure; it does not answer them.

Type II approaches confront the problem directly by quantizing geometry. In canonical formulations, the loss of external time leads to the frozen formalism. A complete deparametrization could, in principle, identify a preferred internal clock and express the remaining variables as evolving relative to it. General relativity does not appear to admit such a construction globally. Time must therefore be recovered locally, relationally, or semiclassically. Different choices of clock may yield inequivalent quantum descriptions, while an approximate clock does not by itself reconstruct causal order or spacetime localization. Type II approaches expose the problem in its sharpest form, but have not produced a unique and fully spacetime-covariant resolution.

Type III approaches retain a classical geometry while coupling it to quantum matter or deriving its dynamics from quantum effects. Their temporal difficulties largely reproduce those of Types I and II. Induced-gravity models require a background causal structure in order to define the quantum matter theory, even though the dynamics of that structure is then supposed to emerge from the matter effective action. Hybrid classical--quantum theories face a different but related challenge: they must explain how the constraint structure of general relativity survives when matter and gravity are treated as fundamentally different kinds of system. If it does not survive, they must specify what replaces it and in what sense the resulting theory still defines a spacetime with a consistent causal structure.

Type IV approaches modify the structures from which geometry or quantum theory is constructed. Those that treat causal relations or histories as fundamental appear, in my view, to address the problem at a more appropriate level. They do not begin by searching for a clock variable within an otherwise timeless formalism. Instead, they ask whether temporal order, causal relations, and complete histories should belong to the foundations of the theory. The price is substantial: probabilities must be assigned to temporally extended alternatives, the notion of time evolution becomes more difficult to formulate, and the standard Hilbert-space framework may no longer be adequate. Type IV approaches are conceptually promising precisely because they accept this greater technical burden, but that burden has also made concrete progress more difficult.

The weak-gravity regime may provide the most promising arena in which these alternatives can be tested. Reaching the Planck energy in an accelerator lies far beyond foreseeable experimental capabilities. By contrast, the preparation of increasingly massive objects in quantum superpositions is already an active experimental programme. Superpositions of objects with masses potentially approaching $10^{10}$ atomic mass units over micrometre scales, for which gravitational quantum effects may be distinguishable, are at least conceivable as an extension of present techniques \cite{ArHo14, BFG25}. Primordial gravitational waves, if they exist and can be shown to have a quantum origin, would provide another important probe; they, too, belong to the weak-field regime.

This changes the experimental meaning of the problem of time. Its observable consequences need not be confined to Planck-scale fluctuations, singularities, or black-hole interiors. Quantum clocks, time-of-arrival observables, multi-event correlations, quantum reference frames, and the temporal ordering of measurements may behave differently when the sources of the gravitational field are placed in macroscopic quantum superpositions. Such experiments could reveal assumptions about time that remain hidden in calculations involving only spatial observables, stationary states, or asymptotic scattering amplitudes.

A decisive experiment need not reveal a large quantum-gravitational correction. It may instead discriminate between different temporal descriptions of the same weak-field process: evolution relative to a fixed background time, or defined relationally by physical clocks coupled to a quantum source, or  organized by an intrinsic causal structure. Competing programmes may agree on conventional matter observables while disagreeing on  causal ordering of events, or clock correlations. Quantum temporal observables may therefore probe foundational differences more directly than tests based only on forces, phases, or entanglement.

The problem of time is neither a single mathematical obstruction nor a question that can be settled by choosing a convenient clock. It expresses a fundamental tension between the way quantum theory describes possible and actual events and the way general relativity identifies temporal and causal structure with dynamical geometry. Any satisfactory resolution must therefore explain not only how a notion of evolution is recovered, but also how events, temporal relations, and causal structure are represented within the full theory. Whether this account will emerge from an existing research programme or it will require new, radical   ideas   remains to be seen.

%For example, such descriptions have been proposed in the following contexts. 
 \chapter*{Acknowledgements}
 
I thank the students of the COST Training School ``Time, Causality and Memory in Quantum Physics'' for their engagement, thoughtful questions, and stimulating discussions, which helped shape these notes. I also thank the local organizing team at Stockholm University for their hospitality and for making the school possible. I am especially grateful to Sara Butler, whose careful editing of the original manuscript, harmonization of the text with the lectures as delivered, and numerous valuable suggestions contributed substantially to the preparation of the final version. She also prepared Figures~2.1 and~2.2.

I thank Maximilian Lock and Niyusha Hosseini for explaining recent developments in the Page--Wootters formalism, and Fabio Costa for several informative discussions about the notion of events in quantum theory.

Over the longer term, my thinking about many of the questions discussed in these notes has been influenced by Ntina Savvidou's research programme and by our countless discussions on the nature of time in physics, for which I am deeply grateful. I also thank Chris Isham, whose review article and subsequent lectures first introduced me to the problem of time when I was a master's student.

This work was supported by COST Action CA23115, ``Relativistic Quantum Information.''

\end{document}